\documentclass[reqno,titlepage]{amsart}

\usepackage[letterpaper,margin=1.1in]{geometry}
\usepackage{anyfontsize}
\usepackage[T1]{fontenc}

\usepackage[scaled]{helvet}
\usepackage{amsmath,amssymb,amsthm}

\usepackage{graphicx}

\usepackage[svgnames]{xcolor}
\usepackage{pgf}
\usepackage{tikz}

\usepackage{soul}
\usepackage{mathrsfs}
\usepackage{stmaryrd}
\SetSymbolFont{stmry}{bold}{U}{stmry}{m}{n}

\usepackage{todonotes}
\usepackage{layout}
\usepackage{mathtools}
\usepackage{paralist}
\usepackage{multirow}
\usepackage{comment}
\usepackage{rotating}

\usepackage{chngcntr}
\usepackage{apptools}

\AtAppendix{\counterwithin{equation}{section}}
\AtAppendix{\counterwithin{table}{section}}
\AtAppendix{\counterwithin{figure}{section}}
\AtAppendix{\counterwithin{assumption}{section}}
\AtAppendix{\counterwithin{lemma}{section}}
\AtAppendix{\counterwithin{theorem}{section}}

\newtheorem{lemma}{Lemma}

\newtheorem{theorem}{Theorem}

\newtheorem{assumption}{Assumption}

\renewcommand{\theassumption}{\Alph{assumption}}
\renewcommand{\thelemma}{\arabic{lemma}}

\usepackage{mathpazo}

\usepackage{natbib}

\usepackage[american]{babel}
\usepackage{csquotes}

\usepackage{xspace}
\usepackage{bm}
\usepackage{setspace}
\usepackage{accents}
\usepackage{enumitem}
\usepackage{footnote}
\usepackage{tabularx}
\usepackage{threeparttable}

\makesavenoteenv{tabular}
\makesavenoteenv{table}

\usepackage[colorlinks=true, linkcolor=Maroon, urlcolor=Maroon, citecolor=Maroon]{hyperref}

\usepackage{cleveref}
\crefname{figure}{figure}{figures}
\creflabelformat{figure}{#2#1#3}

\crefname{equation}{equation}{equations}
\crefrangeformat{equation}{(#3#1#4) to~(#5#2#6)}
\creflabelformat{equation}{(#2#1#3)}

\crefname{lemma}{lemma}{lemmas}
\creflabelformat{lemma}{#2#1#3}
\crefname{condition}{condition}{conditions}
\creflabelformat{condition}{#2#1#3}
\crefname{assumption}{assumption}{assumptions}
\creflabelformat{assumption}{#2#1#3}
\crefname{remark}{remark}{remarks}
\creflabelformat{remark}{#2#1#3}
\crefname{appendix}{appendix}{appendices}
\creflabelformat{appendix}{#2#1#3}

\renewcommand{\Pr}{\mathbb{P}}
\newcommand{\Var}{\mathbb{V}}
\newcommand{\Exp}{\mathbb{E}}
\newcommand{\convp}{\stackrel{p}{\rightarrow}}

\newcommand{\BASELINESTRETCH}{1.5}
\newcommand{\mailingaddressSJ}{Department of Economics, Pennsylvania State University, 619 Kern Graduate Building, University Park, PA 16802}
\newcommand{\mailingaddressSL}{Department of Economics, Columbia University, 1022 International Affairs Building, 420 West 118th Street, New York, NY 10027; 
Centre for Microdata Methods and Practice, Institute for Fiscal Studies, 7 Ridgmount Street, London WC1E 7AE, UK}
\newcommand{\emailaddress}[1]{\href{mailto:#1}{#1}}

\usepackage{subfig}
\newcommand{\one}{\mathbb{1}}
\newcommand{\sP}{\mathcal{P}}
\newcommand{\local}{\textrm{local}}
\newcommand{\mte}{\textrm{marginal}}
\newcommand{\ate}{\textrm{avg}}
\newcommand{\pr}{\textrm{pr}}

\newcommand\raiseT[2]{\raisebox{0.25ex}{$#1#2$}}
\newcommand\tr{{\mathpalette\raiseT{\intercal}}}

\usepackage{fancyhdr}
\allowdisplaybreaks

\begin{document}

\begin{titlepage}
\renewcommand{\thefootnote}{\fnsymbol{footnote}}
\title{Identifying the Effect of Persuasion}
\author[Jun and Lee]{Sung Jae Jun and Sokbae Lee}

\begin{center}
\Large\textsc{Identifying the Effect of Persuasion}\footnote{We would like to thank the editor, four anonymous referees, Eric Auerbach, Stefano DellaVigna, Leonard Goff, Marc Henry, Keisuke Hirano, Joel Horowitz, Charles Manski, Joris Pinkse, Imran Rasul, Roman Rivera, Myunghyun Song, David Yang, and seminar participants at Northwestern University, Rutgers University, Seoul National University and Vanderbilt University for helpful comments. This work was supported in part by the European Research Council (ERC-2014-CoG-646917-ROMIA) and by  the UK Economic and Social Research Council (ESRC) through research grant (ES/P008909/1) to the Centre for Microdata Methods and Practice.  This paper includes applications based on previously published articles, and we would like to thank the authors of these papers for making their data sets and replication files  available via journal archives or personal web pages. 
}
  \vspace{1ex}
\end{center}

\vspace{.5cm}
\begin{center}
	\begin{tabular}{ccc}
	 \large\textsc{Sung Jae Jun\footnote{\mailingaddressSJ. Email:\, \emailaddress{suj14@psu.edu}}}
	 &
	 &
	 \large\textsc{Sokbae Lee\footnote{\mailingaddressSL. Email:\, \emailaddress{sl3841@columbia.edu}}} \\
	 \small\textsc{Penn State Univ.}
	 &
	 &
	 \small\textsc{Columbia Univ. and IFS}		
	\end{tabular}
\end{center}

\begin{center}
\vspace{.7cm} November 30, 2022  \vspace{.7cm}
\end{center}

\noindent
\textbf{Abstract.}
This paper examines a commonly used measure of persuasion whose precise interpretation has been obscure in the literature. By using the potential outcome framework, we define the causal persuasion rate by a proper conditional probability of taking the action of interest with a persuasive message conditional on not taking the action without the message. We then formally study identification under empirically relevant data scenarios and show that the commonly adopted measure generally does not estimate, but often overstates, the causal rate of persuasion.  We discuss several new parameters of interest and provide practical methods for causal inference.

\vspace{.5cm}

\end{titlepage}
\renewcommand{\thefootnote}{\arabic{footnote}\hspace{0.1ex}}
\setcounter{footnote}{0}

\raggedbottom

\section{Introduction}\label{section:intro}
How effectively one can persuade one's audience has been of interest to ancient Greek philosophers in the Lyceum of Athens,\footnote{See \citet{sep-aristotle-rhetoric} for three technical means of persuasion in Aristotle's \textit{Rhetoric}.} early--modern English preachers in St Paul's Cathedral,\footnote{See \citet{kirby2008public} for historic details of the public persuasion at Paul's Cross, the open-air pulpit in St Paul's Cathedral in the 16th century.} and contemporary American news producers at Fox News in New York City.\footnote{\citet{dellavigna2007fox} and \citet{martin2017bias} measure the persuasive effects of slanted news using data on Fox News.}
Recently, economists have been endeavoring  to build theoretical models of persuasion \citep[e.g.,][]{kamenica2011,CDK,gentzkow2017,Prat,bergemann2017information} and to quantify empirically the extent to which persuasive efforts affect the behavior of consumers, voters, donors, and investors \citep[see][for a survey of the recent literature]{dellavigna2010persuasion}.

Since \citet[][DK hereafter]{dellavigna2007fox} proposed a measure of persuasion, it has been used and modified by many authors  \citep[e.g.,][]{EPZ,gentzkow2011effect,DEMPZ,bassi2016persuasion,martin2017bias,CY2019} to quantify the persuasive effects of informational treatment.
However, its precise interpretation has been obscure because of a lack of formal identification analysis. In fact, we show that the commonly used measure of persuasion does not estimate the causal rate of persuasion on any subpopulation in general. Therefore, it is misleading to call DK's measure the persuasion rate, although it is a common practice in the literature; instead, we will reserve the term of the persuasion rate for the population parameter that is properly defined by a conditional probability in the potential outcome framework. 

The flaw of DK's measure arises from failing to distinguish a local average treatment effect \citep[LATE; see][]{imbens1994late} from the the average treatment effect (ATE), where the difference between the two can be substantial for a heterogeneous population; specifically, DK's measure rescales a LATE with a factor that is relevant only for the ATE.  For instance, in DK's example, even if Fox News has a high persuasive effect among those who will watch the channel if and only if it is available through a local cable package, it may not be persuasive at all when other people such as Democrats or Democrat-leaning independents are all included.

Focusing on the case of binary outcomes, we analyze the problem of measuring persuasive effects of informational treatment through the lens of the potential outcome framework. Specifically, we define the persuasion rate by a proper conditional probability of the agent taking an action of interest with a persuasive message given that the agent does not take the action without the conveyed message. We then formally study identification under a few empirically relevant scenarios of data availability. Our analysis will articulate what DK's measure of persuasion estimates, why it is misleading, and how we can fix the problem.

 While the persuasion rate is concerned with the entire population (and hence related with the ATE), we also consider a local persuasion rate that focuses on the group of compliers. 
  Our identification analysis shows that DK's measure estimates neither a local nor an average persuasion rate.

Our identification analyses are based on a few empirically relevant scenarios of data availability. We do this because the problem of data availability is particularly important in the context of measuring persuasive effects. For instance, 
  individual-level partisan vote outcome data rarely exist due to confidentiality issues. Thus, the outcome variable is frequently measured only at an aggregate level. Also, it is not always the case to observe an actual exposure to a persuasive message in the same data set along with the outcome and instrument. Indeed, DK's analysis uses a micro dataset for the treatment and instrument and a separate  aggregate dataset  for the outcome and instrument.
  In order to address these challenges, we consider three different scenarios of data availability explicitly: given the instrument, 
\begin{inparaenum}[(i)]
	\item the outcome and treatment are jointly observed, 
	\item they are observed separately, or 
	\item the treatment is not observed at all. 
\end{inparaenum}
We obtain the sharp bounds on the persuasion rate for the entire population as well as other subpopulations of potential interest. Therefore, our work builds on the econometrics literature on  partial identification \citep[e.g.,][]{Manski:03,Manski:07,Tamer:10} as well as the literature on program evaluation  \citep[see][for surveys of the literature]{Heckman/Vytlacil:07,imbens2009recent}.  

The main findings of this paper can be summarized as follows. If there is no heterogeneity in the population, then DK's measure of persuasion estimates the rate of persuasion, provided that a simple monotonicity assumption is imposed; however, this case is an exception rather than the rule. Indeed, the rate of persuasion is only partially identified as an interval in general, where the sharp lower bound generally corresponds to DK's measure multiplied by the relative size of the complier group regardless of the specific data scenario.
Therefore, DK's measure is strictly larger than the lower bound whenever there is partial compliance. It is also remarkable that the data scenarios matter only for the sharp upper bound. Therefore, the value of observing the treatment and outcome jointly only lies in obtaining a potentially more informative upper bound on the persuasion rate.  We also investigate identification of the local persuasion rate (i.e.,\ the persuasion rate for the group of compliers) under the same three data scenarios. It is point-identified under the most favorable data scenario, but only partially identified under the other scenarios; even in the case of point identification, DK's measure generally differs from the local persuasion rate.  If a continuous instrument is available, then we can target a \emph{marginal persuasion rate} that is akin to the marginal treatment effect \citep[e.g.,][]{Heckman/Vytlacil:05}. Therefore, having a continuous instrument opens up the possibility of point identification of the persuasion rate for a policy-relevant population if the instrument is sufficiently rich.

In order to illustrate our findings, we discuss two empirical examples in the main text, while we provide a few more in the online appendices. First, we revisit \citet{CY2019}, where the interest is in the effect of Chinese students having access to uncensored media on behaviors, beliefs and attitudes; we use the same variables and setup as in the original paper for this exercise. Second, we analyze the voting behavior and newspaper readership using the data from \citet{gerber2009does}. Overall, we show that  DK's measure of persuasion tends to overstate the persuasive effects while masking underlying heterogeneity.

The remainder of the paper is organized as follows. In Section \ref{section:backgrounds}, we recall some backgrounds and specifics of DK's measure of persuasion. Here, we properly define the persuasion rate at the population level by using the potential outcome framework. In Section \ref{section:identification}, we discuss identification of the persuasion rate. In Section \ref{section:late}, we study the local and marginal versions of the persuasion rate. In Section \ref{section:takeaways}, we  discuss our recommendations on what to do in practice, including practical inferential issues,\footnote{Stata commands for estimation and inference based on this paper's identification results are publicly available at \url{https://github.com/persuasio}. Alternatively, they can be installed from within Stata by typing ``\texttt{ssc install persuasio}''. } and clarify the difference between a population version of DK's measure of persuasion and the persuasion rate. In Section \ref{sec:main:example}, we provide two empirical illustrations. We conclude in Section \ref{sec:conclusion}. The online appendices include additional results and examples, including an extension to non-binary outcomes, a detailed discussion about methods for inference, and all of the proofs.

\section{Background}\label{section:backgrounds}

It is helpful to recall DK as a prototypical example, where they study the effect of an exposure to Fox News on the probability of voting for a Republican presidential candidate.  Here, the informational treatment of interest is the viewership of the Fox News channel, and the persuasion rate of the media can be thought of as the proportion of the Fox News viewers who voted for a Republican candidate among those who would not have done so if they had not watched Fox News. It should be noted that the agents' decisions about whether to watch Fox News or not may be correlated with their political orientation. In order to address the endogeneity issue, DK use Fox News availability via local cable in the year of 2000 as an instrumental variable. 

In efforts to measure the persuasion rate as explained above, DK and \citet{dellavigna2010persuasion} propose the (infeasible) estimand $f$ defined as follows: for a binary outcome and by using DK's notation
\begin{equation}\label{def:f:DK}
f = \dfrac{y_\mathbb{T} - y_\mathbb{C}}{e_\mathbb{T} - e_\mathbb{C}}\cdot \dfrac{1}{1 - y_0},
\end{equation}
where $\mathbb{T}$ and $\mathbb{C}$ represent an instrument assignment status such as having Fox News available via local cable or not.\footnote{Therefore, $\mathbb{T}$ and $\mathbb{C}$, which appear to denote treatment and control groups, should be understood as the status of the intent to treat, not the actual treatment status.} Here, for $j\in \{ \mathbb{T}, \mathbb{C} \}$, $y_j$ is the share of group $j$ taking the action of interest (e.g.,\ voting for a Republican candidate), and $e_j$ is the share of group $j$ exposed to a persuasive message. Further, $y_0$ is the share of those who would take the action of interest without listening to the persuasive message.  So, $f$ is a rescaled version of the usual Wald statistic that estimates the LATE, where rescaling is apparently to obtain a ``rate'' that focuses on those who are to be persuaded.  As DK noted, $y_0$ involves a counterfactual that is often unobserved. For this reason, $f$ is generally not a feasible estimand, and DK propose using $y_\mathbb{C}$ in place of $y_0$ as an approximation, which yields a feasible estimand, say $\tilde f$. We will refer to $f$ or its feasible approximation $\tilde f$ by DK's measure of the persuasion rate.

Without a formal justification, it is common in the literature to interpret $f$ as a conditional probability.  For instance, DK explain, on page 1218 of their paper, ``\emph{The key parameter is $f$, the fraction of the audience that is convinced by Fox News to vote Republican.}'' Similar interpretations are prevalent in the literature, as the following quotations demonstrate:
\begin{quote}
	\citet[][p.645]{dellavigna2010persuasion}: \emph{Whenever possible, we report the results in terms of the persuasion rate (DellaVigna \& Kaplan 2007), which estimates the percentage of receivers that change the behavior among those that receive a message and are not already persuaded.} \\
	\citet[][p.3003]{gentzkow2011effect}: \emph{We can also translate our estimates into a “persuasion rate” (DellaVigna and Kaplan 2007) -- the number of eligible voters who changed their voting behavior as a result of the introduction of the newspaper, as a fraction of all those who could have changed their behavior.}
	\\
	\citet[][p.125]{DEMPZ}: \emph{The persuasion rate is the fraction of the audience of a media outlet who are convinced to change their behavior (in this case, their vote) as a result of being exposed to this media outlet.}
	\\ 
	\citet[][p.2590]{martin2017bias}: \emph{Finally, we computed estimates of DellaVigna and Kaplan's (2007) concept of persuasion rates: the success rate of the channels at converting votes from one party to the other. The numerator in the persuasion rate here is the number of, for example, the Fox News Channel (FNC) viewers who are initially Democrats but by the end of an election cycle change to supporting the Republican party. The denominator is the number of FNC viewers who are initially Democrats.}
\end{quote}
Also, in their survey, \citet{dellavigna2010persuasion} use $\tilde f$, i.e., a feasible approximation of $f$, as a key summary statistic to compare persuasive effects across different studies. 

However, as we mentioned in the introduction, neither $f$ nor $\tilde f$ estimates the persuasion rate, i.e., the intended conditional probability, on any subpopulation in general. Therefore, it is misleading to label $f$ or $\tilde f$ as a persuasion rate, and it is generally invalid to make comparisons across different studies. Indeed, $f$ or $\tilde f$ may not even be a proper conditional probability in a heterogeneous population: e.g., the approximation $\tilde f$ can even be larger than one.  We will articulate under what assumptions $\tilde f$ turns out to be a proper conditional probability, and how we should interpret it in its relationship with the persuasion rate.

In order to facilitate our discussion, we start with formally defining the persuasion rate at the population level by using the potential outcome framework. Let $T_i$ denote the binary indicator that equals $1$ if individual $i$ is exposed to persuasive information such as Fox News. Let $Y_i(t)$ be a binary indicator, which shows agent $i$'s potential action when $T_i$ is set to $t \in \{0,1\}$. For example, $Y_i(1)$ equals $1$ if individual $i$ votes for a Republican candidate after watching Fox News. The econometrician never observes both $Y_i(0)$ and $Y_i(1)$ but can only observe either of the two: that is,  $Y_i = T_iY_i(1) +(1-T_i) Y_i(0)$.\footnote{So, both $Y_i$ and $T_i$ are binary. In online \Cref{sec:nonbinary}, we extend our results to the case where the potential outcomes are multinomial.} Then, the fraction of the people who take the action of interest with an exposure to a persuasive message, among those who would not without it, is given by
\begin{equation}\label{eq:persuasion}
\theta_\pr := \Pr\{ Y_i(1) = 1 \mid Y_i(0) = 0 \},
\end{equation}
provided that the conditional probability is well-defined: $\theta_\pr$ is the persuasion rate at the population level. 
Using conditional probability is to rule out the case of ``preaching to the converted'';\ if $Y_i(0) = 1$, then those individuals are already ``persuaded'' to take the action of interest even without the persuasive treatment, and therefore we do not count them in defining the persuasion rate.\footnote{The idea of using conditional probability to define a parameter of interest can also be found in \citet{heckman1997making}, though their context is quite different from ours.}

The common estimand $\tilde f$ (or even $f$) does not estimate $\theta_\pr$ in a heterogeneous population; in fact, even in a homogeneous population, $\tilde f$ or $f$ cannot be (asymptotically) equated with $\theta_\pr$ without an extra monotonicity assumption. However, the rescaled quantity $(e_\mathbb{T} - e_\mathbb{C} ) \tilde f $, which is always no greater than $\tilde f$, does provide valid information about $\theta_\pr$ in that it corresponds to the sharp lower bound of the identified interval of $\theta_\pr$ in general. The best way to clarify all the issues is to conduct a rigorous analysis on the identification of $\theta_\pr$, which is our next topic. For quick takeaways, see \Cref{section:takeaways}.

\section{Identification of the Persuasion Rate}\label{section:identification}

Identification of $\theta_\pr$ is challenging for various reasons, including the fact that $\theta_\pr$ depends on the joint distribution of the potential outcomes and that $T_i$ can be endogenous, and it is often difficult to observe $T_i$ and $Y_i$ jointly. To allow for potential endogeneity, we use a binary instrument, $Z_i$, throughout the paper unless otherwise specified. Exogenous covariates, $X_i$, could be observed, but we suppress $X_i$ from our identification analysis. In other words, we implicitly assume throughout the paper that all assumptions and results are conditional on the value of $X_i$. Therefore, the observed variables are $Y_i, T_i, Z_i$, all of which are binary throughout the main text. See \Cref{sec:media} for how to deal with $X_i$ in practice. Also, see \Cref{sec:nonbinary} for an extension to the case of multinomial outcomes. 

The data issue on $T_i$ is addressed by considering three scenarios of data availability. Specifically, for the purpose of the identification analysis, we assume that for $(y,t,z)\in \{0,1\}^3$, (i) $\Pr(Y_i = y, T_i = t\mid Z_i=z)$ is known, (ii) $\Pr(Y_i = y\mid Z_i=z)$ and $\Pr(T_i = t\mid Z_i=z)$ are known, or (iii) $\Pr(Y_i = y\mid Z_i=z)$ is all that is known, depending on the specific scenario of interest. For example, DK use town-level election data to estimate $\Pr(Y_i = y \mid Z_i=z)$ and micro-level media audience data to infer $\Pr(T_i = t\mid Z_i=z)$, which corresponds to Case (ii).\footnote{Throughout the discussion, we  assume that  $T_i$ is correctly measured if it is observed. See \citet{latewithmismeasurement}, \citet{nguimkeu2016estimation}, and  \citet{Ura2018miclassification} for the issues of mismeasured treatment. Their subject matter is distinct from ours.} 

It requires an additional assumption to address the challenge that $Y_i(1)$ and $Y_i(0)$ are never simultaneously observed while $\theta_\pr$ depends on their joint distribution. Before we present our identification results, we discuss our key assumptions in the following subsection.

\subsection{The Key Assumptions}

Our first key assumption is that the persuasive message is directional, which will be important to decouple $\theta_\pr$ by the marginals of the potential outcomes.  

\begin{assumption}[Monotonic Treatment Response]\label{ass:binary-monotone}
	The potential outcomes $Y_i(1)$ and $Y_i(0)$ are binary, and they satisfy $Y_i(0) \leq Y_i(1)$ with probability one.
\end{assumption}

\Cref{ass:binary-monotone} is a binary version of the monotonic treatment response (MTR) assumption used in \citet{manski1997monotone} and \citet{manski2000mtr}.\footnote{In online Appendix A, we present a simple economic model that motivates \Cref{ass:binary-monotone}.} \Cref{ass:binary-monotone} allows $Y_i(0) = Y_i(1)$ with probability one, and therefore it does not rule out the possibility that `watching Fox News' has no impact on the agent's behavior at all. The inequality in \Cref{ass:binary-monotone} means that the messages Fox News delivers are \emph{biased or directional} in favor of Republican candidates; i.e.,\ if a voter is going to vote for a Republican candidate without watching Fox News, then watching Fox News will not change that. In other words, \Cref{ass:binary-monotone} rules out the possibility that the level of distrust that a voter has on Fox News is so high that she takes actions based on the opposite of the messages Fox News delivers.

Since \cref{ass:binary-monotone} is a key assumption in the paper, we first clarify how much we can hope for with and without \Cref{ass:binary-monotone}.

\begin{lemma}\label{lem:direction}
	We generally have
	\begin{equation}\label{lemma1-result0}
	\max\Bigr[ 0,\ \dfrac{ \Pr\{ Y_i(1) = 1 \} - \Pr\{ Y_i(0) = 1 \} }{ 1 - \Pr\{ Y_i(0) = 1\} } \Bigr]
	\leq \theta_\pr \leq
	\min\Bigl[ \dfrac{\Pr\{ Y_i(1) = 1 \}}{1-\Pr\{Y_i(0)=1\}}  ,\ 1 \Bigr],
	\end{equation}
	where the bounds are sharp in that $\theta_\pr$ can be anything between the bounds without changing the marginals $\Pr\{ Y_i(1) = 1\}$ and $\Pr\{ Y_i(0) = 1\}$.	Further, \Cref{ass:binary-monotone} holds if and only if $\theta_\pr = \theta_\ate$, where
	\begin{equation}\label{lemma1-result}
		\theta_\ate := \dfrac{ \Pr\{ Y_i(1) = 1 \} - \Pr\{ Y_i(0) = 1 \} }{ 1 - \Pr\{ Y_i(0) = 1\} }.
	\end{equation} 
\end{lemma}

\Cref{lemma1-result0} is a consequence of the Fr\'{e}chet--Hoeffding inequality on the probability of a joint event. Since the two potential outcomes are never observed simultaneously, all we can ever hope to identify is their marginals and \Cref{lemma1-result0} expresses the (sharp) bounds on $\theta_\pr$ in terms of the marginal probabilities of the potential outcomes. 

Suppose that there are some voters who have such a high level of distrust on Fox News that rather not watching Fox News would help them to take favorable actions to a Republican candidate but that those voters are only minority and on average we still have a strict stochastic dominance relationship between $Y_i(1)$ and $Y_i(0)$ (i.e., $\Pr\{ Y_i(1) = 1 \} > \Pr\{ Y_i(0) = 1 \}$). Then, the lower bound on $\theta_\pr$ will be ensured to be nontrivial. 

\Cref{ass:binary-monotone} is stronger than the stochastic dominance, but it delivers a stronger result. In fact, \Cref{ass:binary-monotone} is necessary and sufficient to express $\theta_\pr$ in terms of the marginal probabilities of the counterfactual outcomes. In this case, the conditional probability $\theta_\pr$ is the average treatment effect (ATE) divided by $\Pr\{  Y_i(0) = 0\}$.  Throughout the rest of the paper, we present most of our results by using \Cref{ass:binary-monotone} because it is not only convenient but also being biased or directional seems to be the nature of persuasive effort. However, we emphasize that $\theta_\ate$, the rescaled version of the ATE, is always a valid lower bound on $\theta_\pr$ as \Cref{lemma1-result0} shows.

The next assumption is concerned about the treatment assignment $T_i$ and the instrument $Z_i$. 

\begin{assumption}[No Defiers and an Exogenous Instrument]\label{ass:instrument}
	 The binary treatment $T_i$ has a threshold structure, i.e.,
	 \begin{equation} \label{eq:threshold}
	 	T_i = \one\{ V_i \leq e(Z_i) \},
	 \end{equation}
	 where $V_i$ is an unobserved random variable that is uniformly distributed on $[0,1]$. The binary instrument $Z_i$ is independent of $\bigl(Y_i(t), V_i \bigr)$ for $t= 0,1$. Finally, we have $0\leq e(0) < e(1) \leq 1$. 
\end{assumption}

\Cref{ass:instrument} is standard for causal inference using instrumental variables. The intent-to-treat (ITT), $Z_i$, is randomly assigned; however,  $T_i$ can be endogenous via the dependence between $V_i$ and $Y_i(t)$. The function $e(\cdot)$ is the propensity score or, more descriptively in our context, it can be referred to as the \emph{exposure rate}. 

If $T_i(z)$ denotes counterfactual treatment when a binary instrument takes value $z$, then agent $i$ is called a complier when $T_i(1) = 1$ and $T_i(0)=0$; never-takers ($T_i(z) = 0$ for all $z$), always-takers ($T_i(z)=1$ for all $z$), and defiers ($T_i(1) = 0, T_i(0) = 1$) are similarly understood. Under \Cref{ass:instrument}, $i$ is a complier if and only if $e(0)<V_i\leq e(1)$. Similarly, $i$ is an always-taker when $V_i\leq e(0)$, and she is a never-taker when $V_i> e(1)$. Therefore, under \Cref{ass:instrument}, there are only three groups in the population: i.e.,\ always-takers, never-takers and compliers. Indeed, as \citet{vytlacil2002independence} has shown, the threshold structure in \Cref{eq:threshold} is equivalent to assuming the absence of defiers, which is a popular assumption in econometrics to identify the local average treatment effect.

In the following two subsections, we present identification results for $\theta_\pr$, which is the same as $\theta_\ate$ under \cref{ass:binary-monotone}. \Cref{section:sharp} covers the simplest case, where everybody complies so that there is no difference between the actual treatment and the intent to treat (ITT), i.e., $T_i = Z_i$ for each $i$: this case is referred to as the \emph{sharp persuasion design}, where there is no distinction among different data scenarios. In this case, not surprisingly, $\theta_\ate$ is point identified from the distribution of $Y_i$ given $Z_i$. However, when $T_i$ and $Z_i$ are different, which we call the \emph{fuzzy persuasion design}, we have only partial identification of $\theta_\ate$, where each of the three data scenarios become relevant. Before we move on, we define
\begin{equation}\label{eq:thetaL}
	\theta_L
	:=
	\dfrac{\Pr(Y_i = 1\mid Z_i = 1)-\Pr(Y_i = 1\mid Z_i = 0)}{1-\Pr(Y_i = 1\mid Z_i = 0)},
\end{equation}
provided that $\Pr(Y_i = 1\mid Z_i=0)<1$: $\theta_L$ is an identified parameter from the distribution of $Y_i$ given $Z_i$. It turns out that first, $\theta_L$ is equal to $\theta_\ate$ under the sharp persuasion design, and second, it is the sharp lower bound on $\theta_\ate$ in the fuzzy persuasion design regardless of which of the three data scenarios applies.

\subsection{The Sharp Persuasion Design} \label{section:sharp}


\begin{theorem}\label{thm:sharp}
	Suppose that \Cref{ass:binary-monotone,ass:instrument} hold. If $e(1) - e(0) = 1$ (i.e.,\ $T_i = Z_i$ with probability one), then for $z\in\{0,1\}$, we have $\Pr\{ Y_i(z) = 1\} = \Pr( Y_i = 1 \mid Z_i = z)$, and hence $\theta_\pr = \theta_\ate = \theta_L$.
\end{theorem}

The condition of $e(1) - e(0) = 1$ means that everybody is a complier, and hence there is essentially no difference between $T_i$ and $Z_i$; thus, the sharp design is equivalent to a situation where $T_i$ is observed and randomized. However, this is rather an exceptional situation in social sciences. The key identification question should be how far we can go when the design is not sharp (i.e.,\ not everybody is a complier). We answer this question in the following subsection.

Without \Cref{ass:binary-monotone}, the general sharp identified bounds on $\theta_\pr$ are given by
\begin{equation}
\max\{0,\ \theta_L\}\leq \theta_\pr \leq \min\Bigl\{ \dfrac{\Pr(Y_i=1 \mid Z_i = 1)}{1-\Pr(Y_i = 1 \mid Z_i = 0)},\ 1 \Bigr\}.
\end{equation}
Therefore, even in the sharp persuasion design, $\theta_\pr$ is only partially identified and its sharp bounds can be trivial without \Cref{ass:binary-monotone}. Even so, it is worth noting that $\theta_L$ remains a valid lower bound.

\subsection{The Fuzzy Persuasion Design}\label{section:fuzzy}

In the fuzzy design, the three scenarios of data availability we mentioned earlier become pertinent.

\subsubsection{Identification with the Joint Distribution of $(Y_i, T_i, Z_i)$}
Even the full joint distribution of $(Y_i,T_i,Z_i)$ does not point-identify the ATE. Recall that those with $Z_i = 0$ and $T_i =0$ comprise the compliers and the never-takers, while those with $Z_i = 0$ and $T_i = 1$ are the always-takers. Similarly, those with $Z_i = 1$ and  $T_i = 0$ are the never-takers, while those with $Z_i = 1$ and $T_i=1$ consist of the compliers and the always-takers. Therefore, these four cases correspond to different subpopulations, and the only subpopulation that we can study for both $T_i=0$ and $T_i = 1$ in common is that of compliers, which explains why the Wald statistic estimates the LATE, not ATE. For the same reason, $\theta_\ate$ cannot be point identified; however, we can derive its sharp bounds.

\begin{assumption}[Full Observability] \label{ass:fuzzy-best}
	The joint distribution of $(Y_i, T_i, Z_i)$ is known, where $\Pr(Y_i = 0 \mid Z_i = 0 ) > 0$.
\end{assumption}

\begin{theorem}\label{thm:fuzzy-best}
	Suppose that \Cref{ass:binary-monotone,ass:instrument,ass:fuzzy-best} are satisfied. Then, the sharp identified interval of $\theta_\pr = \theta_\ate$ is given by $[\theta_L, \theta_U]$, where $\theta_L$ is given in \Cref{eq:thetaL} and
	\begin{equation*}
		\theta_U := \dfrac{\Pr(Y_i = 1, T_i = 1\mid Z_i = 1)-\Pr(Y_i = 1, T_i = 0\mid Z_i = 0) + 1-e(1)}{1-\Pr(Y_i = 1, T_i = 0 \mid Z_i = 0)}.
	\end{equation*}	
\end{theorem}

To prove \Cref{thm:fuzzy-best}, we first derive the sharp identified bounds for  $\Pr\{Y_i(1) = 1 \}$ and $\Pr\{Y_i(0) = 1 \}$ separately; we denote them by the intervals $[m_a, M_a]$ and $[m_b,M_b]$, respectively.  These bounds are special cases of  \citet{manski2000mtr} under the MTR assumption coupled with the exogeneity of the instrument.  Then, letting $a := \Pr\{ Y_i(1) = 1 \}$ and $b := \Pr\{ Y_i(0) = 1 \}$, we obtain the upper bound of the identified interval of $\theta_\ate$ by solving
\begin{equation}\label{LP-bound-eq}
	\max_{a,b}\quad  
	\dfrac{a-b}{1-b}
	\quad\text{subject to}\quad
	a \in [m_a,\ M_a],\
	b \in [m_b,\ M_b],\
	a\geq b,
\end{equation}
while the lower bound can be found by doing minimization instead of maximization. We then appeal to continuity and the intermediate value theorem for the sharpness result.  

It is proved in online \Cref{app:proofs} that $m_a = \Pr(Y_i=1\mid Z_i = 1)$ and $M_b = \Pr(Y_i=1\mid Z_i = 0)$. Therefore, an examination of \eqref{LP-bound-eq} reveals that the lower bound is attained when $a = m_a$ and  $b = M_b$. To develop intuition behind \Cref{thm:fuzzy-best}, we discuss what $a=m_a$ and $b = M_b$ means, for which the behavior of the never-takers and that of the always-takers matter. Since $m_a = \Pr\{Y_i(1) = 1, T_i = 1 \mid Z_i=1\} + \Pr\{Y_i(0) = 1, T_i=0 \mid Z_i=1\}$, we know that $a = m_a$ holds when $\Pr\{ Y_i(0)=1, T_i = 0\mid Z_i = 1\} = \Pr\{ Y_i(1)=1, T_i = 0\mid Z_i = 1\}$. Here, the event $Z_i = 1, T_i=0$ means that $i$ is a never-taker, because defiers are assumed to be non-existent. Therefore, $a=m_a$ means that the treatment has no effect on the group of never-takers, unless there are no never-takers at all. Similarly, $b = M_b$ holds when $\Pr\{Y_i(1)=1,T_i=1 \mid Z_i=0 \} = \Pr\{ Y_i(0)=1,T_i=1 \mid Z_i = 0 \}$. Since $Z_i = 0, T_i = 1$ means that $i$ is an always-taker, we know that $b=M_b$ holds when the treatment does not affect the behavior of the always-takers, unless there are no always-takers at all.  Therefore, $\theta_\ate$, which is the same the persuasion rate $\theta_\pr$ under \cref{ass:binary-monotone}, is smallest when there are null treatment effects for both the never-takers and the always-takers.

Intuition for the upper bound can also be obtained by considering the non-complier groups. The upper bound corresponds to the case where $a = M_a$ and $b = m_b$, where it is shown in \Cref{app:proofs} that $M_a = \Pr(Y_i = 1, T_i=1 \mid Z_i=1)+1-e(1)$ and $m_b = \Pr(Y_i=1,T_i=0 \mid Z_i=0)$. Here, note that $a=M_a$ is equivalent to $\Pr\{Y_i(1)=0,T_i=0 \mid Z_i=1 \} =0$, and $b=m_b$ is to $\Pr\{ Y_i(0) = 1, T_i = 1 \mid Z_i=0\} = 0$. Therefore, we can see that the persuasion rate $\theta_\ate$ equals the upper bound when every never-taker has $Y_i(1) = 1$ and none of the always-taker has $Y_i(0)=1$: e.g.,\ all those who never watch Fox News (whether it is available or not) would actually have voted for a Republican candidate if they had watched it and all those who always watch Fox News would not have voted for a Republican without watching the channel.

The bounds in \Cref{thm:fuzzy-best} shrink to a singleton as $\bigl( e(0), e(1) \bigr)$ approaches $(0,1)$, which is consistent with the result in \Cref{thm:sharp}. Also, it is worth noting that the lower bound $\theta_L$  only depends on the distribution of $(Y_i, Z_i)$: observing $T_i$ along with $(Y_i, Z_i)$ helps only for the upper bound.  If $e(1)$ is too small, then the upper bound will not be very informative: $\theta_U$ converges to $1$ as $e(1)$ approaches $0$; that is,  if nobody reads a newspaper when they receive free subscriptions, then we do not learn much about how ``persuading'' the newspaper is.  However, even if $e(1)$ approaches  $1$, the upper bound does not necessarily shrink to the lower bound; e.g.,\ we do not necessarily pin down the persuasion rate of reading the newspaper even if everybody who has free subscriptions actually reads it.

We now establish partial identification of $\theta_\pr$ without \Cref{ass:binary-monotone}, i.e., $\theta_\pr \neq \theta_\ate$, for the sake of completeness. Let $NT = \{V_i > e(1)\}$ and $AT = \{ V_i\leq e(0)\}$ be the event of $i$ being a never-taker and an always-taker, respectively.

\begin{theorem}\label{thm:fuzzy-best-wo-mono}
	Suppose that \Cref{ass:instrument,ass:fuzzy-best} are satisfied. 
	\begin{enumerate}[label=(\roman*)]
		\item\label{part1-non-MTR} 	
		If $\Pr(Y_i = 0\mid Z_i = z)>0$ for $z=0,1$, then the sharp identified interval of $\theta_\pr$ is given by
	\begin{multline*}
	\max\Bigl\{0,\ \frac{\Pr(Y_i=1,T_i=1\mid Z_i=1)-\Pr(Y_i=1,T_i=0\mid Z_i=0)-e(0)}{1-\Pr(Y_i=1,T_i=0\mid Z_i=0)-e(0)} \Bigr\} \\
    \leq
	\theta_\pr 
	\leq
	\min\Bigl\{  \frac{\Pr(Y=1,T=1\mid Z=1) +1-e(1)}{1-\Pr(Y=1,T=0\mid Z=0)},\ 1 \Bigr\}. 
    \end{multline*}
    
\item\label{part2-non-MTR}     
If $\Pr(Y_i = 0\mid Z_i = 0)>0$ and
$\Pr\{ Y_i(1)=1,\ \mathcal{E} \} \geq \Pr\{ Y_i(0)= 1,\ \mathcal{E}\}$ for $\mathcal{E} \in \{ NT, AT\}$, then the sharp identified interval of $\theta_\pr$ is given by 
\[
\max\{ 0,\ \theta_L\} \leq \theta_\pr \leq     	\min\Bigl\{  \frac{\Pr(Y=1,T=1\mid Z=1) +1-e(1)}{1-\Pr(Y=1,T=0\mid Z=0)},\ 1 \Bigr\}.
\]
\end{enumerate}
\end{theorem}

\Cref{thm:fuzzy-best-wo-mono} shows that $\theta_L$  continues to be the sharp lower bound, provided that $\theta_L \geq 0$, i.e., $\Pr(Y_i = 1\mid Z_i = 1) \geq \Pr(Y_i = 1\mid Z_i = 0)$, and a stochastic dominance condition holds for the never-takers and always-takers. However, the upper bound will be larger than that in \Cref{thm:fuzzy-best} in general.

\subsubsection{Identification with the Knowledge of the Exposure Rates}
As in the case of DK, the researcher may not directly observe $T_i$ along with $(Y_i,Z_i)$ but may have auxiliary data from which the exposure rates $e(1)$ and $e(0)$ can be estimated.\footnote{The case in which the outcome and the treatment are separately observed belongs to an identification problem called the \textit{ecological inference problem}. For instance,  \citet{CrossManski} and \citet{Manski2017longshort} discuss bounding a ``long regression'' by using information from a ``short regression''. Their substantive concerns are distinct from ours.} In this case, the sharp identified bounds on $\theta$ become generally wider than those of \Cref{thm:fuzzy-best}.

\begin{assumption}[Observability of Two Marginals]\label{ass:fuzzy-e}
	Only the distribution of $(Y_i, Z_i)$ and the exposure rates $\{ e(0), e(1) \}$ are known, where $\Pr(Y_i = 0 \mid Z_i = 0) > 0$. 
\end{assumption}

\begin{theorem}\label{thm:fuzzy-e}
	Suppose that \Cref{ass:binary-monotone,ass:instrument,ass:fuzzy-e} are satisfied. Then, the sharp identified interval of $\theta_\pr = \theta_\ate$ is given by $[\theta_L,\ \theta_{U_e}]$, where $\theta_L$ is given in \Cref{eq:thetaL} and
	\begin{equation}\label{upper-bound-Ue}
		\theta_{U_e}
		:=
		\dfrac{ \min\{ 1, \Pr(Y_i = 1\mid Z_i = 1)+1-e(1) \} - \max\{0,\ \Pr(Y_i = 1\mid Z_i = 0)-e(0) \}}{1 - \max\{0,\ \Pr(Y_i = 1\mid Z_i = 0)-e(0) \}}.
	\end{equation}
\end{theorem}

Therefore, the upper bound in this case is nontrivial if and only if $e(1) > \Pr(Y_i = 1 \mid Z_i = 1)$.\footnote{The trivial case can occur in applications. See \Cref{tb:turnout-results} for such cases.} Note that it is the relative size of the take-up rate $e(1)$ (e.g.,\ the probability of reading a newspaper when a free  subscription to it is offered) that determines how much we can hope to learn about the persuasion rate.  For example, if the probability of watching Fox News is too small relative to the probability of voting for a Republican candidate when Fox News was introduced in the local cable, then it becomes difficult to pin down how successfully Fox News persuaded their audience to vote for a Republican candidate. Also, it is worth noting that $e(0) = 0$ is not uncommon as \Cref{tb:turnout-results} and \Cref{sec:main:example} show. In this case, the maximum in the expression of the upper bound is unnecessary. Intuition for the lower bound is the same as the case of \Cref{thm:fuzzy-best}, because $\theta_L$ requires only the distribution of $Y_i$ given $Z_i$.

\subsubsection{Identification with No Information Associated with $T_i$}

The final scenario is the least informative one, where $T_i$ is not observed at all. This is an almost trivial case, but we state it in a separate theorem for the sake of completeness.

\begin{assumption}[Limited Observability] \label{ass:fuzzy-worst}
	No information associated with $T_i$ is available (i.e.,\ the distribution of $(Y_i, Z_i)$ is all that is known), where $\Pr(Y_i = 0\mid Z_i = 0) > 0$. 
\end{assumption}

\begin{theorem}\label{thm:fuzzy-worst}
	Suppose that \Cref{ass:binary-monotone,ass:instrument,ass:fuzzy-worst} are satisfied. Then, the sharp bound of $\theta_\pr = \theta_\ate$ is given by $[\theta_L,1]$, where $\theta_L$ is given in \Cref{eq:thetaL}.
\end{theorem}
The lower bound from \Cref{thm:fuzzy-e} depends only on the distribution of $(Y_i,Z_i)$, and therefore $\theta_L$ continues to be the lower bound in this case as well. Further, since no information for $e(1)$ and $e(0)$ is available, it suffices to note that the upper bound in \Cref{thm:fuzzy-e} equals one whenever $e(0)>\Pr(Y_i=1\mid Z_i=0)$ and $e(1)<\Pr(Y_i=1\mid Z_i=1)$.

\section{The Local and Marginal Persuasion Rates} \label{section:late}

In this section, we consider rates of persuasion on other subpopulations that have been considered in econometrics: i.e.,\ 
\begin{equation} \label{eq:local and mte}
\left\{
\begin{aligned}
\theta_\local &:= \Pr\{ Y_i(1) = 1 \mid Y_i(0) = 0, e(0) < V_i \leq e(1) \},  \\
\theta_\mte(v) &:= \Pr\{ Y_i(1) = 1 \mid Y_i(0) = 0, V_i = v \} \; \text{ for $0 < v < 1$},
\end{aligned}
\right.
\end{equation}
provided that the conditional probabilities are well-defined: $\theta_\local$ is the persuasion rate for the compliers \citep[e.g.,][]{imbens1994late}, whereas $\theta_\mte(v)$ is for the subpopulation such that $V_i = v$ \citep[e.g.,][]{Heckman/Vytlacil:05}.

First, we obtain identification results for $\theta_\local$ under the three sampling scenarios in the fuzzy persuasion design. The first step for this purpose is to note that the same reasoning as \Cref{lem:direction} yields
\begin{equation}\label{eq:local2}
	\theta_\local
	=
	\frac{\Pr\{ Y_i(1) = 1 \mid  e(0) < V_i \leq e(1) \} - \Pr\{ Y_i(0) = 1 \mid  e(0) < V_i \leq e(1) \}}{1 - \Pr( Y_i(0) = 1 \mid  e(0) < V_i \leq e(1) \} },
\end{equation}
where the numerator is the LATE, which has received great attention in the econometrics literature \citep[see][for a debate]{Deaton2010, Heckman2010, Imbens2010}. The denominator that rescales the LATE is also conditioned on the same subpopulation of the compliers. 

\begin{theorem}\label{thm:local}
	Suppose that \Cref{ass:binary-monotone,ass:instrument} are satisfied.
	\begin{enumerate}[label=(\roman*)]
		\item\label{part1-local} Under \Cref{ass:fuzzy-best}, $\theta_\local$ is point identified by $\theta_\local = \theta^*$, where
		\[
		\theta^* := \dfrac{\Pr(Y_i = 1 \mid Z_i = 1) - \Pr(Y_i = 1 \mid Z_i  = 0)}{\Pr(Y_i = 0, T_i = 0 \mid Z_i = 0) - \Pr( Y_i = 0, T_i = 0 \mid Z_i = 1)}.
		\]
		\item\label{part2-local} Under \Cref{ass:fuzzy-e}, the sharp identified interval of $\theta_\local$ is given by $[\theta_L^*, 1]$, where
		\[
		\theta_L^*
		:=
		\max\Bigl\{ \theta_L,\
		\dfrac{\Pr(Y_i = 1 \mid Z_i = 1) - \Pr(Y_i = 1\mid Z_i  = 0)}{e(1)-e(0)}
		\Bigr\}.
		\]
		\item\label{part3-local} Under \Cref{ass:fuzzy-worst}, the sharp identified interval of $\theta_\local$ coincides with that of $\theta_\pr=\theta_\ate$, i.e.,\ $\bigl[ \theta_L,\ 1 \bigr]$.
	\end{enumerate}
\end{theorem}
Recall that the identification of the LATE requires the joint distribution of $(T_i, Z_i)$ and that of $(Y_i, Z_i)$ separately but not the full joint distribution of $(Y_i, T_i, Z_i)$. Unlike the LATE,  the point identification in  \Cref{thm:local}\ref{part1-local} demands the knowledge of the joint  distribution of $(Y_i, T_i, Z_i)$.\footnote{The denominator of \Cref{eq:local2} requires that we know the marginal distribution of $Y_i(0)$ for the compliers. \citet{imbens1997estimating} show that the marginal distributions of $Y_i(1)$ and $Y_i(0)$ for the compliers are identified if the joint distribution of $(Y_i, T_i, Z_i)$ is known; however, they did not consider the local persuasion rate.}
\Cref{thm:local}\ref{part2-local} shows that this requirement is not only sufficient but also necessary to achieve the point identification of $\theta_\local$.

Just like the LATE, it may be contentious whether or not $\theta_\local$ should be the parameter of interest, because the compliers are concerned with  an unidentified subgroup of the  population. However, we take a practical view that the identification results on $\theta_\local$ can complement the results obtained in \Cref{section:identification}. 

The local persuasion rate $\theta_\local$ represents the average persuasive effect for a population that is different from the entire population. Given this caveat, it is interesting to note that, in \Cref{thm:local}\ref{part2-local}, the upper bound on $\theta_\local$ is always trivial in contrast to $\theta_\ate$, but the lower bound of $\theta_\local$ can never be worse than that of $\theta_\ate$.  Therefore, in principle, the length of the identified interval of $\theta_\ate$ can be smaller than that of $\theta_\local$. If $T_i$ is not observed at all, then there is no advantage in focusing on the compliers. \Cref{thm:local}\ref{part3-local} confirms the intuition that the bounds for $\theta_\local$ are identical to those for $\theta_\ate$ if the distribution of $(Y_i, Z_i)$ is the only piece of information available. This corresponds to an uninteresting case for $\theta_\local$ though, as we have no information on compliers.

Data requirements for the identification of $\theta_\mte(v)$ are generally quite demanding: e.g.,\ a continuous instrument is needed. However, identifying $\theta_\mte(v)$ for various values of $v$ can open up the possibility of point identification of $\theta_\ate$. Therefore, it is worth understanding what is sufficient for the identification of $\theta_\mte$.   

If $Y_i$ and $T_i$ are jointly observed along with a \emph{continuous} instrument $Z_i$, then $\theta_\mte(v)$ can be point identified as in \citet{Heckman/Vytlacil:05} and \citet{carneiro2011}. Examples of continuous instruments can be found in the literature on the media effects on voting. For instance, \citet{EPZ} and \citet{DEMPZ} use the signal strength of NTV and Serbian radio as instruments, respectively; in both of the papers, $(Y_i, T_i, Z_i)$ are jointly observed. The following assumption describes the situation in which we can obtain point identification of $\theta_\mte(v)$.  We use the standard results in the literature \citep[e.g.][]{Heckman/Vytlacil:05} for the subsequent theorem.

\begin{assumption}[Marginal Treatment Effects]\label{ass:mte}
	\begin{enumerate}[label=(\roman*)]
		\item The joint distribution of $(Y_i, T_i, Z_i)$ is known.
		\item $T_i$ has the threshold structure in \Cref{eq:threshold}, where $V_i$ is uniformly distributed on $[0,1]$, and $Z_i$ is independent of $\bigl(Y_i(t), V_i \bigr)$ for $t= 0,1$.
		\item The distribution of $e(Z_i)$ is absolutely continuous with respect to Lebesgue measure.
	\end{enumerate}
\end{assumption}

\begin{theorem}\label{thm:mte}
	Suppose that \Cref{ass:binary-monotone,ass:mte} are satisfied. Then, for $v$ such that $v$ is in the interior of the support of $e(Z_i)$,
	 $\theta_\mte(v)$ is point identified by
	\begin{align}\label{mte-formula}
		\theta_\mte(v) = \dfrac{\partial \Pr \{ Y_i = 1 \mid e(Z_i) = e \}/\partial e \big|_{e=v}}{1 + \partial \Pr\{ Y_i = 1, T_i = 0 \mid e(Z_i) = e \}/\partial e \big|_{e=v}},
	\end{align}
	provided that $\Pr \{ Y_i = 1\mid e(Z_i) = e \}$ and $\Pr \{ Y_i = 1, T_i = 0 \mid e(Z_i) = e \}$ are continuously differentiable with respect to $e$.
\end{theorem}
Similarly to the case of $\theta_\ate$ or $\theta_\local$, \Cref{ass:binary-monotone} enable us to rewrite $\theta_\mte(v)$ as $\Exp\{ Y_i(1) - Y_i(0) \mid V_i = v \} / \Pr\{Y_i(0) = 0\mid V_i = v\}$; i.e.,\ $\theta_\mte(v)$ is a rescaled version of the marginal treatment effect of \citet{Heckman/Vytlacil:05}. \Cref{thm:mte} is a direct consequence of that.  

\Cref{thm:mte} does not consider the other two scenarios of data availability. This is mainly because  continuous instruments are relatively infrequent in the context of persuasion, and we are not aware of any applications where continuous instruments are available while the outcome and treatment are not jointly observed.

If the support of the exposure rate $e(Z_i)$ is equal to the unit interval $[0,1]$, then \Cref{thm:mte} shows the identification of $\theta_\mte(v)$ for all $v$ in the unit interval. Then, we can use $\theta_\mte(v)$ to construct different policy-oriented quantities as in \citet{Heckman/Vytlacil:05} and \citet{carneiro2011}. For instance, the persuasion rate of the entire population can be obtained by $\int_0^1 \theta_\mte(v) dF\{v\mid Y_i(0) = 0\}$, which is equal to 
\begin{equation*}
\theta_\ate
= 
\frac{\int_0^1 \partial\Pr(Y_i = 1\mid e(Z_i)=e )/\partial e\big|_{e=v} dv}{1 + \int_0^1 \partial \Pr\{ Y_i = 1, T_i = 0\mid e(Z_i) = e \} / \partial e\big|_{e=v} dv}
\end{equation*}
by Bayes' theorem.

\section{Discussion}\label{section:takeaways}
In this section we articulate the relationships between $\theta_\ate, \theta_\local$, and DK's measures $f$ and $\tilde f$ defined in \Cref{section:backgrounds}, and we summarize the main takeaways of our identification results. We assume that \Cref{ass:binary-monotone,ass:instrument} hold throughout this section, so we have $\theta_\pr = \theta_\ate$. Also, in order to be consistent with the identification analysis, we work with the population versions of $f$ and $\tilde f$: i.e.,\
\begin{align}
f\convp 
\theta_{DK}
&:=
\dfrac{\Pr(Y_i = 1\mid Z_i = 1) - \Pr(Y_i = 1\mid Z_i  = 0)}{e(1) - e(0)} \dfrac{1}{1- \Pr\{ Y_i(0) = 1 \}}, \label{eq:DK}  
\\
\tilde f \convp 
\tilde \theta_{DK}
&:=
\dfrac{\Pr(Y_i = 1\mid Z_i = 1) - \Pr(Y_i = 1\mid Z_i  = 0)}{e(1) - e(0)} \dfrac{1}{1- \Pr( Y_i = 1 \mid Z_i = 0 ) }.\label{eq:DKtilde}
\end{align}

First, neither $\theta_{DK}$ nor $\tilde \theta_{DK}$ is generally equal to the persuasion rate $\theta_\ate$, or that for the compliers, $\theta_\local$, which is clear from the fact that  
\[
\Pr\{ Y_i(0)=0 \} \theta_{DK}
=
\Pr\{ Y_i=0\mid Z_i = 0 \} \tilde \theta_{DK}
=
\Pr\{Y_i(0) = 0\mid e(0)<V_i\leq e(1) \} \theta_\local
\]
is equal to the LATE, while $\Pr\{ Y_i(0) = 0\} \theta_\ate$ is equal to the ATE.\footnote{Recall that the population version of the Wald statistic is equal to the LATE under \Cref{ass:instrument}.} For example, $\theta_{DK}$ rescales the LATE with an unconditional probability and hence it does not render a well-defined conditional probability in general. Similarly, $\tilde\theta_{DK}$ is not necessarily a `rate' in spite of the rescaling factor. So, making comparisons across different studies based on $\theta_{DK}$ or $\tilde\theta_{DK}$ can be misleading, although it is a common practice \citep[e.g.,][]{dellavigna2010persuasion}.

There are some special cases of exception though:
\begin{inparaenum}[(i)] 
\item\label{a} everybody is a complier as in the sharp persuasion design, 
\item\label{b} $T_i$ is independent of the potential outcomes $Y_i(t)$ for $t=0,1$ conditional on $Z_i$, or 
\item\label{c} there is no heterogeneity in the treatment effect in that $Y_i(1) - Y_i(0)$ is a constant. 
\end{inparaenum}
That is, in cases (\ref{a}) and (\ref{b}), there is no endogeneity issue, whereas in case (\ref{c}), no one is affected by the persuasive message (i.e.,\ $Y_i(1) - Y_i(0) = 0$ for all $i$), or everybody is persuaded (i.e.\ $Y_i(1) - Y_i(0) = 1$ for all $i$). Note that the LATE is the same as the ATE when any of the three cases applies.
As a result, we have that in case (\ref{a}), $\theta_{DK} = \tilde\theta_{DK} = \theta_\ate = \theta_\local$ holds; in case (\ref{b}), $\theta_{DK} = \theta_\ate = \theta_\local \leq \tilde\theta_{DK}$; and, in case (\ref{c}), $\theta_{DK} = \theta_\ate = \theta_\local = 1\leq \tilde\theta_{DK}$ if everybody is persuaded, or $\theta_{DK}=\tilde\theta_{DK} =\theta_\ate = \theta_\local = 0$, unless any of them are ill-defined, if no one is affected by the informational treatment. Therefore, the feasible version $\tilde f$ of DK's measure of persuasion does estimate the persuasion rate $\theta_\ate$ in case (\ref{a}), although, as DK correctly pointed out, $\tilde\theta_{DK}$ does approximate $\theta_{DK}$ in case (\ref{b}) as well if either $e(0)$ or $\theta$ is close to zero.\footnote{In case (\ref{b}), we have $ \Pr(Y_i = 1 \mid Z_i = 0) =  \Pr( Y_i = 1, T_i = 1 \mid Z_i = 0 ) + \Pr( Y_i = 1, T_i = 0\mid Z_i = 0 ) = \Pr\{ Y_i(0) = 1\} + \bigl[ \Pr\{ Y_i(1) = 1\} - \Pr\{ Y_i(0) = 1\} \bigr] e(0)$.}

Our identification results in the previous sections show that $\theta_L$ is the sharp lower bound of the identified interval of $\theta_\ate$ in the fuzzy design regardless of whether the full joint distribution of the outcome, treatment, and instrument is available or not. The parameter $\theta_L$ has been reported in the literature without understanding that it is the sharp lower bound on $\theta_\ate$. For instance, \citet{dellavigna2010persuasion} extensively estimate $\tilde\theta_{DK}$ by using many examples but they report $\theta_L$ as a lower bound of $\tilde\theta_{DK}$ when $T_i$'s are unobserved and hence $e(1)$ and $e(0)$ are unknown. Our results show that $\theta_L$ is \emph{always} a meaningful parameter, but $\tilde\theta_{DK}$ may not. Therefore, even when information about $e(0)$ and $e(1)$ is available, $\theta_L$ is a better parameter to estimate than $\tilde\theta_{DK}$. 

Indeed, if the full joint distribution of $(Y_i,T_i,Z_i)$ is available, then we recommend reporting $[\theta_L,\ \theta_U]$ along with $\theta^*$; these can be consistently estimated by their sample analogs. If $(Y_i,Z_i)$ is observed with some auxiliary information for $e(0)$ and $e(1)$ available, then $[\theta_L,\ \theta_{U_e}]$ and $[\theta_L^*,\ 1]$ should be reported. If $T_i$ is not observed at all, then the interval $[\theta_L,\ 1]$ is the best we can hope for to study either $\theta_\ate$ or $\theta_\local$. 

Note that $\theta_L$ should be estimated all the time; it only requires data on $(Y_i,Z_i)$. Because the actual $T_i$ can be difficult to observe, researchers have used an extra micro-level survey to obtain auxiliary data on $T_i$, which seems quite costly. However, the value of an attempt to observe $T_i$ can be limited, depending on which parameter the researcher wants to learn about. For instance, if the researcher cares about the persuasion rate of the entire population, then observing $T_i$ does not add any information for the lower bound, while it can potentially improve the upper bound. 
If the group of compliers is of interest, then whether we observe $T_i$ or not, and how we observe it, can be relevant issues; we have $\theta_L^*\geq \theta_L$ in the second data scenario and $\theta^*$ is point identified if $(Y_i, T_i, Z_i)$ is jointly observed. If $Z_i$ is continuously distributed, the value of observing $(Y_i, T_i, Z_i)$ jointly  increases dramatically as well. In summary, our identification analysis shows that the value of observing $T_i$ depends crucially on which population the researcher is 
interested in.

In order to illustrate the difference between DK's measure and our bounds, we have calculated them in \Cref{tb:turnout-results}. We focus on the results reported in \citet[their Table 1]{dellavigna2010persuasion} when the outcome variable is voter turnout. We have chosen this type of study as the turnout is among the most studied outcome variables in the literature and it is naturally a binary measure. 
\Cref{tb:turnout-results} provides estimates of 
$\Pr(Y_i=1 \mid Z_i=z)$ and $e(z)$ for $z=0,1$, thereby enabling us to obtain the bounds based on \cref{thm:fuzzy-e} and \cref{thm:local}~(ii).
It can be seen that using DK's persuasion rates alone may lead to misleading conclusions because the bounds on $\theta_\ate$ as well as those on $\theta_\local$ are in fact wide. Moreover, the results in \Cref{tb:turnout-results} suggest that
identification power under \Cref{ass:binary-monotone,ass:instrument,ass:fuzzy-e} in this example is limited, especially for the upper bounds on $\theta_\ate$ and $\theta_\local$. We further illustrate these points with empirical examples in \Cref{sec:main:example}.

Finally, since the parameters are partially identified, inference should also account for that. The method proposed by \citet{Stoye:07} is useful for that purpose, at least in the most favorable data scenario, in which case the sample analog principle and the delta method show that we can construct the estimators $\hat \theta_L$ and $\hat \theta_U$ that are asymptotically jointly normal. Therefore, by \citet{Stoye:07}, a $(1-\alpha)$ confidence interval for $\theta_\ate$ can be obtained by $[\hat\theta_L-c_\alpha\hat\sigma_L,\, \hat \theta_U + c_\alpha \hat \sigma_U]$, where $\hat\sigma_L$ and $\hat \sigma_U$ are the estimated standard errors of $\hat \theta_L$ and $\hat \theta_U$, respectively, and $c_\alpha$ is chosen by solving
\[
\Phi\Bigl( c_\alpha + \frac{\hat \Delta}{\max(\hat\sigma_L,\hat\sigma_U)}   \Bigr) - \Phi(-c_\alpha) = 1-\alpha,
\]
where $\Phi$ is the distribution function of the standard normal and $\hat \Delta$ is the estimated length of the identified interval.

The second data scenario is slightly more complicated, because $\theta_{U_e}$ and $\theta_L^*$ contain the min or max function that is not smooth; so, the delta method does not apply. In online Appendices \ref{inference-apr} and \ref{inference-lpr}, we propose a two-step method for inference to overcome this problem, which we have applied to the empirical example we discuss in \Cref{sec:main:example}.  In the third data scheme, confidence intervals for $\theta_\ate$ and $\theta_\local$ always coincide, and they can be obtained by using a one-side critical value on $\hat \theta_L$. Specifically, they are given by $[\hat\theta_L - z_{1-\alpha}\hat \sigma_L,\, 1]$, where $z_{1-\alpha}$ is the $(1-\alpha)$ quantile of the standard normal distribution.  Online Appendices  \ref{inference-apr} and \ref{inference-lpr} provide a more detailed discussion on inference.
Furthermore, see online \Cref{sec:inference} for semiparametrically efficient estimation of the two key parameters, i.e.,\ $\theta_L$ and $\theta^*$, when exogenous covariates $X_i$ are present and integrated out. 

In sum, this paper clarifies identification issues when we insert exposure as a choice variable and employ a proper causal framework that is used in policy evaluation to model two causal links (i.e.,\ $Z_i \rightarrow T_i$ and $T_i \rightarrow Y_i$).\footnote{We are grateful to an anonymous referee who provided us with  insightful comments.}

\section{Empirical Examples}\label{sec:main:example}

\subsection{Effects of Uncensored Media}
In this subsection, we revisit \citet[CY hereafter]{CY2019}, who conducted a field experiment in China to measure the effects of providing students with internet access to the uncensored media on various outcome variables. Excluding the existing users, the subjects in their experiments consist of four groups: 
(i) the control group;
(ii) the control group who were encouraged to visit foreign news websites blocked by the Great Firewall;
(iii) students who received free access to uncensored internet;
and
(iv) students who received both the access and encouragement treatments.
They followed the subjects over 18 months to collect outcomes on media-related behaviors, beliefs, and attitudes among other things. It turns out that there were no differences between groups (i) and (ii) and the effects were the largest for group (iv), i.e., the access plus encouragement group (the \emph{Group-AE} students from now on). To benchmark their findings, CY computed DK's measure of persuasion for the \emph{Group-AE} students (see online Appendix Table A.13 of their paper for the details) and commented that ``their estimated persuasion rates are of a similar magnitude to those found in authoritarian regimes that typically have highly regulated media markets'' (see pp. 2323--24 in CY).

In this section, we use the data from CY to illustrate how the common practice of reporting DK's measure  can lead to misleading conclusions by contrasting DK's measure of persuasion with our proposed approaches. 
As in CY, we focus on the \emph{Group-AE} students. That is, $Z_i = 1$ if the $i$th subject 
is randomly assigned to the \emph{Group-AE} group, and $Z_i = 0$ if 
the $i$th subject is randomly assigned to the control or control-encouragement group, while dropping the access only group and the existing users. 
In CY's two-stage least squares analysis (Table 3 in their paper), the treatment variable is 
$T_i = 1$ if the $i$th subject is an active user of the censorship circumvention tool 
and and $T_i = 0$ otherwise.
We use the same treatment variable in our analysis. 
To replicate the results in CY in a representative but succinct way, we focus on the 11 outcome variables listed in Panel A of online Appendix Table A.13 in their paper.
They represent media-related behaviors, beliefs, and attitudes and are transformed to binary variables by CY. 

Recall the population version of DK's measure $f$; see \Cref{eq:DK}. In their online Appendix Table A.13,  
CY measure $\Pr(Y_i = 1\mid Z_i = 1) - \Pr(Y_i = 1\mid Z_i  = 0)$ by the \emph{intent-to-treat} effects of  
the \emph{Group-AE} assignment,
and approximate  
$\Pr\{ Y_i(0) = 1 \}$
using variables collected at the time of the baseline survey or by the estimates of 
$\Pr( Y_i = 1 \mid Z_i = 0 )$ as in $\tilde \theta_{DK}$, if the former is unavailable. Table 2 in CY and the data provided by CY indicate that the change in the exposure rate is 45.5\% if the treatment status is measured by being active users, and therefore we use $e(1) - e(0) = 0.455$ for our subsequent calculations.  

\Cref{tb:CY} summarizes the empirical results. In Column (2) of Table~\ref{tb:CY}, we recompute CY's persuasion rates for the 11 outcome variables listed in Panel A of online Appendix Table A.13 in their paper. As we explained above, these estimates are based on $e(1) - e(0) = 0.455$. Out of the 11 outcome variables, the median persuasion rate is 101\%, and therefore these ``persuasion rates'' cannot be understood as conditional probabilities.  Columns (3) and (4) report our estimates of the average and local persuasion rates along with the 95\% confidence intervals (in curly braces) that were obtained via 10,000 bootstrap replications.  
Our estimates show that 
(i) the average persuasion rates are only partially identified and the widths of the identified intervals are 
substantial for most of the outcomes,
(ii) the point-identified local persuasion rates are contained by the identified intervals for the average persuasion rates
and are typically closer to the upper end points of the intervals.
The persuasive effects are of a relatively large magnitude in that the smallest lower end point of the confidence interval for the average persuasion is 16\%. However, the original estimates overstate the magnitude by a substantial factor
and mask under-identification of average persuasion rates.    
In short, we find that the subjects in the experiments responded to exposure to uncensored internet highly heterogeneously, indicating that it is important to go beyond the benchmark measures of DK type.

\subsection{Effects of Political News}
We now illustrate our proposed methods by using data from \citet[][GKB hereafter]{gerber2009does}, who report findings from a field experiment to measure the effect of political news. We have chosen this example because it contains a credible binary instrument from the field experiment and we can also illustrate all of the three sampling scenarios as well as the case of nonbinary outcomes; for the theory on the multinomial outcome case, see online \Cref{sec:nonbinary}.  In GKB, there are three statuses in the intention to treat: a control group, an offer of free subscription to \textit{The Washington Post}, and one to \textit{The Washington Times}.  To illustrate the usefulness of our paper, we focus on \textit{The Washington Post} and drop all observations from \textit{The Washington Times} subscription. That is, $Z_i = 1$ if the $i$th individual received free subscription to \textit{The Washington Post}, and $Z_i = 0$ if not.

Focusing on the ITT analysis, GKB have reported ITT estimates for various outcomes $Y_i$. \citet{dellavigna2010persuasion} compute persuasion rates for GKB, for which they simply set $T_i = 1$ if the $i$th individual opted into the free subscription and $T_i = 0$ if they opted out of it.\footnote{We provide the empirical results of bound analysis using the opting-into-the-free-subscription treatment variable in 
online \Cref{sec:gkb:further}.}  In this section, for the purpose of illustrating our identification results, we consider a different treatment variable: $T_i = 1$ if the $i$th individual read a newspaper at least several times per week and $T_i = 0$ otherwise, which is a variable that GKB kept track of in a follow-up survey.  Therefore, the relevant treatment we consider differs from that of \citet{dellavigna2010persuasion}, but it is whether individuals have \emph{actually} read the newspaper or not.  The outcome variables we consider are as follows. For the binary case, $Y_i = 1$ if the $i$th individual reported voting for the Democratic candidate in the 2005 gubernatorial election, and $Y_i = 0$ if the subject did not vote for the Democratic candidate or did not vote at all. For the multinomial case, not voting at all is treated as an outside option.  We use only a subsample of the GKB data with those who responded to the follow-up survey to use information on $(Y_i, T_i, Z_i)$ jointly.  After dropping observations for \textit{The Washington Times} subscription and removing missing data, we summarize the GKB data in \Cref{tb:GKB-sum-data}. Although the joint distribution of $(Y_i, T_i, Z_i)$ is observed in this example, we also consider using the two marginals of $(Y_i,Z_i)$ and $(T_i, Z_i)$ separately, to make a comparison. The estimates are summarized in \Cref{tb:GKB-results}. Because the size of the sample extract we use is relatively modest ($n=701$) for an interval-identified object,  we report the 80\% confidence intervals obtained by the inference methods described in \Cref{section:takeaways} as well as in online \Cref{inference-apr,inference-lpr}.

First, we discuss the case where the full joint distribution of $(Y_i, T_i, Z_i)$ is used.  In this data scenario, the average effect of persuasion by reading the newspaper is bounded between $7\%$ and $63\%$.  In contrast, the persuasion rate for the group of compliers is point-estimated by $81\%$.  It is interesting to note that the estimate of $\theta_\local$ is so large that it is greater than the upper bound of $\theta_\ate$. This suggests that individuals are highly heterogeneous in this example, indicating that $\tilde \theta_{DK}$ might not be a well-defined conditional probability here. Indeed, the estimate of $\tilde \theta_{DK}$ in \Cref{eq:DKtilde} is $\hat{\tilde \theta}_{DK} = 1.1027$, which is greater than one.

When the marginals of $(Y_i, Z_i)$ and $(T_i, Z_i)$ are used separately, the upper bound on $\theta_\ate$ increases from $63\%$ to $78\%$. Further, $\theta_\local$ is no longer point estimated but we only know that it is bounded between $78\%$ and $100\%$.  This difference illustrates the loss of identification power if we do not observe the joint distribution of $(Y_i, T_i, Z_i)$.

Finally, we estimate the lower bound on the average persuasion rate by additionally conditioning on those who would vote even without reading the newspaper (see online \Cref{sec:nonbinary} for details). The resulting lower bound on the average persuasion increases from 0.0707 (0.0289) to 0.0975  (0.0554), where the numbers in the parentheses are the left-end points of the 80\% confidence intervals. Therefore, the (point identified) ITT effect is 5\%, while the lower bound of the average persuasion rate is about 7\%, or 10\% if we further focus on those who would vote without reading the newspaper.


\section{Conclusions}\label{sec:conclusion}

We have set up a simple econometric model of persuasion, introduced several parameters of interest, and analyzed their identification.  The empirical examples in \Cref{section:takeaways,sec:main:example} as well as the examples in online \Cref{sec:media,sec:vcm} demonstrate that the persuasive effects are highly heterogeneous in the settings of media and fundraising.

We have focused on the case of binary outcomes and binary treatments. 
In online \Cref{sec:nonbinary}, we extend our analysis to nonbinary outcomes.
If the outcome is nonbinary, then we can condition on those who would not choose the outside option without the treatment. For instance, suppose that we have three options of voting for a Republican, voting for a Democrat, or not voting at all. Then, the persuasive effect of a message supporting a Republican can be measured in a couple of different ways: focusing on those who would not have voted for a Republican without the message is one way and conditioning on those who would have voted for a Democrat (i.e.,\ voted but not voted for a Republican) is the other. In the latter case, we show that the resulting lower bound is always no smaller than that of the binary outcome case.

In general, treatments are multivalued: 
unordered treatments (e.g.,\ watching Fox News, CNN or MSNBC) and ordered treatments (e.g.,\ numbers of hours watching Fox News) arise naturally in applications. It would be fruitful to build on recent developments in multivalued treatments \citep[e.g.,][]{HUV2006,HV2007-handbook-2,HUV2008,HeckmanPinto, LeeSalanie} to investigate identification of persuasive effects.  It would  also  be interesting to estimate deep parameters in an economic model of persuasion by using a more structural approach in the set-up of multivalued treatments. These are topics for future research.




\begin{table}[htbp]
	\centering
	\begin{threeparttable}
	\caption{Persuasion Rates: Papers on Voter Turnout\label{tb:turnout-results}}
	\begin{tabularx}{17cm}{XcX}
	&
\begin{tabular}{lccccccc}
\hline \hline
    Paper     & 	$\hat{y}(1)$	&	$\hat{y}(0)$	&	$\hat{e}(1)$ & $\hat{e}(0)$ 	&	$\tilde f$	&	$[\hat{\theta}_L,\hat{\theta}_{U_e}]$	&	$[\hat{\theta}_L^*,1]$ \\
    &(1) & (2) & (3) & (4) & (5) & (6) & (7) \\
\hline
\citet{Gerber:Green:00}	    &	.472	&	.448	&	.279	& 0	&	.156	&	[.043,1]	    &	[.086,1]	\\
\citet{Green:et-al:03}	    &	.310	&	.286	&	.293	& 0	&	.115	&	[.034,1]	    &	[.082,1]	\\
\citet{Gerber:Green:2001}	&	.711	&	.660	&	.737	& 0	&	.204	&	[.150,.924]	&	[.150,1]	\\
\citet{Gerber:Green:2001}	&	.416	&	.405	&	.414	& 0	&	.045	    &	[.018,1]	    &	[.027,1]	\\
\citet{Gentzkow:06}	        &	.455	&	.435	&	.800	& 0	&	.044	    &	[.035,.389]	    &	[.035,1]	\\
\citet{gentzkow2011effect}	&	.700	&	.690	&	.250	& 0	&	.129	&	[.032,1]	    &	[.040,1]	\\
\hline
\end{tabular}
	&
	\end{tabularx}
	\vspace{1ex}
	\begin{tablenotes}
	{\small
	\item[a] The outcome variable is voter turnout except for \citet{Gentzkow:06}, where  exposure to television discouraged voters to go to the polls. Thus, to have positive persuasive effects in all rows, the outcome variable for  \citet{Gentzkow:06} is not to vote.
		\item[b] In columns (1) and (2),
	$\hat{y}(z)$ denotes  estimates of $\Pr(Y_i = 1\mid Z_i = z)$ for $z=1,0$.
		\item[c] $\tilde f$ is the persuasion rate reported in \citet[table 1]{dellavigna2010persuasion}.
		\item[d]
		$[\theta_L, \theta_{U_e}]$ and $[\theta_L^*,1]$ are the sharp lower and upper bounds on the average and local persuasion rates, respectively, under \Cref{ass:binary-monotone,ass:instrument,ass:fuzzy-e}.
		\item[e] The third row corresponds to the row in Table 1 of \citet{dellavigna2010persuasion} under the treatment labeled ``phone calls by youth vote''.
		\item[f] The fourth row corresponds to the row in Table 1 of \citet{dellavigna2010persuasion} under the treatment labeled ``phone calls 18--30-year-olds''.
		\item[g] Table 1 of \citet{dellavigna2010persuasion} gives $\hat{e}(1)-\hat{e}(0)$ but $\hat{e}(0) \equiv 0$
		in each row by study design.  
					}	
	\end{tablenotes}
	\end{threeparttable}
\end{table}

\clearpage

\begin{table}[htbp]
	\centering
	\begin{threeparttable}
		\caption{Persuasion Rates of Exposure to Uncensored Internet\label{tb:CY}}
		\begin{tabularx}{15cm}{XcX}
			&	
			\begin{tabular}{lccc}
				\hline\hline \noalign{\smallskip}
				(1) & (2) & (3) & (4) \\
				\noalign{\smallskip}\hline \noalign{\smallskip}
				Outcome & CY19 & \multicolumn{2}{c}{Persuasion Rates}  \\
				&  & Average  & Local  \\
				\noalign{\smallskip}\hline \noalign{\smallskip}
				A.1.2 Ranked high: foreign websites 
				& 35.5 & [21.9, 68.4] & 41.0  \\
				&      & \{15.6, 72.9\} & \{29.2, 51.1\} \\ 
				\noalign{\smallskip}\hline \noalign{\smallskip}
				A.1.6  Freq. of visiting foreign websites for info.  
				& 168.5 & [55.5, 76.6] & 70.4  \\
				&  		& \{50.6, 80.2\} & \{64.6, 75.7\} \\
				\noalign{\smallskip}\hline \noalign{\smallskip}
				A.2.1 Purchase discounted tool we offered
				& 45.6 & [22.7, 64.2] & 38.7  \\
				& 	   & \{19.3, 67.9\} & \{32.6, 44.9\} \\
				\noalign{\smallskip}\hline \noalign{\smallskip}
				A.2.2 Purchase any tool	
				& 101.2	& [49.5, 85.0] & 76.8  \\
				&  		& \{45.5, 87.7\} & \{71.9, 81.6\} \\
				\noalign{\smallskip}\hline \noalign{\smallskip}
				A.3 Valuation of access to foreign media outlets
				& 121.7 & [49.3, 82.1] & 73.4  \\
				& 		& \{42.8, 85.8\} & \{65.8, 80.0\} \\
				\noalign{\smallskip}\hline \noalign{\smallskip}
				A.4 Trust in non-domestic media outlets
				& 158.1 & [62.9, 85.8] & 81.6  \\
				& 		& \{57.8, 89.0\} & \{76.2, 86.4\} \\
				\noalign{\smallskip}\hline \noalign{\smallskip}
				A.5.1 Degree of censorship on domestic news outlets
				& 82.3 	& [35.1, 73.5] & 57.0  \\
				& 		& \{29.6, 77.4\} & \{48.7, 64.3\} \\
				\noalign{\smallskip}\hline \noalign{\smallskip}
				A.5.2 Degree of censorship on foreign news outlets	
				& 114.6 & [39.1, 74.4] & 60.4  \\
				&		& \{33.8, 78.2\} & \{53.1, 67.3\} \\
				\noalign{\smallskip}\hline \noalign{\smallskip}
				A.6 Censorship unjustified
				& 77.6 	& [31.6, 66.6] & 48.6  \\
				& 		& \{23.5, 72.0\} & \{35.8, 59.2\} \\
				\noalign{\smallskip}\hline \noalign{\smallskip}
				A.7.1 Domestic cens. driven by govt. policies
				& 192.7	& [75.0, 88.6] & 86.8  \\
				& 		& \{64.4, 94.2\} & \{76.5, 94.4\} \\
				\noalign{\smallskip}\hline \noalign{\smallskip}
				A.7.2 Foreign cens. driven by govt. policies
				& 54.1	& [36.9, 76.3] & 61.0  \\
				&		& \{19.2, 84.1\} & \{32.4, 77.4\} \\
				\noalign{\smallskip}\hline
			\end{tabular}
			&
		\end{tabularx}
		\vspace{1ex}
		\begin{tablenotes}
			{\small
				\item[a] In Column (2), we recompute CY's persuasion rates for the 11 outcome variables listed in Panel A of online Appendix Table A.13 in their paper. For these estimates, we use $e(1) - e(0) = 0.455$. 
				\item[b] Column (3) reports the bound estimates of the average persuasion rates in brackets.
				\item[c] Column (4) reports the point estimates of the local persuasion rates.
				\item[d] The 95\% confidence intervals are given in curly braces, obtained via 10,000 bootstrap replications.
				\item[e] The persuasion rates are expressed in percentage. 		
			}	
		\end{tablenotes}
	\end{threeparttable}
\end{table}%

\clearpage

\begin{table}[htbp]
	\centering
	\caption{Summary Statistics of the GKB Data\label{tb:GKB-sum-data}}
	\begin{tabularx}{11cm}{XcX}
	&	
	\begin{tabular}{crrr}
		\hline\hline
		\multicolumn{4}{l}{\textit{The Washington Post} ($Z_i=1$)} \\
		& \multicolumn{2}{c}{Read a newspaper} & Total \\
		Voted for Democrat   & $T_i = 0$ &    $T_i = 1$ &  \\
		$Y_i = 0$ &    94 &   93 &  187   \\
		$Y_i = 1$ &    31 &   68 &    99   \\
		Total        &  125 &  161 &  286   \\
		\hline
		\multicolumn{4}{l}{Control ($Z_i=0$)} \\
		& \multicolumn{2}{c}{Read a newspaper} & Total \\
		Voted for Democrat   & $T_i = 0$ &    $T_i = 1$ &  \\
		$Y_i = 0$ &    162 &   130 &  292   \\
		$Y_i = 1$ &     46 &    77  &   123   \\
		Total        &   208 &  207 &  415   \\
		\hline
	\end{tabular}
	&
	\end{tabularx}
	\vspace{1ex}
	\begin{tablenotes}
		{\small
			\item[a] The GKB data are used after dropping observations for \textit{The Washington Times} subscription and removing missing data.
		}	
	\end{tablenotes}
\end{table}%

\begin{table}[htbp]
	\centering
	\begin{threeparttable}
	\caption{Estimates of the Key Parameters\label{tb:GKB-results}}
	\begin{tabularx}{11cm}{XcX}
	&
	\begin{tabular}{c|c|c|c}
	\hline\hline
								& $(Y_i, T_i, Z_i)$			&		$(Y_i, Z_i)$ and $(T_i, Z_i)$ &     $(Y_i, Z_i)$ only     \\  \hline
	ITT  & \multicolumn{3}{c}{0.0498}  \\	
		&	\multicolumn{3}{c}{$\{0.0036, \    0.0959 \}$} \\ \hline				
	$\theta_\ate$					& $[0.0707,\ 0.6343]$		&		$[0.0707,\ 0.7832]$  &      $[0.0707,\ 1]$    \\
	                                                 &    $\{ 0.0289, \ 0.6610 \}$            &  $\{0.0286, \ 0.8143 \}$           &  $\{0.0288,\ 1\}$ \\ \hline
	LATE  & \multicolumn{2}{c}{0.7759}  & $[0.0498,\ 1]$  \\	
		&	\multicolumn{2}{c}{$\{0, \    1 \}$} & $\{0.0195,\ 1\}$ \\ \hline	
	    $\theta_\local$			   &   $0.8067$					&		$[0.7759,\ 1]$ &      $[0.0707,\ 1]$   \\
	                                                 &    $\{0.1243, \  1 \}$                            &     $\{0.0069, \  1 \}$       & $\{0.0288,\ 1\}$ \\
	\hline
	\end{tabular}
	&
	\end{tabularx}
	\vspace{1ex}
	\begin{tablenotes}
	{\small
		\item[a]
		The first and third rows show the  estimates of ITT and LATE, respectively.
		The second row corresponds to $[\hat \theta_L,\, \hat \theta_U]$, $[\hat \theta_L,\, \hat \theta_{U_e}]$, and $[\hat \theta_L,\, 1]$, respectively.
		The fourth row shows $\hat\theta^*$, $[\hat\theta_L^*,\, 1]$, and $[\hat \theta_L,\, 1]$, respectively.
		\item[b] 80\% confidence intervals are given in curly braces.
		\item[c] The sharp identified interval for LATE is [ITT, 1] when only the joint distribution of $(Y_i,Z_i)$ is available. This is because $e(1) - e(0)$ is unknown and can be everywhere between 0 and 1 in this case.  
	}	
	\end{tablenotes}
	\end{threeparttable}
\end{table}

\clearpage
\appendix

\renewcommand{\thepage}{A-\arabic{page}}
\setcounter{page}{1}

\renewcommand{\theequation}{\thesection\arabic{equation}}
\setcounter{equation}{0}
\renewcommand{\thetable}{\thesection\arabic{table}}
\setcounter{table}{0}
\renewcommand{\thefigure}{\thesection\arabic{figure}}
\setcounter{figure}{0}

\renewcommand{\theassumption}{\thesection\arabic{assumption}}
\setcounter{assumption}{0}
\renewcommand{\thelemma}{\thesection\arabic{lemma}}
\setcounter{lemma}{0}
\renewcommand{\thetheorem}{\thesection\arabic{theorem}}
\setcounter{theorem}{0}


\section*{\Large Online Appendices to ``Identifying the Effect of Persuasion''}


\section*{\Large Introduction}

The online appendices of the paper are grouped into four parts.
Part I includes additional results that are omitted from the main text.
In \Cref{sec:economic_model}, we present a simple economic model to motivate our setup. Focusing on a binary treatment and a binary outcome, we formulate a model of persuasion within the framework of expected utility maximization. This formulation naturally leads to a potential outcome setup with a certain monotonicity restriction. 
In \Cref{sec:nonbinary}, we give identification results with nonbinary outcomes.

Part II provides additional empirical examples. 
 In \Cref{sec:gkb:further}, we provide further discussion on \Cref{sec:main:example} by analyzing the case where treatment is whether individuals opted into the free subscription to \emph{The Washington Post}.  
 In \Cref{sec:media}, we revisit the empirical literature on the effects of news media on voting, where we apply our identification results to two published articles.
In particular,  when we revisit \citet[DK hereafter]{dellavigna2007fox} using their original data, we find that the identification region for the persuasion rate $\theta_\ate$ is between 1\% and 100\% and that the lower bound for the local persuasion rate $\theta_\local$ is either 12\% or 38\%, depending on the specification of the fixed effects.  These results suggest that the persuasive effect of Fox News is fairly large for compliers, i.e., those who would watch Fox News if and only if it is available via the local cable, but that DK's data are uninformative about the general population.  In \Cref{sec:vcm}, we look at the literature on door-to-door fundraising and we illustrate the usefulness of our results by applying them to two published papers.

Part III deals with estimation and inference problems. In \Cref{inference-apr}, we explain methods for  Inference on the average persuasion rate;
in \Cref{inference-lpr}, we describe methods for  inference on the local persuasion rate. In \Cref{sec:inference}, we consider semiparametrically efficient estimation of the two key parameters, i.e.,\ the lower bounds on $\theta_\ate$ and $\theta_\local$, and we provide an empirical illustration.

Part IV, which is \Cref{app:proofs}, contains all the proofs.

\section*{\Large Part I. Additional Theoretical Results}

\section{A Microeconomic Foundation for \texorpdfstring{\Cref{ass:binary-monotone}}{assumption-binary-monotone}}\label{sec:economic_model}

In this section, we consider a binary choice problem under binary states. The states are unknown to the agents at the time of the decision and they rely on their subjective belief about the states to make a decision.
Suppose that  two possible states are denoted by $S \in \mathscr{S} = \{\text{High}, \text{Low}\}$.  Let $T_i \in \{0,1\}$ indicate individual $i$'s status of the informational treatment. Further, let $q_i(t)$ describe individual $i$'s subjective belief about the state when $T_i$ is  set to $t \in \{0,1\}$: i.e., $q_i(t) = \Pr( S = \text{High} \mid T_i = t, I_i)$, where $I_i$ denotes all other information available to individual $i$.  \Cref{tb:util} describes the utility individual $i$ receives from each choice conditional on the state.
The payoffs matrix in \Cref{tb:util} is from  \citet[][see matrix (5)]{bergemann2017information}.

\begin{table}[htbp]
\begin{center}
	\caption{Utility by choice and state \label{tb:util}}
	\vspace{1ex}
	\begin{tabular}{l|cc}
		\hline\hline
								&   $S=\text{Low}$			&  $S=\text{High}$ \\
		\hline
		Vote ($1$)  			& 	$-1$			&  $U_i \geq 0$  \\
		Not vote ($0$)	 &	$0$				&	$0$	\\
		\hline
	\end{tabular}
\end{center}
\end{table}

The utility from option $0$ is normalized to be $0$ for each state.  Since the expected utility is all that matters for the decision, the utility from option $1$ when the state is ``low'' is normalized to be $-1$: the sign restrictions are to make the choice nontrivial.
 The utility term $U_i$ is not observed by the econometrician.

Suppose that individual $i$ maximizes expected utility. Then, individual $i$ chooses option 1 if and only if expected utility, $-\{ 1-q_i(t)\} + q_i(t) U_i$ for $t\in \{0,1\}$, is positive with subjective belief $q_i(t)$ about the state. Therefore, when the informational treatment is set to be $t \in \{0,1\}$, the potential outcome $Y_i(t)$ can be written as follows:
\begin{equation*}
 Y_i(t) = \one \bigl[ - \bigl\{ 1 - q_i(t) \bigr\} + q_i(t) U_i \geq 0 \bigr],
\end{equation*}
where $\one [ \cdot ]$ is the usual indicator function.
We now make the following assumptions.
\begin{assumption}\label{ass:dist}
$U_i$ has a conditional density $f\{ \cdot\mid q_i(0), q_i(1) \}$ such that $f\{u \mid q_i(0), q_i(1)\} > 0$ for all $u \in [0, \infty)$ with probability one.
\end{assumption}

\begin{assumption}\label{ass:direction}
	$q_i(0) \leq q_i(1)$ with probability one. 
\end{assumption}

\Cref{ass:dist} says that $U_i$ is continuously distributed given $q_i(0)$ and $q_i(1)$. However, it does not rule out the possibility that $U_i$ and $q_i(t)$ are dependent on each other. \Cref{ass:direction} simply means that the informational treatment may shift an agent's belief \emph{only in one direction}.

\begin{lemma}\label{lem:direction:app}
	Under \Cref{ass:dist}, \Cref{ass:direction} is equivalent to $Y_i(0) \leq Y_i(1)$ with probability one.
\end{lemma}
Therefore, \Cref{lem:direction:app} provides a microeconomic foundation for \Cref{ass:binary-monotone}.

\section{An Extension to Nonbinary Outcomes}\label{sec:nonbinary}

\subsection{The Setup and the sharp lower bound}\label{sec:iden:multi}
When outcomes are not binary, one might want to treat an outside option differently. In the voting example, those who would go out and vote even without any informational treatment may be the relevant subpopulation to consider in defining the rate of persuasion. Below we formalize this idea.

Suppose that $\mathcal{S} = \{ 0,1,-1\}$, where $0$ is an outside option, $1$ is the target action of persuasion, and $-1$ represents taking any other action.  For instance, taking action $0$ can mean that the agent does not vote at all, whereas taking action $1$ means that the agent votes for a candidate from party $1$ and taking $-1$ means that the agent votes for a candidate from any other party. We then denote agent $i$'s potential outcomes by the vector of binary variables $Y_i(t) = \bigl( Y_{i0}(t), Y_{i1}(t), Y_{i,-1}(t) \bigr)$ for $t\in \{0,1\}$. Finally, we assume that the choices in $\mathcal{S}$ are exclusive and exhaustive so that $\sum_{j\in \mathcal{S}} Y_{ij}(t) = 1$ for $t\in \{0,1\}$.\footnote{Here, we note that there is no loss of generality in assuming that $\mathcal{S}$ has only three options; if not, we can simply define $Y_{i,-1}(t) = \sum_{j\in \mathcal{S}\backslash \{0,1\}} Y_{ij}(t)$.}  Similarly to the binary case, we impose monotonicity on the target action of persuasion (i.e.,\ $Y_{i1}(1) \geq Y_{i1}(0)$ with probability one). The following assumption summarizes our set-up.

\begin{assumption}[Multinomial Outcomes]\label{ass:mult1}
	$Y_{i1}(1) \geq Y_{i1}(0)$ and $\sum_{j\in\mathcal{S}} Y_{ij}(t) = 1$ for $t=0,1$ with probability one. 
\end{assumption}

If we only focus on the target action $Y_{i1}(t)$, then the persuasion rate $\theta_\pr$ can be defined as in \Cref{section:backgrounds} (i.e.,\ $\theta_\pr = \Pr\{ Y_{i1}(1) = 1 \mid Y_{i1}(0) = 0\}$).  However, if one wants to be explicit about the presence of the outside option, then conditioning on those who would not choose the outside option without the treatment seems appropriate to define the rate of persuasion; that is,
\begin{equation}\label{def:mult2}
\theta_\mathrm{mult} := \Pr\{ Y_i(1) = (0, 1,0) \mid Y_i(0) = (0,0,1) \} = \Pr\{ Y_{i1}(1) = 1 \mid Y_{i0}(0) = 0, Y_{i1}(0) = 0 \},
\end{equation}
provided that the conditional probability is well-defined: the second equality uses \Cref{ass:mult1}. Note that $\theta_\mathrm{mult}$ is a different parameter from $\theta_\pr$, where $\theta_\mathrm{mult}$ now measures the fraction of the people who would vote for the candidate of interest among those who would still vote but for somebody else without the persuasive treatment. The advantage of having a multinomial model is that we can pay attention to this extra layer of conditioning that comes from the presence of an outside option.

Similarly to the binary case, \Cref{ass:mult1} enables us to express $\theta_{\textrm{mult}}$ in terms of the marginal distributions of the potential outcomes.

\begin{lemma}\label{thm:mult}
	Under \Cref{ass:mult1}, we have
	\begin{equation*}
	\theta_\mathrm{mult}
	=
	\frac{\Pr\{ Y_{i1}(1) = 1 \} - \Pr\{Y_{i1}(0) = 1 \}}{1 - \Pr\{ Y_{i0}(0) = 1 \} - \Pr\{ Y_{i1}(0) = 1 \}}. 
	\end{equation*}
\end{lemma}

\Cref{thm:mult} shows that the conditional probability $\theta_\mathrm{mult}$ can be obtained by simply rescaling the ATE as before. One complication (compared with the binary case) is that we have an extra term in the denominator (i.e.,\ $\Pr\{ Y_{i0}(0) = 1\}$), which substantially complicates the identification analysis. 

The assumption of $Y_{i1}(1) \geq Y_{i1}(0)$ is sufficient for \Cref{thm:mult}, but it is not necessary. Indeed, \Cref{ass:mult1} rules out the possibility of having $Y_i(1) = (1,0,0)$ and $Y_i(0) = (0,1,0)$, but this is an irrelevant event for $\theta_\mathrm{mult}$ because $\theta_\mathrm{mult}$ focuses only on the case where $Y_{i0}(t) = 0$ for both $t=0,1$.  However, \Cref{ass:mult1} turns out to be quite convenient to obtain informative bounds of $\theta_\mathrm{mult}$.

\begin{assumption}[No Defiers and an Exogenous Instrument]\label{ass:mult2}
	$T_i$ satisfies \Cref{ass:instrument}. Further, for $t=0,1$ and $j\in \{0,1,-1\}$, $\bigl( Y_{ij}(t), V_i \bigr)$ is independent of $Z_i$.
\end{assumption}

\Cref{ass:mult2} is a trivial extension of \Cref{ass:instrument}.  In the following subsection, we provide a complete characterization of the sharp identified set of $\theta_\mathrm{mult}$ under \Cref{ass:mult1,ass:mult2} in each of the three scenarios of data availability.  However, we present here only the sharp lower bound on $\theta_\mathrm{mult}$ when the joint distribution of $(Y_i, T_i, Z_i)$ is available, as it seems to be the most useful result in practice. Here, $Y_i = ( Y_{i0}, Y_{i1}, Y_{i,-1} ) = T_iY_i(1) + (1-T_i) Y_i(0)$ is now a three-dimensional vector of observed binary variables.

\begin{theorem} \label{thm:multlow}
	Suppose that \Cref{ass:mult1,ass:mult2} are satisfied and that the joint distribution of $(Y_i, T_i,Z_i)$ is known, where $\Pr(Y_{i,-1}=1\mid Z_i=0)>0$. If $\Pr(Y_{i1}=1 \mid Z_i=1) + \Pr(Y_{i0}=1,T_i = 0 \mid Z_i = 0) < 1$, then the lower bound of the sharp identified interval of $\theta_\mathrm{mult}$ is given by
	\[
	\theta_{L,\mathrm{mult}}
	:=
	\frac{\Pr(Y_{i1} = 1 \mid Z_i = 1) - \Pr(Y_{i1}=1 \mid Z_i = 0)}{1 - \Pr(Y_{i0}=1,T_i = 0 \mid Z_i=0) - \Pr(Y_{i1} = 1 \mid Z_i = 0)}.
	\] 
\end{theorem}

\Cref{thm:multlow} shows the sharp lower bound of $\theta_\mathrm{mult}$ in the most favorable data scenario; the complete characterization of the sharp identified set of $\theta_\mathrm{mult}$ in the other data scenarios can be found in online \Cref{app:add-iden-nonbinary}.  The condition of $\Pr(Y_{i1}=1 \mid Z_i=1) + \Pr(Y_{i0}=1,T_i = 0 \mid Z_i = 0) \geq 1$ represents an extreme situation, in which case $\theta_\mathrm{mult}$ can be shown to be equal to $1$. The intuition is clear; if there are too many people who do not vote while they are untreated, then there are too few people to ``persuade'' as $\theta_\mathrm{mult}$ focuses only on the group of people who would vote even without the treatment. Including this trivial situation, the sharp lower bound $\theta_{L,\mathrm{mult}}$ is always no smaller than that of the binary persuasion rate. Therefore, ignoring the presence of an outside option can lead to underestimating the persuasive effect of the treatment if $\theta_\mathrm{mult}$ is the parameter of interest. In fact, in the empirical example in \Cref{sec:main:example}, we estimate the lower bound of $\theta_\mathrm{mult}$ in \Cref{def:mult2} with the outside action of not voting at all. We find that the estimated lower bound on the average persuasion increases from 0.0707  to 0.0975  
if we condition on those who would vote without reading the newspaper.

Unlike the lower bound $\theta_L$ in the  binary case, $\theta_{L,\mathrm{mult}}$ depends on the joint distribution of $(Y_i, T_i)$ given $Z_i$. Therefore, if the sampling scheme does not allow the econometrician to access the full joint distribution, then even the lower bound will change.  By applying a version of the Fr\'{e}chet--Hoeffding bounds, we can derive the sharp identified bounds of $\theta_{\textrm{mult}}$ under the other two sampling schemes (i.e.,\ \Cref{ass:fuzzy-e,ass:fuzzy-worst}). Not surprisingly, in the least informative case of \Cref{ass:fuzzy-worst}, the sharp lower bound of $\theta_{\textrm{mult}}$ becomes identical to that of $\theta_\pr$ (i.e.,\ $\theta_L$).

\subsection{Complete Characterization of the Identified Sets}\label{app:add-iden-nonbinary}

In this section we present a complete characterization of the sharp identified set with nonbinary outcomes under the three scenarios of data availability. In order to present the results, we introduce the following notation. Let
\begin{equation}
p_j(y\mid z) := \Pr( Y_{ij} = y \mid Z_i = z)
\quad\text{and}\quad
p_j(y,t \mid z) := \Pr( Y_{ij} = y, T_i = t \mid Z_i = z).
\end{equation}

The first theorem is a complete characterization of the sharp identified set of $\theta_\textrm{mult}$ when $(Y_i,T_i,Z_i)$ is jointly observed.

\begin{theorem} \label{thm:mult id}
	Suppose that \Cref{ass:mult1,ass:mult2} are satisfied and that the joint distribution of $(Y_i, T_i, Z_i)$ is known, where $p_{-1}(1\mid 0)>0$. Then, the sharp identified set of $\theta_\mathrm{mult}$ is given as follows.
	\begin{enumerate}
		\item If $p_1(1 \mid 1) + p_0(1,0 \mid 0) \geq 1$, then $\theta_\mathrm{mult} = 1$;
		\item If $p_1(1 \mid 1) + p_0(1,0 \mid 0)< 1 \leq p_1(1,1 \mid 1) + p_0(1,0 \mid 0)+ 1-e(1) + e(0)$, then
		\[
		\frac{p_1(1 \mid 1) - p_1(1 \mid 0)}{1-p_0(1,0 \mid 0) - p_1(1 \mid 0) }
		\leq
		\theta_\mathrm{mult}
		\leq 1; 
		\]
		\item If $p_1(1,1 \mid 1) + p_0(1,0 \mid 0)+ 1 - e(1) + e(0) < 1$, then
		\[
		\frac{p_1(1\mid 1) - p_1(1\mid 0)}{1-p_0(1,0\mid 0) - p_1(1\mid 0) }
		\leq
		\theta_\mathrm{mult}
		\leq
		\frac{p_1(1,1\mid 1)+1-e(1)-p_1(1,0\mid 0)}{1-p_0(1,0\mid 0)-e(0)-p_1(1,0\mid 0)}.
		\]
	\end{enumerate}
\end{theorem}

We now present the results under \Cref{ass:fuzzy-e}.

\begin{theorem}\label{thm:mult id2}
	Suppose that \Cref{ass:mult1,ass:mult2} are satisfied and that the distribution of $(Y_i, Z_i)$ and $e(1),e(0)$ are known, where $p_{-1}(1\mid 0)>0$. Then, the sharp identified set of $\theta_\mathrm{mult}$ is given as follows.
	\begin{enumerate}
		\item If $p_1(1\mid 1) + \max\{0,\ p_0(1\mid 0)-e(0) \} \geq 1$, then $\theta_\mathrm{mult} = 1$;
		\item If $p_1(1\mid 1) + \max\{0,\ p_0(1\mid 0)-e(0) \} < 1 \leq \min\{ p_1(1\mid 1),\ e(1)  \} + \min\{ p_0(1\mid 0),\ 1-e(0) \} + 1-e(1) + e(0)$, then
		\[
			\max \left\{ \frac{p_1(1\mid 1) - p_1(1\mid 0)}{1-p_1(1\mid 0)},  \frac{p_1(1\mid 1) - p_1(1\mid 0)}{1 -  p_0(1\mid 0)+e(0) -p_1(1\mid 0) }  \right\}
				\leq \theta_\mathrm{mult}\leq 1;
		\]
		\item If $\min\{ p_1(1\mid 1),\ e(1)  \} + \min\{ p_0(1\mid 0),\ 1-e(0) \} + 1-e(1) + e(0) < 1$, then
		\begin{multline*}
				\max \left\{ \frac{p_1(1\mid 1) - p_1(1\mid 0)}{1-p_1(1\mid 0)},  \frac{p_1(1\mid 1) - p_1(1\mid 0)}{1-  p_0(1\mid 0)+e(0) -p_1(1\mid 0) }  \right\}
				\leq
				\theta_\mathrm{mult} \\
				\leq
				\frac{\min\{p_1(1\mid 1),\ e(1) \}+1-e(1) - \max\{0,\ p_1(1\mid 0)-e(0) \}}{1- \min\{ p_0(1\mid 0),\ 1-e(0) \} - e(0)  -\max\{0,\ p_1(1\mid 0)-e(0) \}}.
		\end{multline*}
	\end{enumerate} 
\end{theorem}
The three cases in \Cref{thm:mult id2} are exclusive and exhaustive. The condition describing the third case can be simplified to $p_1(1,1\mid 1) + p_0(1\mid 0) < e(1) - e(0)$ because the inequality can hold only if $\min\{ p_1(1\mid 1), e(1) \} = p_1(1\mid 1)$ and $\min\{p_0(1\mid 0), 1-e(0)\} = p_0(1\mid 0)$.    We finally give the result under the assumption that $T_i$ is not observed at all.

\begin{theorem}\label{thm:mult id3}
	Suppose that \Cref{ass:mult1,ass:mult2} are satisfied and that only the distribution of $(Y_i, Z_i)$ is known, where $p_{-1}(1\mid 0)>0$. Then, the sharp identified set of $\theta_\mathrm{mult}$ is the same as the binary case, i.e.,\
	\[
		\frac{p_1(1\mid 1) - p_1(1\mid 0)}{1-p_1(1\mid 0)} \leq \theta_\mathrm{mult}\leq 1,
	\]
	provided that the denominator is different from zero. 
\end{theorem}

Therefore, unlike the binary case, the sharp lower bound on $\theta_\mathrm{mult}$ depends on the sampling scheme.  If $T_i$ is not observed at all, then observing the choices of the outside option does not help to infer the persuasion rate with the extra conditioning on those who would participate without the treatment.

\section*{\Large Part II. Additional Empirical Results}

In this part of the appendices, we provide additional empirical results. 
Before going through details, \Cref{app-empirical-results-tab} provides  a concise summary of additional empirical results that report 
either interval or point estimates of both the average and local persuasion rates ($\theta_\ate$ and $\theta_\local$). 
\Cref{app-empirical-results-tab} shows that the persuasive effect is highly heterogeneous.

\begin{table}[htbp]
\centering
\begin{threeparttable}
\caption{Summary of Additional Empirical Results: Estimates of Persuasion Rates}\label{app-empirical-results-tab}
\begin{tabularx}{11cm}{XcX}
	&
\begin{tabular}{ccc}
\hline \hline
  Treatment & $\theta_\ate$ & $\theta_\local$  \\
\hline
\multicolumn{3}{l}{\citet{dellavigna2007fox}}    \\
& &  \\
  Watching Fox News & $[0.005,0.997]$ &  $[0.117,1]$  \\
  Watching Fox News & $[0.011,0.997]$ & $[0.375,1]$  \\
 \hline
\multicolumn{3}{l}{\citet{landry2006}}    \\ 
& &  \\
  VCM & [0.095, 0.719]  & 0.253   \\
 VCM with seed money   &  [0.052,  0.699]  & 0.148   \\
 Single-prize lottery   & [0.171,  0.794]  &  0.455  \\
 Multiple-prize lottery   & [0.126, 0.775]   & 0.359   \\
\hline
\multicolumn{3}{l}{\citet{DLM}}  \\
\multicolumn{3}{l}{La Rabida Children's Hospital}  \\
& &  \\
  Baseline &   [0.071,  0.666]  & 0.175  \\
  Flyer      &   [0.068, 0.704]  & 0.188   \\
 Opt--Out &   [0.054,  0.749]  & 0.177   \\
\hline
\multicolumn{3}{l}{\citet{DLM}} \\
\multicolumn{3}{l}{East Carolina Hazard Center} \\
& &  \\
  Baseline &   [0.047,  0.617]  & 0.109  \\
  Flyer      &   [0.051, 0.655]  & 0.129   \\
 Opt--Out &   [0.030,  0.686]  & 0.086   \\
\hline
\end{tabular}
&
	\end{tabularx}
	\vspace{1ex}
	\begin{tablenotes}
	{\small
		\item[a]
The first row for \citet{dellavigna2007fox} refers to estimates when U.S. House district fixed effects are controlled for; the second row corresponds to those with county fixed effects.
\item[b] VCM stands for voluntary contributions mechanism.
}
	\end{tablenotes}
	\end{threeparttable}
\end{table}%

\section{Further Discussion on \texorpdfstring{\Cref{sec:main:example}}{the Empirical Example}}\label{sec:gkb:further}

\subsection{Opting into the Free Subscription to The Washington Post}

In \Cref{sec:main:example}, we provide an empirical example using the data from \citet{gerber2009does}.
As we describe in the main text, the main difference between our example and that in \citet{dellavigna2010persuasion} is what persuasive treatment is concerned with. In the latter, binary treatment is $T_i = 1$ if the $i$th individual opted into the free subscription and $T_i = 0$ if they opted out of it. We believe \citet{dellavigna2010persuasion}'s analysis is closer to an intent-to-treat (ITT) analysis because merely opting into the free subscription does not mean that individuals were exposed to the persuasive messages of the newspaper unless they actually read them. Furthermore, the difficulty of using the treatment variable as in \citet{dellavigna2010persuasion} is that there is no individual-level information regarding who opted out of the free subscription  from the original dataset of \citet{gerber2009does}.  \citet{dellavigna2010persuasion} computed the changes in the exposure rate $e(1) - e(0) = 94\%$ from \citet[page 38]{gerber2009does}, which  states that ``6 percent of households in the treatment groups opted out of the free subscription.'' Given the lack of micro-level information, we can only compute the bounds using summary statistics on $\Pr(Y_i = 1 \mid Z_i = 1)$, $\Pr(Y_i = 1 \mid Z_i = 0)$, $e(1) = \Pr(T_i = 1\mid Z_i = 1)$, and $e(0) = \Pr(T_i = 1\mid  Z_i = 0)$. This case corresponds to \Cref{thm:fuzzy-e} in the paper. It follows from Table 1C of \citet{gerber2009does} that 49\% and 41\% voted for a Democrat in 2005 VA election in the Washington Post treatment and control group, respectively: i.e.,\ $\widehat\Pr(Y_i = 1 \mid Z_i = 1) = 0.49$ and $\widehat\Pr(Y_i = 1\mid Z_i = 0) = 0.41$. Since the control group in \citet{gerber2009does} did not receive an offer to the free subscription, we have $\widehat e(1) = 0.94$ and $\widehat e(0) = 0$. These numbers result in the bounds $[0.136, 0.237]$ for the average persuasion rate $\theta_\ate$ 
and $[0.136, 1]$ for the local persuasion rate $\theta_{local}$, respectively. On the other hand, the estimate of $\widetilde{\theta}_{DK}$ is $0.144$. The difference between $\theta_L$ and $\widetilde{\theta}_{DK}$
is smaller here because the size of compliers is large (that is, 94\%) when treatment refers to opting into the free subscription.

\section{Additional Examples: The Effects of Media on Voting}\label{sec:media}

In this section we revisit the recent empirical literature on the effects of media on voting  and apply our identification results.

\subsection{The Effect of Fox News: \texorpdfstring{\citet{dellavigna2007fox}}{dellavigna2007fox} Revisited}\label{DK-empirical}

In DK, the entry of Fox News in cable markets plays a role of an instrument conditional on a set of covariates. That is, $Z_i$ is a binary variable that equals one if Fox News was a part of the local cable package in the town where the $i^{th}$ individual was living in 2000. To apply our result to DK, let  $Y_i$ be the binary dependent variable that equals one if individual $i$ voted for the Republican candidate in the 2000 presidential election. As DK argue in their paper,  Fox News availability in 2000 is likely to be idiosyncratic, only after controlling for a set of covariates. We will be explicit about conditioning on covariates $X_i$  to apply our identification results: i.e., using Bayes' theorem,
\begin{multline*}
	\theta_\ate
    =
   \int \frac{\Pr\{ Y_i(1) = 1\mid X_i = x \} - \Pr\{ Y_i(0) = 1\mid X_i = x \}}{\Pr\{ Y_i(0) = 0\mid X_i = x \}} dF\{x\mid Y_i(0) = 0 \} \\
    =
    \frac{ \int \Pr\{ Y_i(1) = 1\mid X_i=x \} dF(x) - \int \Pr\{ Y_i(0) = 1\mid X_i =x\} dF(x) }{1 - \int \Pr\{ Y_i(0) = 1\mid X_i = x \} dF(x) },
\end{multline*}
where $F$ denote the marginal distribution of $X_i$. Since the sharp identified bounds of $\Pr\{ Y_i(t) = 1\mid X_i =x\}$ for $t=0,1$ can be obtained in the same way as \cref{lem:Manski,lem:Manski2}, depending on the data availability scenario, the same reasoning as the proof of \Cref{thm:fuzzy-best,thm:fuzzy-e,thm:fuzzy-worst} leads to the sharp bounds on $\theta_\ate$ when we have $X_i$ in the data. 

To be more specific, let $e(z,x) = \Pr(T_i = 1 \mid Z_i = z, X_i = x)$, and we focus on the second data scenario, i.e., the case of \Cref{ass:fuzzy-e}, given the type of DK's data. Then, similarly to \Cref{lem:Manski2}, the sharp identified bounds of $\Pr\{ Y_i(t) =1 \mid X_i\}$ are given by 
\begin{align*}
&\Pr(Y_i=1\mid Z_i = 1, X_i) \leq \Pr\{ Y_i(1) = 1\mid X_i \} \leq \min\bigl\{ 1,\ \Pr(Y_i = 1\mid Z_i = 1,X_i) + 1-e(1,X_i) \bigr\}, \\
&\max\{0,\ \Pr(Y_i=1\mid Z_i = 0,X_i) - e(0, X_i) \} \leq \Pr\{ Y_i(0) = 1\mid X_i\} \leq \Pr( Y_i = 1\mid Z_i= 0, X_i), 
\end{align*}
from which 
\begin{multline*}
\int\Pr( Y_i = 1\mid Z_i = 1, X_i = x) dF(x) \\
\leq 
\Pr\{ Y_i(1) = 1 \} 
\leq 
\int \min \bigl\{1,\ \Pr(Y_i = 1\mid Z_i = 1,X_i =x) + 1-e(1,x) \bigr\} dF(x),  
\end{multline*}
and 
\begin{multline*}
    \int \max\bigl\{ 0,\ \Pr(Y_i=1\mid Z_i = 0,X_i=x) - e(0, x) \bigr\} dF(x) \\
	\leq 
	\Pr\{ Y_i(0) = 1 \} 
	\leq 
	\int \Pr( Y_i = 1\mid Z_i= 0, X_i = x) dF(x).   
	\end{multline*}
Therefore, following the proof of \cref{thm:fuzzy-best} shows that the sharp identified bounds of $\theta_\ate$ when $X_i$ is in the data can be obtained by integrating both the numerators and the denominators of the ``conditional'' sharp identified bounds with respect to the marginal distribution of $X_i$. 

In order to implement this idea, we write the lower bound as a function of the values of $X_i$:  i.e.,\ 	
	\begin{equation}\label{conditional-lower-bound}
		\tilde \theta_L(x)
		:=
		\dfrac{\Pr(Y_i = 1 \mid Z_i = 1,X_i = x)-\Pr(Y_i = 1\mid Z_i = 0, X_i = x)}{1-\Pr(Y_i = 1\mid Z_i = 0, X_i = x)},
	\end{equation}
which is the sharp lower bound on the conditional persuasion rate, i.e., $\Pr\{ Y_i(1) =1 \mid Y_i(0) = 0, X_i = x\}$. Then, to obtain the lower bound for the persuasion rate in the population, we integrate the numerator and the denominator of \eqref{conditional-lower-bound} with respect to the distribution $F$ of $X_i$, so that
\begin{equation}\label{eq:theta_L with X}
	\theta_L =	\frac{\int \Pr(Y_i = 1 \mid Z_i = 1,X_i = x) dF(x) - \int \Pr(Y_i = 1\mid Z_i = 0, X_i = x) dF(x)}{1-\int \Pr(Y_i = 1\mid Z_i = 0, X_i = x)dF(x)}.
\end{equation}
Note that $X_i$ is first controlled for and is averaged out.

To estimate $\theta_L$ in this way by using DK's data,\footnote{The  data used in DK are available at \url{http://eml.berkeley.edu/~sdellavi/index.html}.} we adopt similar specifications as in DK. They estimated $\Pr(Y_i = 1 \mid Z_i, X_i)$ using a town--level linear regression model, where the dependent variable is the Republican two--party vote share for the 2000 presidential election minus the same variable for the 1996 election. To be consistent with our econometric framework, we modify the dependent variable to be the votes cast for the Republican candidate in the 2000 presidential election divided by the population of age 18 and older. Recall that in our setup, $Y_i = 0$ if individual $i$ did not vote for the Republican candidate. This event includes the case of voting for different candidates or that of not voting for any candidate at all.  As for the town--level covariates, we include the Republican vote share as a share of the voting--age population in the 1996 election, census controls for both 1990 and 2000, cable system controls, and U.S. House district fixed effects (or county fixed effects). These specifications correspond to the main specifications of DK (see columns (4) and (5) of table IV in DK). In the regression, the town--level observations are weighted by the population of age 18 and older in 1996.

DK used two different data sources for $(Y_i, Z_i, X_i)$ and $(T_i, Z_i, X_i)$. Hence, we can look at the upper bound for $\theta_\ate$ and the lower bound for $\theta_\local$ using these. Again, making use of the covariates explicitly, we use the conditional version of \cref{lem:Manski2} and follow the same reasoning as the proof of \Cref{thm:fuzzy-e} to obtain
\[
\theta_{U_e} =	\frac{\int \tilde \theta_{U_e, n} (x) dF(x)}{\int \tilde \theta_{U_e, d} (x) dF(x)},
\]
where  
\begin{align*}
\tilde \theta_{U_e,n}(X_i) &:= \min\{ 1, \Pr(Y_i = 1 \mid Z_i = 1, X_i)+1-e(1,X_i) \} - \max\{0,\ \Pr(Y_i = 1 \mid Z_i = 0, X_i)-e(0,X_i) \}, \\
\tilde\theta_{U_e,d}(X_i) &:= 1 - \max\{0,\ \Pr(Y_i = 1 \mid Z_i = 0, X_i)-e(0,X_i) \}.
\end{align*}

The case of $\theta_\local$ when we have $X_i$ in the data is similar. Let $\mathcal{C}_i = \{ e(0,X_i)< V_i\leq e(1,X_i)\}$ denote the event of individual $i$ being a complier. We note that
\begin{align} 
\theta_\local
&=
\int \Pr\{ Y_i(1) = 1\mid Y_i(0) = 0, X_i = x, \mathcal{C}_i \} dF( x\mid Y_i(0) = 0, \mathcal{C}_i ) \notag \\
&=
\int \left( \frac{\Pr\{ Y_i(1) = 1\mid X_i = x, \mathcal{C}_i \} - \Pr\{ Y_i(0) = 1\mid X_i = x, \mathcal{C}_i \}}{\Pr\{ Y_i(0) = 0\mid X_i = x, \mathcal{C}_i \}}  \right) dF( x\mid Y_i(0) = 0, \mathcal{C}_i )  \notag \\
&=
\frac{\int \Pr\{ Y_i(1) = 1\mid X_i = x, \mathcal{C}_i \} - \Pr\{ Y_i(0) = 1\mid X_i = x, \mathcal{C}_i \}dF( x\mid \mathcal{C}_i )}{\Pr\{ Y_i(0) = 0\mid \mathcal{C}_i \}}  \label{eq:local X}
\end{align}
by Bayes' theorem. Here, the numerator of \eqref{eq:local X} is equal to
\begin{equation}\label{eq:num-local x}
    \int  \frac{\Pr(Y_i = 1\mid Z_i = 1,X_i=x) - \Pr(Y_i = 1\mid Z_i  = 0,X_i=x)}{e(1,x)-e(0,x)} dF( x\mid \mathcal{C}_i ),
\end{equation}
and the denominator of \eqref{eq:local X} can be rewritten as 
\begin{equation}\label{eq:den-local x}
    \Pr\{ Y_i(0) = 0\mid \mathcal{C}_i\}
    =
    \int \Pr\{ Y_i (0) = 0\mid X_i = x, \mathcal{C}_i \} dF(x\mid \mathcal{C}_i).
\end{equation}
Since  
\begin{equation*}
dF( x\mid \mathcal{C}_i )
=
\frac{e(1,x) - e(0,x)}{\int \{ e(1,x) - e(0,x) \} dF(x)} dF(x). 
\end{equation*}
by Bayes' theorem again, we can combine \cref{eq:local X,eq:num-local x,eq:den-local x} to obtain 
\begin{equation}\label{eq:local x 1}
\theta_\local
=
\frac{\int \bigl\{ \Pr(Y_i = 1\mid Z_i = 1,X_i=x) - \Pr(Y_i = 1\mid Z_i  = 0,X_i=x) \bigr\} dF(x)}{\int \Pr\{ Y_i (0) = 0,\ e(0,x) < V_i\leq e(1,x) \mid X_i = x  \} dF(x) }
\end{equation}
The numerator of \eqref{eq:local x 1} is directly identified from data in all of the three data scenarios we discussed. The denominator of \eqref{eq:local x 1} is also identified under \Cref{ass:fuzzy-best} by
\[
\int \Pr(Y_i = 0, T_i = 0\mid Z_i = 0, X_i = x) - \Pr(Y_i = 0, T_i = 0\mid Z_i = 1, X_i = x) dF(x),   
\]
but it is not under \Cref{ass:fuzzy-e}. The sharp identified bounds on the denominator of \eqref{eq:local x 1} under \Cref{ass:fuzzy-e} can be obtained exactly in the same reasoning as in the proof of part \ref{part2-local} of \Cref{thm:local}. Specifically, the sharp bounds on $\theta_\local$ are now given by
\begin{equation}\label{lower-bound-late-x}
\dfrac{ \int \bigl\{ \Pr(Y_i = 1\mid Z_i = 1,X_i=x) - \Pr(Y_i = 1\mid Z_i  = 0,X_i=x) \bigr\} dF(x) }{\int \min \bigl\{ 1- \Pr(Y_i = 1 \mid Z_i =0,X_i=x),\ e(1,x)-e(0,x) \bigr\} dF(x) } 
\leq \theta_\local \leq 1
\end{equation}
under \Cref{ass:fuzzy-e}. The lower bound in \eqref{lower-bound-late-x} is always no smaller than
\begin{multline*}	
    \max\Biggl\{
	\frac{\int \bigl\{ \Pr(Y_i = 1\mid Z_i = 1,X_i=x) - \Pr(Y_i = 1\mid Z_i  = 0,X_i=x) \bigr\}  dF(x)}{1- \int \Pr(Y_i = 1 \mid Z_i =0,X_i=x) dF(x)}, \\
	\frac{\int \bigl\{ \Pr(Y_i = 1\mid Z_i = 1,X_i=x) - \Pr(Y_i = 1\mid Z_i  = 0,X_i=x) \bigr\} dF(x)}{\int \{ e(1,x) - e(0,x) \} dF(x)} 
    \Biggr\},
\end{multline*}
where the first expression in the max function is the sharp lower bound $\theta_L$ given in \eqref{eq:theta_L with X} and the second one is the average of the conditional local average treatment effect over the group of compliers.

DK estimated $e(z,x)$ using the microlevel Scarborough data on television audiences.
We focus on ``diary audience'' measure in DK\footnote{The microlevel Scarborough data contain the ``recall'' measure  regarding whether a respondent watched a given channel in the past seven days and the  ``diary'' measure on whether a respondent watched a channel for at least one full half-an-hour block according to the seven-day diary.} and take the same specifications as in columns (2) and (3) of table VIII from DK.

\begin{table}[htbp]
\centering
\caption{Persuasion Rates: Fox News Effects \label{DK-results} }
\begin{tabular}{ccc}
\hline \hline
                              & (1)                      & (2) \\
                              & U.S. House district  & County \\
                              & fixed effects & fixed effects \\
\hline
$\theta_\ate$       & [0.005,0.997] &    [0.011,0.997]   \\
$\theta_\local$ & [0.117,1] &    [0.375,1]   \\
\hline
\end{tabular}
\end{table}%

\Cref{DK-results} summarizes our empirical results.\footnote{To estimate the unconditional bounds reported in the table, the conditional ones are weighted by the number of respondents in a town for the Scarborough data. In addition, the predicted probabilities are truncated to be between 0 and 1.}
Column (1) shows estimation results when U.S. House district fixed effects are controlled for and column (2) displays corresponding results for county fixed effects.

The bounds for $\theta_\ate$ are wide and uninformative. However, the lower bounds for $\theta_\local$ are sizable and also comparable to the estimates of the persuasion rates reported in DK (0.11 and 0.28, respectively).  In sum, we conclude that the persuasive effect of Fox News seems fairly large for  the compliers, that is,  those who would watch the Fox News channel if and only if it is randomly available, although the data do not say much about the entire population.

\subsection{The NTV Effect: \texorpdfstring{\citet{EPZ}}{EPZ} Revisited}\label{sec:NTV}

As mentioned in the main text, \citet[][EPZ hereafter]{EPZ} used a continuous instrument, i.e.,\ the signal strength of NTV, to measure the persuasive effect of watching
 NTV (the anti-Putin TV station)  on a parliamentary election in 1999. Further, in the individual--level survey data in EPZ, $(Y_i, T_i, Z_i)$ are jointly observed. Therefore, in this subsection, we apply the identification result of the marginal persuasion rate to this example using the EPZ data.

We look at two parties: the progovernment party ``Unity'' and the most popular opposition party OVR (``Fatherland--All Russsia''). During the 1999 election campaign, Unity was opposed by NTV, while OVR were supported by NTV. Thus, EPZ presumed a negative persuasion rate for voting for Unity but a positive persuasion rate for OVR. To be consistent with our theoretical framework and other empirical examples, $Y_i$ is either $Y_{\text{Unity},i}$ or $Y_{\text{OVR}, i}$, depending on which party we consider. Specifically, we let $Y_{\text{Unity}, i} = 1$ if an individual did not vote for Unity and $Y_{\text{Unity}, i} = 0$ otherwise;  $Y_{\text{OVR}, i} = 1$ if an individual voted for OVR and $Y_{\text{OVR}, i} = 0$ otherwise.
As in the previous section, it is necessary to condition on covariates. We take the baseline covariates as in columns (1) and (2) of tables 6 and 7 in EPZ. They include individual demographic characteristics such as gender, age, marital status, and education, and subregional variables such as population size and average wage.

\begin{figure}[htbp]
\caption{Estimates of Marginal and Average Persuasion Rates \label{figure-example-NTV}	}
\vspace*{-2ex}	
\begin{center}
\makebox{
\includegraphics[origin=bl,scale=.6,angle=0]{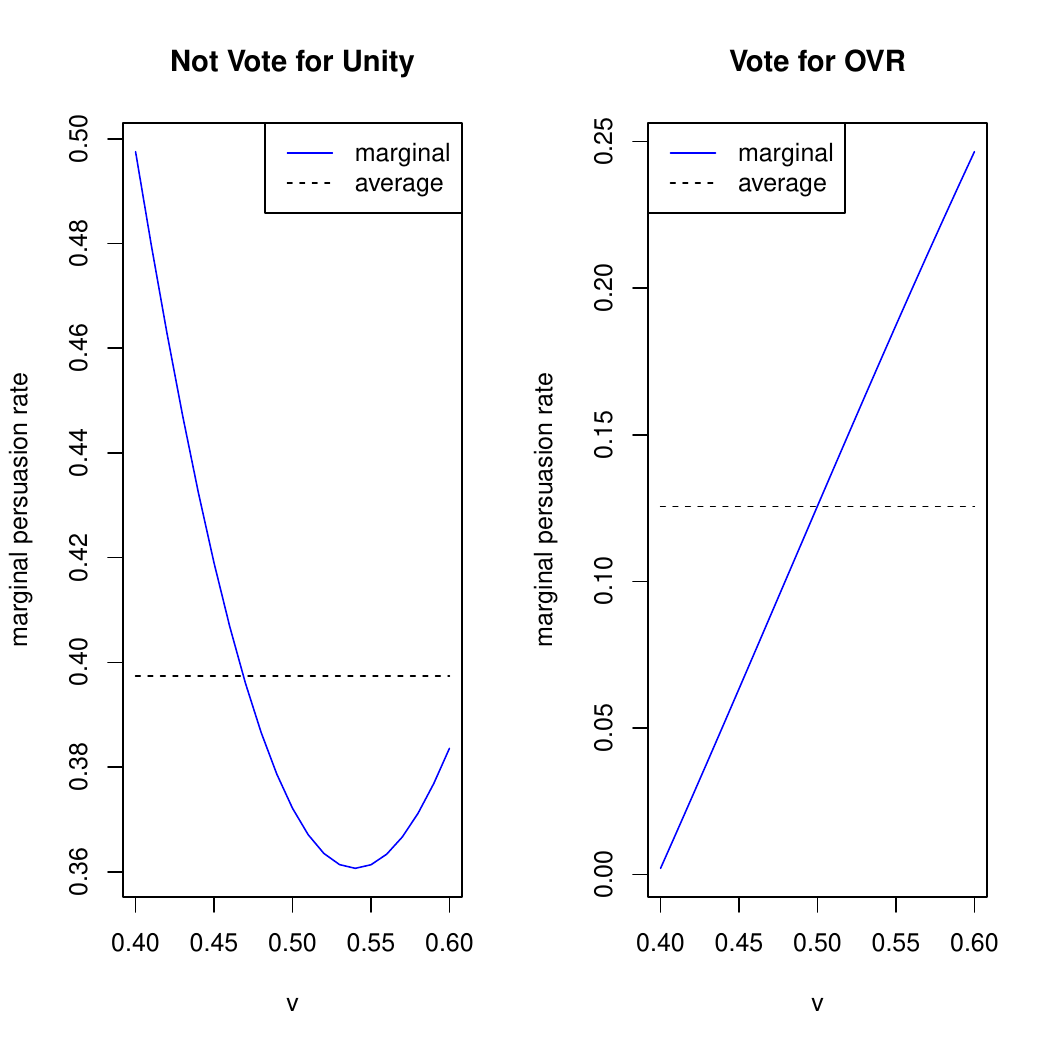}
}
\end{center}
\par
\vspace*{-1ex}
\parbox{6in}{
Notes: The left and right panels of the figure show estimates of the marginal and average persuasion rates for not voting for Unity and voting for OVR, respectively. The domain of $V_i$ is limited to $[0.4,0.6]$ and the average persuasion rates here refer to the averages of the marginal persuasion rates between $0.4$ and $0.6$.}
\end{figure}

Since we have covariates $X_i$, $\theta_\mte(v)$ is equal to 
\begin{align*} 
    \int &\Pr\{ Y_i(1) = 1\mid Y_i(0) = 0, V_i = v, X_i = x\} dF\{x\mid Y_i(0) = 0, V_i = v \} \\
    &=
    \int \frac{\Pr\{ Y_i(1) = 1\mid V_i = v, X_i = x\} - \Pr\{ Y_i(0) = 1\mid V_i = v, X_i = x\}}{\Pr\{ Y_i(0) = 0\mid V_i = v, X_i = x\}} dF\{x\mid Y_i(0) = 0, V_i = v \} \\
    &=
    \frac{\int \Pr\{ Y_i(1) = 1\mid V_i = v, X_i = x\} - \Pr\{ Y_i(0) = 1\mid V_i = v, X_i = x\} dF(x)}{1 - \int \Pr\{ Y_i(0) = 1\mid V_i =v, X_i = x \} dF(x)},
\end{align*}
where the last equality is by Bayes' theorem and the independence between $V_i$ and $X_i$. Therefore, $\theta_\mte(v)$ is obtained by integrating both the numerator and the denominator of the conditional marginal persuasion rate with respect to the marginal distribution of $X_i$: i.e.,
\begin{equation} \label{eq:theta_mte x}
\theta_\mte(v)
=
\frac{\int \partial \Pr\{Y_i = 1\mid e(Z_i,X_i) = e, X_i = x \}/\partial e \big|_{e=v} dF(x) }{1+\int \partial \Pr\{ Y_i = 1, T_i = 0\mid e(Z_i,X_i) = e, X_i = x \}/\partial e \big|_{e=v} dF(x) }. 	
\end{equation}

For the sake of simplicity, we estimate $\theta_\mte(v)$ via polynomial sieves: the relevant conditional probabilities, i.e., $e(z,x)$, $\Pr(Y_i = 1\mid e(Z_i, X_i) = e, X_i = x)$ and $\Pr(Y_i = 1, T_i = 0\mid e(Z_i, X_i) = e, X_i = x)$, are all estimated by regression models that are linear in $X_i$ and cubic in either $Z_i$ or $e(Z_i,X_i)$, depending on the conditional probability to be estimated, while interacting demographic variables in $X_i$ with  $Z_i$ or $e(Z_i, X_i)$, respectively.\footnote{The exposure rate $e(z,x)$ is first estimated and its fitted values, truncated to be between 0 and 1, are included as a regressor to estimate the other two conditional probabilities.} 
Then, we obtain the conditional marginal persuasion rate $\theta_\mte(v,x)$, i.e., the estimates of \Cref{mte-formula} with extra conditioning on $X_i = x$. Finally, we integrate $x$ out of the numerator and the denominator of $\theta_\mte(v,x)$ by using the sample marginal distribution of $X_i$ as we illustrated in \eqref{eq:theta_mte x}.

\Cref{figure-example-NTV} presents the estimation results. 
As it is more dificult to estimate $\theta_\mte(v)$ for high or low values of $v$, we focus on the middle range of $v$, i.e., $v\in [0.4, 0.6]$: the average persuasion rate on the middle range of $v$ is computed by
\begin{equation*}
\theta_\ate [0.4, 0.6]
= 
\frac{\int_{0.4}^{0.6} \partial\Pr(Y_i = 1\mid e(Z_i)=e )/\partial e\big|_{e=v} dv}{\int_{0.4}^{0.6} \big[ 1 + \partial \Pr\{ Y_i = 1, T_i = 0\mid e(Z_i) = e \} / \partial e\big|_{e=v} \big] dv}.
\end{equation*}
In the left panel,   $\theta_\mte(v)$ and $\theta_\ate$ are plotted as a function of $v$, when the outcome variable is not to vote for Unity. It can be seen that the marginal persuasive rate is about 50\% at $v=0.4$ but just 40\% for $v > 0.5$. In view of \Cref{eq:threshold}, $V_i$ can be interpreted as the unobserved cost of watching NTV. The estimation results suggest that the persuasive effect for not voting for Unity is stronger for those whose unobserved cost of watching NTV is lower. In the right panel, corresponding results are shown for OVR. In this case, the persuasive effect is weaker for those with lower values of $v$.

A striking pattern we can learn from \Cref{figure-example-NTV} is that persuasive effects are highly heterogeneous. This may partially answer the puzzle reported in EPZ.  They found relatively modest positive persuasive effects for opposition parties but much stronger persuasive effects for Unity using aggregate voting outcomes, while the magnitudes are similar using individual survey data.\footnote{EPZ estimated the persuasion rate using a continuous version of DK. See equations (3) and (4) in EPZ for their formulae.} Our estimation results indicate that the marginal persuasive effects are highly heterogeneous, thereby implying that different aggregate averages can be substantially different from each other. The average persuasive effect $\theta_\ate  [0.4, 0.6]$ is plotted as a horizontal line in each panel of \Cref{figure-example-NTV}: it is 39.7\% against Unity  and 12.5\% for OVR. In short, this application exemplifies the identification power of continuous instruments that can uncover the patterns of heterogeneity in persuasive effects.

\section{Additional Examples: Door-to-Door Fundraising}\label{sec:vcm}

\citet*{landry2006} and \citet{DLM} designed field experiments of door--to--door fundraising to examine various aspects of charity giving. In this section, we use their data to illustrate the usefulness of our identification results.

The common data structure in both papers is that for each type of experimental treatments,  we observe $(Y_i, T_i, Z_i)$:
\begin{itemize}
\item $Y_i = 1$ if a household made a contribution to door--to--door fundraising,
\item $T_i = 1$ if a household answered the door and spoke to a solicitor,
\item $Z_i = 1$ if a household was approached by a solicitor.
\end{itemize}
If $Z_i = 0$ (a household was not approached by a solicitor), then $T_i = 0$ and furthermore it is very likely that $Y_i=0$. Hence, in this section, we assume that
$\Pr(Y_i = 1, T_i = 0\mid Z_i = 0) = \Pr(Y_i = 1\mid Z_i = 0) = 0$. In addition, we assume that if $Y_i = 1$, it must be the case that $T_i = 1$. In other words, we assume that  it is  impossible to have both $Y_i = 1$ and $T_i = 0$ (making a contribution without answering the door). Thus,  $\Pr(Y_i = 1, T_i = 1\mid Z_i = 1) = \Pr(Y_i = 1\mid Z_i = 1)$. These assumptions were  also used in computation of the persuasion rates for donors in \citet{dellavigna2010persuasion}. Under these assumptions, we have the bound for $\theta_\ate$ as
\begin{equation*}
{\theta}_L	= {\Pr}(Y_i = 1\mid Z_i = 1)
\quad\text{and}\quad
{\theta}_U
= {\Pr}(Y_i = 1\mid Z_i = 1) + 1 - e(1).
\end{equation*}
In addition,
\begin{equation*}
{\theta}_{\textrm{local}}
=
\Pr(Y_i = 1\mid Z_i = 1)/e(1);
\end{equation*}
$\theta_\local$ is the same as the usual LATE.

\subsection{\texorpdfstring{\citet{landry2006}}{landry2006} Revisited}

In this study, there were four treatments: VCM (voluntary contributions mechanism), VCM with seed money, single-prize lottery, and multiple-prize lottery.
Using Table II of \citet{landry2006}, we compute the persuasive effects by treatment and report results in \Cref{LLLPR-sum-data}.

\begin{table}[htbp]
\caption{Persuasive Effect by Treatment in \citet{landry2006}}
\begin{center}
\begin{tabular}{cccccc}
\hline\hline
Treatment & ${\Pr}(Y_i = 1 \mid Z_i = 1)$ &  $e(1)$ & ${\theta}_L$ & ${\theta}_U$ &  ${\theta}_{\textrm{local}}$ \\
\hline
VCM &  9.5\% &  37.6\%  & 9.5\%  & 71.9\%  & 25.3\%  \\
VCM with seed money  & 5.2\% & 35.3\% & 5.2\%  & 69.9\%  & 14.8\%  \\
Single-prize lottery  & 17.1\%  & 37.7\% & 17.1\%  & 79.4\%  &  45.5\% \\
Multiple-prize lottery  & 12.6\% & 35.2\% & 12.6\%  & 77.5\%  & 35.9\%  \\
All  & 10.8\% & 36.3\% &  10.8\% & 74.5\%  & 29.7\%  \\
\hline
\end{tabular}
\end{center}
\label{LLLPR-sum-data}
\end{table}%

Based on the lower bound and the LATE parameter, it seems that the single--prize lottery is the most effective fundraising tool, whereas the VCM with seed money is the least effective. However, the identification regions for $\theta_\ate$ of all four treatments overlap and there is no clear ranking based on those. This suggests that if one cares about the persuasive effect for the population, the evidence is inconclusive.

\subsection{\texorpdfstring{\citet{DLM}}{DLM} Revisited} In their study of charity giving, \citet[][DLM hereafter]{DLM} designed both fundraising and survey treatments to test for altruism and social pressure in charity giving. In this section, we focus only on  three fundraising treatments: namely, the baseline treatment, the flyer treatment, and the opt-out treatment. The baseline treatment is the standard door--to--door funding raising campaign, the flyer treatment is with the flyer that provided information on fundraising  the date before the solicitation, and the opt--out treatment is with the flyer that had an additional feature of a ``Do Not Disturb'' checkbox.  There were two charities in each of the fundraising treatments: La Rabida Children's Hospital and
the East Carolina Hazard Center.

\begin{table}[htbp]
\caption{Persuasive Effect by Treatment in DLM}
\begin{center}
\begin{tabular}{cccccc}
\hline\hline
 Treatment & ${\Pr}(Y_i = 1\mid Z_i = 1)$ &  $e(1)$ & ${\theta}_L$ & ${\theta}_U$ &  ${\theta}_{\textrm{local}}$ \\
\hline
\multicolumn{6}{l}{La Rabida Children's Hospital} \\
Baseline &  7.1\% &  40.5\%  & 7.1\%  & 66.6\%  & 17.5\%  \\
Flyer      &  6.8\% &  36.4\%  & 6.8\%  & 70.4\%  & 18.8\%  \\
Opt--Out &  5.4\% &  30.4\%  & 5.4\%  & 74.9\%  & 17.7\%  \\
\hline
\multicolumn{6}{l}{East Carolina Hazard Center} \\
Baseline &  4.7\% &  43.0\%  & 4.7\% & 61.7\%  & 10.9\%  \\
Flyer      &  5.1\% &  39.6\%  & 5.1\%  & 65.5\%  & 12.9\%  \\
Opt--Out &  3.0\% &  34.4\%  & 3.0\%  & 68.6\%  & 8.6\%  \\
\hline
\end{tabular}
\end{center}
\label{DLM-sum-data}
\end{table}%

DLM pointed out that treatments were randomized within a date--solicitor time block and estimated linear probability models with covariates: solicitor fixed effects, date--town fixed effects, hourly time block fixed effects, and area rating dummies. We use the same specification as in DLM, estimate ${\Pr}(Y_i = 1 \mid Z_i = 1,X_i = x)$ and $e(1,x)$, and then average out the conditional estimates as in \Cref{DK-empirical}.\footnote{The data collected in DLM are available at \url{http://eml.berkeley.edu/~sdellavi/index.html}. As before, the predicted probabilities are truncated to be between 0 and 1, when they are averaged out.} The resulting estimates are reported in \Cref{DLM-sum-data}, where we report the persuasive effect by treatment/charity.

The local persuasion rate is point identified and is higher for the in--state charity, La Rabida Children's Hospital. The estimates of $\theta_\local$ are the highest for the flyer treatment in both charities. This does not mean that the flyer treatment is the most effective in fundraising for the general population. Note that the compliers of the baseline treatment are different from those of the flyer treatment. For example, it could be the case that households at the margin of giving might have decided not to answer the door after they noticed the flyer. Unlike $\theta_\local$, $\theta_L$ and $\theta_U$ are comparable across different treatments. However, as in the previous section, it is difficult to see whether there is a significant difference across  treatments if we focus on the bounds for $\theta_\ate$.\footnote{In addition to the fundraising treatments, DLM relied on survey treatments and structural estimates to draw conclusions in their paper.}

\section*{\Large Part III. Inference and Semiparametric Estimation}

\section{Inference on the Average Persuasion Rate}\label{inference-apr}

In this section, we provide methods for carrying out  inference on the average persuasion rate, for which we assume that the data are independent and identically distributed.

\subsection{The case of Theorem \ref{thm:fuzzy-best}}

In this subsection, we  consider Theorem \ref{thm:fuzzy-best}, where
 the sharp identified interval of $\theta_\ate$ is given by
$[\theta_L, \theta_U]$.
Recall that
\begin{align*}
\theta_L
&= \dfrac{\Pr(Y_i = 1\mid Z_i = 1) -\Pr(Y_i = 1\mid Z_i = 0)}{1-\Pr(Y_i = 1\mid Z_i = 0)}, \\
\theta_U
&= \dfrac{\Pr(Y_i = 1, T_i = 1\mid Z_i = 1)-\Pr(Y_i = 1, T_i = 0\mid Z_i = 0) + 1-e(1)}{1-\Pr(Y_i = 1, T_i = 0\mid Z_i = 0)}.
\end{align*}
To define the sample analog estimators of $\theta_L$ and  $\theta_U$, define
\begin{align*}
\widehat{\Pr}(Y_i = 1\mid Z_i = 1)
&= \frac{\sum_{i=1}^n \one(Y_i = 1, Z_i = 1)}{\sum_{i=1}^n \one(Z_i = 1)}, \\
\widehat{\Pr}(Y_i = 1\mid Z_i = 0)
&= \frac{\sum_{i=1}^n \one(Y_i = 1, Z_i = 0)}{\sum_{i=1}^n \one(Z_i = 0)}, \\
\widehat{\Pr}(Y_i = 1, T_i = 1\mid Z_i = 1)
&= \frac{\sum_{i=1}^n \one(Y_i = 1, T_i = 1, Z_i = 1)}{\sum_{i=1}^n \one(Z_i = 1)}, \\
\widehat{\Pr}(Y_i = 1, T_i = 0\mid Z_i = 0)
&= \frac{\sum_{i=1}^n \one(Y_i = 1, T_i = 0, Z_i = 0)}{\sum_{i=1}^n \one(Z_i = 0)}, \\
\widehat{e}(1)
 &= \frac{\sum_{i=1}^n \one(T_i = 1, Z_i = 1)}{\sum_{i=1}^n \one(Z_i = 1)}.
\end{align*} 	
Then we define
\begin{align*}
\widehat{\theta}_L 
&=
\dfrac{\widehat{\Pr}(Y_i = 1\mid Z_i = 1)-\widehat{\Pr}(Y_i = 1\mid Z_i = 0)}{1-\widehat{\Pr}(Y_i = 1\mid Z_i = 0)}, \\
\widehat{\theta}_U
&= 
\dfrac{\widehat{\Pr}(Y_i = 1, T_i = 1\mid Z_i = 1)-\widehat{\Pr}(Y_i = 1, T_i = 0\mid Z_i = 0) + 1-\widehat{e}(1)}{1-\widehat{\Pr}(Y_i = 1, T_i = 0\mid Z_i = 0)}.
\end{align*}
Since $\widehat{\theta}_L$ and $\widehat{\theta}_U$ are asymptotically jointly normal,
we follow \citet{Imbens/Manski:04} and  \citet{Stoye:07} to construct a confidence interval for $\theta_\ate$.
Standard arguments based on the delta method yield that $\widehat{\theta}_L$ and $\widehat{\theta}_U$ have asymptotically linear approximations:
\begin{align*}
\widehat{\theta}_L - \theta_L
&= n^{-1} \sum_{i=1}^n \varphi_{L,i} + o_p (n^{-1/2}), \\
\widehat{\theta}_U - \theta_U
&= n^{-1} \sum_{i=1}^n \varphi_{U,i} + o_p (n^{-1/2}),
\end{align*}
where $\varphi_{L,i}$ and $\varphi_{U,i}$ are  influence functions that can be approximated by the following sample analogs:
\begin{align*}
\widehat{\varphi}_{L,i} &=
\frac{1}{[ 1-\widehat{\Pr}(Y_i = 1\mid Z_i = 0) ]}
\frac{1}{\widehat{\Pr}(Z_i = 1)} \left\{
 \one(Y_i = 1, Z_i = 1) - \widehat{\Pr}(Y_i = 1, Z_i = 1) \right\} \\
&- \frac{1}{[ 1-\widehat{\Pr}(Y_i = 1\mid Z_i = 0) ]} \frac{\widehat{\Pr}(Y_i = 1\mid Z_i = 1)}{\widehat{\Pr}(Z_i = 1)}
\left\{ \one(Z_i = 1) - \widehat{\Pr}(Z_i = 1) \right\}
 \\
&+ \frac{\widehat{\theta}_L - 1}{[ 1-\widehat{\Pr}(Y_i = 1\mid Z_i = 0) ]}
\frac{1}{\widehat{\Pr}(Z_i = 0)} \left\{
 \one(Y_i = 1, Z_i = 0) - \widehat{\Pr}(Y_i = 1, Z_i = 0) \right\} \\
&- \frac{\widehat{\theta}_L - 1}{[ 1-\widehat{\Pr}(Y_i = 1\mid Z_i = 0) ]} \frac{\widehat{\Pr}(Y_i = 1\mid Z_i = 0)}{\widehat{\Pr}(Z_i = 0)}
\left\{ \one(Z_i = 0) - \widehat{\Pr}(Z_i = 0) \right\}, \\
\widehat{\varphi}_{U,i}
&=
\frac{1}{[ 1-\widehat{\Pr}(Y_i = 1, T_i = 0\mid Z_i = 0) ]}
\frac{1}{\widehat{\Pr}(Z_i = 1)} \\
&\times
  \left\{ \one(Y_i = 1, T_i = 1, Z_i = 1) - \widehat{\Pr}(Y_i = 1, T_i = 1, Z_i = 1) + 1(T_i = 0, Z_i = 1) - \widehat{\Pr}(T_i = 0, Z_i = 1) \right\} \\
&- \frac{1}{[ 1-\widehat{\Pr}(Y_i = 1, T_i = 0\mid Z_i = 0) ]} \frac{\widehat{\Pr}(Y_i = 1, T_i = 1, Z_i = 1) +  \widehat{\Pr}( T_i = 0, Z_i = 1)}{\widehat{\Pr}(Z_i = 1)} \\
& \times
 \left\{ \one(Z_i = 1) - \widehat{\Pr}(Z_i = 1) \right\}
 \\
&+ \frac{\widehat{\theta}_U - 1}{[ 1-\widehat{\Pr}(Y_i = 1, T_i = 0\mid Z_i = 0) ]}
\frac{1}{\widehat{\Pr}(Z_i = 0)} \left\{
 \one(Y_i = 1, T_i = 0, Z_i = 0) - \widehat{\Pr}(Y_i = 1, T_i = 0, Z_i = 0) \right\} \\
&- \frac{\widehat{\theta}_U - 1}{[ 1-\widehat{\Pr}(Y_i = 1, T_i = 0\mid Z_i = 0) ]} \frac{\widehat{\Pr}(Y_i = 1, T_i = 0\mid Z_i = 0)}{\widehat{\Pr}(Z_i = 0)}
\left\{ \one(Z_i = 0) - \widehat{\Pr}(Z_i = 0) \right\}.
\end{align*}
Now define
\begin{equation*}
\widehat{\sigma}^2_L
=
n^{-2} \sum_{i=1}^n \widehat{\varphi}_{L,i}^2,\quad
\widehat{\sigma}^2_U
=
n^{-2} \sum_{i=1}^n \widehat{\varphi}_{U,i}^2, \quad\text{and}\quad
\widehat{\Delta} = \widehat{\theta}_U - \widehat{\theta}_L.
\end{equation*}
That is, $\widehat{\sigma}_L$ and $\widehat{\sigma}_U$ are standard errors of $\widehat{\theta}_U$ and  $\widehat{\theta}_L$, respectively, and $\widehat{\Delta}$ is the estimated length of the identification region. Let
\begin{align}\label{CI-type1}
\text{CI}_\alpha^{\text{Theorem \ref{thm:fuzzy-best}}} 
=
\left[\widehat{\theta}_L - c_\alpha \widehat{\sigma}_L , \widehat{\theta}_U + c_\alpha \widehat{\sigma}_U  \right],
\end{align}
where $c_\alpha$ solves
\begin{align*}
\Phi \left(  c_\alpha + \frac{\widehat{\Delta}}
{\max \{\widehat{\sigma}_L,\widehat{\sigma}_U \}} \right)  -  \Phi \left(  - c_\alpha \right) =  1 - \alpha.
\end{align*}
Since $\widehat{\theta}_L \leq \widehat{\theta}_U$ by construction, Lemma 3 and Proposition 1 of \citet{Stoye:07} imply that $\theta_\ate \in \text{CI}_\alpha^{\text{Theorem \ref{thm:fuzzy-best}}}$ with probability $1-\alpha$ uniformly as $n \rightarrow \infty$, provided that the data generating process satisfies mild regularity conditions given in Assumption 1 (i) and (ii) of \citet{Stoye:07}.

 \subsection{The case of Theorem \ref{thm:fuzzy-e}}

Recall that in this case,   the sharp identified interval of $\theta_\ate$ is given by $[\theta_L,\ \theta_{U_e}]$, where
	\begin{equation*}
		\theta_{U_e}
		=
		\dfrac{ \min\{ 1, \Pr(Y_i = 1\mid Z_i = 1)+1-e(1) \} - \max\{0,\ \Pr(Y_i = 1\mid Z_i = 0)-e(0) \}}{1 - \max\{0,\ \Pr(Y_i = 1\mid Z_i = 0)-e(0) \}}.
	\end{equation*}
It is convenient to introduce additional notation. Let
\begin{equation*}
\xi_1 = \Pr(Y_i = 1\mid Z_i = 1)+1-e(1)
\quad\text{and}\quad
\xi_2 = \Pr(Y_i = 1\mid Z_i = 0)-e(0).
\end{equation*}	
Then $\xi_1$ and $\xi_2$ can be estimated by their sample analogs as before: i.e.,\
\begin{equation*}
\widehat{\xi}_1 = \widehat{\Pr}(Y_i = 1\mid Z_i = 1)+1-\widehat{e}(1)
\quad\text{and}\quad
\widehat{\xi}_2 = \widehat{\Pr}(Y_i = 1\mid Z_i = 0)-\widehat{e}(0).
\end{equation*}	
Furthermore, let $\widehat{\sigma}(\xi_1)$ and $\widehat{\sigma}(\xi_2)$ denote respective standard errors.\footnote{The standard errors can be constructed differently, depending on the sampling assumptions regarding the two marginal distributions of $(Y_i, Z_i)$ and $(T_i, Z_i)$.}  For any $\alpha \in (0,1/2)$, define $z_\alpha$ such that  $ \Phi(z_\alpha) = 1-\alpha$. Hence,  $z_{\alpha/2}$ is the two-sided standard normal critical value. We now choose $\overline{\alpha}$ that is close to zero, say $\overline{\alpha} = 0.001$, by which we define the following three possibilities.
\begin{itemize}
\item[(i)] $\widehat{\xi}_1  - z_{\overline{\alpha}/4} \cdot \widehat{\sigma}(\xi_1) \geq 1$;
\item[(ii)] $\widehat{\xi}_1  + z_{\overline{\alpha}/4} \cdot \widehat{\sigma}(\xi_1) \leq 1$ and $\widehat{\xi}_2  + z_{\overline{\alpha}/4} \cdot \widehat{\sigma}(\xi_2) \leq 0$;
\item[(iii)] neither (i) nor (ii).
\end{itemize}
The critical value $z_{\overline{\alpha}/4}$ is used here to reflect the fact that the two equalities are tested jointly against both positive and negative directions, which can be viewed as pretesting.
Case (i) corresponds to the case that  $\widehat{\xi}_1$ is much larger than 1, implying that the upper bound is 1.
Then we recommend using the confidence set such that
$\left[\widehat{\theta}_L - z_{\alpha-\overline{\alpha}} \widehat{\sigma}_L , 1  \right]$.
To accommodate the error in the pretesting stage, we recommend using the $z_{\alpha-\overline{\alpha}}$ critical value here.
Case (ii) suggests that  $\widehat{\xi}_1$ is much smaller  than 1 and  $\widehat{\xi}_2$ is sufficiently less than zero. Therefore, in this subcase, the upper bound reduces to $\theta_{U_e} = \xi_1$. Then the estimators of both lower and upper bounds are asymptotically linear, implying that one can use the confidence interval similar to \eqref{CI-type1}. Again, to accommodate the error in the pretesting stage, we recommend using the $z_{\alpha-\overline{\alpha}}$ critical value in applying  \eqref{CI-type1}.
This ensures that the asymptotic coverage probability is at least as large as $1-\alpha$ in applying  \eqref{CI-type1}.
It turns out that case (ii) was the relevant case for the empirical example reported \Cref{sec:main:example}.

Case (iii) is more complicated. We rely on a simple projection method to construct a valid confidence set.\footnote{A similar inference method is adopted in \citet{Horowitz:Lee:2020}; however, the subject matter is different here.}
It follows from the proofs of \Cref{lem:Manski2} and \Cref{thm:fuzzy-e} that
the two end points of the sharp identified interval of $\theta_\ate$ have the form:
\begin{equation}\label{idset-appendix}
	 \max_{a,b \in [0,1]^2} \text{ and } \min_{a,b \in [0,1]^2}\ \  \dfrac{a-b}{1-b}
	 \quad\text{s.t.}\quad
	 \left\{
	 \begin{aligned}
	& a\geq b, \\
	& a \in \bigl[ \Pr(Y_i = 1\mid Z_i = 1),\ \Pr(Y_i = 1\mid Z_i = 1)+1-e(1) \bigr], \\
	& b \in \bigl[ \Pr(Y_i = 1\mid Z_i = 0)-e(0),\ \Pr(Y_i = 1\mid Z_i = 0) \bigr].
	\end{aligned}
	\right.
\end{equation}

Rewrite the constraints above as
\begin{align*}
	& a \in \bigl[ \Exp( \one\{Y_i = 1\}  \mid Z_i = 1 ),\  \Exp (\one\{Y_i = 1\}  \mid Z_i = 1) + \Exp ( \one\{T_i=0\} \mid Z_i = 1)\bigr] =: [m_a, M_a], \\
	& b \in \bigl[ \Exp( \one\{Y_i = 1\}  \mid Z_i = 0)- \Exp(  \one \{T_i = 1 \} \mid Z_i = 0),\ \Exp( \one\{Y_i = 1\}  \mid Z_i = 0) \bigr]  =: [m_b, M_b].
\end{align*}
Since $a$ and $b$ are bounded by conditional expectations, it is simple to construct a joint confidence set for $a$ and $b$ in the form of the Cartesian product (e.g., using  a Bonferroni correction). That is, we can find the confidence set such that the following holds asymptotically with probability at least $1-\alpha$:
\begin{equation*}
	 a \in \bigl[ \hat{m}_a,\  \hat{M}_a \bigr]
	 \quad\text{and}\quad
	 b \in \bigl[ \hat{m}_b,\  \hat{M}_b \bigr].
\end{equation*}
Consequently, the following optima include the true identified set of $\theta_\ate$ defined in \Cref{idset-appendix}, asymptotically with probability at least $1-\alpha$:
\begin{equation*}
	 \max_{a,b \in [0,1]^2} \text{ and } \min_{a,b \in [0,1]^2}\ \  \dfrac{a-b}{1-b} \quad\text{s.t.}\quad
	 \left\{
	\begin{aligned}
	& a\geq b, \\
	& a \in \bigl[ \hat{m}_a,\  \hat{M}_a \bigr], \\
	& b \in \bigl[ \hat{m}_b,\  \hat{M}_b \bigr].
	\end{aligned}
	\right.
\end{equation*}

 \subsection{The case of Theorem \ref{thm:fuzzy-worst}}\label{inference-apr-worst}

In this case, the sharp bound of $\theta_\ate$ is given by $[\theta_L,1]$. Thus,
the one-sided confidence interval for $\theta_L$ provides the valid confidence set for $\theta_\ate$. That is,
$\theta_\ate \in \left[\widehat{\theta}_L - z_\alpha \widehat{\sigma}_L , 1  \right]$ holds asymptotically with probability at least $1-\alpha$,
where $z_\alpha$ satisfies  $ \Phi(z_\alpha) = 1-\alpha$.

\section{Inference on the Local Persuasion Rate}\label{inference-lpr}

\subsection{Case under \texorpdfstring{\Cref{ass:fuzzy-best}}{ass:fuzzy-best}}

Recall that in this case, $\theta_\local$ is point identified by $\theta_\local = \theta^*$, where
		\[
		\theta^* := \dfrac{\Pr(Y_i = 1\mid Z_i = 1) - \Pr(Y_i = 1\mid Z_i  = 0)}{\Pr(Y_i = 0, T_i = 0 \mid Z_i = 0) - \Pr( Y_i = 0, T_i = 0\mid Z_i = 1)}.
		\]
Thus, in this case, one can use the standard delta method to construct the two-sided confidence interval based on the asymptotic normality of the sample analog estimator of $\theta^*$.   Let $\widehat{\theta}^*$ and $\widehat{\sigma}^*$ denote the sample analog estimator of $\theta^*$ and its standard error, respectively. Then the random interval $[ \widehat{\theta}^* - z_{\alpha/2} \widehat{\sigma}^*, \widehat{\theta}^* + z_{\alpha/2} \widehat{\sigma}^*]$ includes $\theta_\local$ asymptotically with probability $1-\alpha$.

\subsection{Case under \texorpdfstring{\Cref{ass:fuzzy-e}}{ass:fuzzy-e}}

The sharp identified interval of $\theta_\local$ is given by $[\theta_L^*, 1]$, where $\theta_L^* = \max(\mathrm{Wald},\ \theta_L)$ with
\[
\mathrm{Wald} := \dfrac{\Pr(Y_i = 1\mid Z_i = 1) - \Pr(Y_i = 1\mid Z_i  = 0)}{e(1) - e(0)}.
\]
Thus, the lower bound on $\theta_\local$ has the form of the intersection bound \citep{CLR}. To describe how to conduct inference, define the sample analog estimator of $\widehat{\mathrm{Wald}}$:
\[
\widehat{\mathrm{Wald}} := \dfrac{\widehat{\Pr}(Y_i = 1\mid Z_i = 1) - \widehat{\Pr}(Y_i = 1\mid Z_i  = 0)}{\widehat{e}(1) - \widehat{e}(0)}.
\]
Let $\widehat{\sigma}_{\mathrm{Wald}}$ be its standard error. Then, the random interval $[ \widehat{\theta}_L^*(\alpha), 1 ]$ includes
$\theta_\local$ with probability at least $1-\alpha$, where
\[
\widehat{\theta}_L^*(\alpha)
:=
\max \{0,\  \widehat{\mathrm{Wald}} - z_{\alpha/2} \widehat{\sigma}_{\mathrm{Wald}},\  \widehat{\theta}_L - z_{\alpha/2} \widehat{\sigma}_L \}.
\]
Here, the critical value $z_{\alpha/2}$ is based on simple  Bonferroni correction.  An adaptive inequality selection proposed in \citet{CLR} can be adopted to construct a sharper critical value than $z_{\alpha/2}$.

\subsection{Case under \texorpdfstring{\Cref{ass:fuzzy-worst}}{ass:fuzzy-worst}}

In this case, the sharp identified interval of $\theta_\local$ coincides with that of $\theta_\ate$.  Hence, one can use the  inference method presented in \Cref{inference-apr-worst}.

\section{Semiparametric Estimation}\label{sec:inference}

\subsection{Efficient Estimation}\label{subsec:inference}
In this section, we are explicit about the vector $X_i$ of exogenous covariates and consider semiparametrically efficient estimation of the two key parameters, i.e.,\ $\theta_L$ and $\theta^*$.\footnote{Efficient estimation of $\theta_L^*$ or $\theta_{U_e}$ is significantly more challenging because of the min/max function. It also requires a careful construction of an estimator for the marginal persuasion rate. These are beyond the scope of the current paper but interesting topics for future research.}
We focus on independent and identically distributed (i.i.d.)\ data again.  For $\theta_L$ we work with the dataset $\{ (Y_i, Z_i, X_i^\tr)^\tr:\ i=1,2,\cdots n\}$, whereas we use $\{ (Y_i, T_i, Z_i, X_i^\tr)^\tr:\ i=1,2,\cdots n\}$ for $\theta^*$.
Since we are now explicit about $X_i$, the objects of interest will be defined by integrating $X_i$ out as we explained in \cref{DK-empirical}: when we analyze the local persuasion rate, we will use the conditional distribution of $X_i$ given the group of compliers. Let $\tilde\theta_L(x)$ be the sharp lower bound conditional on $X_i = x$ and let $\tilde \theta^*(x)$ be the local persuasion rate conditional on $X_i = x$: i.e., $\tilde \theta_L(x) = \tilde \theta_N(x) / \{ 1-\tilde\theta_{L,D}(x)\}$ and $\tilde\theta^*(x) = \tilde \theta_N (x) / \tilde \theta_D^*(x)$, where
\begin{align*}
\tilde \theta_N(x)
&:=
\Pr(Y_i=1\mid Z_i = 1,X_i=x) - \Pr(Y_i=1\mid Z_i=0,X_i=x), \\ 
\tilde\theta_{L,D}(x)
&:= 
\Pr(Y_i = 1\mid Z_i = 0, X_i = x), \\
\tilde \theta_D^*(x)
&:= 
\Pr(Y_i=0,T_i=0\mid Z_i=0,X_i=x) - \Pr(Y_i=0,T_i=0\mid Z_i=1,X_i=x). 
\end{align*}
Now, we take care of $X_i$ as we explained in \cref{DK-empirical} to obtain the parameters we consider estimating in this section: i.e.,
\begin{equation} \label{eq:theta_L theta^*}
	\theta_L
	=
	\frac{\int \tilde\theta_N(x) dF(x)}{1-\int \tilde\theta_{L,D}(x)dF(x)},
    \quad
	\theta^* 
	= 
	\frac{\int \tilde\theta_N(x) dF(x)}{\int \tilde \theta_D^*(x) dF(x)},
\end{equation}
provided that $\int \tilde\theta_D^*(x) dF(x)\neq 0$ and $\int \tilde\theta_{L,D}(x) dF(x) < 1$. 

Semiparametric estimation of $\theta_L$ and $\theta^*$ is straightforward: i.e.,
\begin{equation}\label{eq:estimators}
\hat\theta_L
:=
\frac{n^{-1}\sum_{i=1}^n \hat{\tilde{\theta}}_N(X_i)} {1-n^{-1}\sum_{i=1}^n \hat{\tilde{\theta}}_{L,D}(X_i)}
\quad\text{and}\quad
\hat\theta^*
:=
\frac{\sum_{i=1}^n \hat{\tilde{\theta}}_N(X_i)}{\sum_{i=1}^n \hat{\tilde{\theta}}_D^*(X_i)},
\end{equation}
where $\hat{\tilde{\theta}}_N(x), \hat{\tilde{\theta}}_{L,D}(x)$, $\hat{\tilde{\theta}}_D^*(x)$ are defined by replacing the probabilities in the definition of $\tilde \theta_N(x), \tilde\theta_{L,D}(x)$, and $\tilde \theta_D^*(x)$ with their consistent estimators, respectively.

Semiparametric estimators like the ones in \Cref{eq:estimators} converge at the usual $\sqrt{n}$ rate. Instead of listing all regularity conditions, which are well understood in the literature \citep[see, e.g.,][]{newey1994asymptotic, Ai2003, Chen2003, Ai2012, Ackerberg2014}, we will derive the pathwise derivatives of $\theta_L$ and $\theta^*$. The theorem stated below is the first main result of this section. 

\begin{theorem}\label{thm:efficiency}
	Suppose that $e(0,x)<e(1,x)$ with $\inf_x e(0,x)>0$ and $\sup_x e(1,x)<1$, and that $0<\inf_x \Pr(Z_i = 1\mid X_i = x) \leq \sup_x \Pr(Z_i = 1\mid X_i = x) < 1$. Then, the numerators and denominators of $\theta_L$ and $\theta^*$ are all pathwise differentiable in the sense of \citet{newey1994asymptotic}.
\end{theorem}

\Cref{thm:efficiency} does not display the specific forms of the pathwise derivatives simply because their expressions are too long and distracting; they are provided in \Cref{lem:F1F0 eff,lem:FnFd eff}.\footnote{The regularity conditions imposed on the exposure rates in \Cref{thm:efficiency} may not be necessary but we do not explore minimal conditions here.} Below we discuss the relevance of \Cref{thm:efficiency}.

The pathwise differentiability can tell us a couple of things about the semiparametric estimators $\hat \theta_L$ and $\hat \theta^*$. Specifically, we can combine the pathwise derivatives of the numerators and denominators of $\theta_L$ and $\theta^*$ with the delta method, and we can find out what the influence functions $F_L(Y,Z,X)$ and $F^*(Y,T,Z,X)$ of $\hat\theta_L$ and $\hat \theta^*$ should be if they are asymptotically linear: see \Cref{lem:pathwise theta_L,lem:pathwise theta*} for the explicit expressions of $F_L(Y,T,Z,X)$ and $F^*(Y,T,Z,X)$. Therefore, the asymptotic variance of $\hat \theta_L$ and $\hat \theta^*$ will be $\Var\bigl\{ F_L(Y_i,Z_i,X_i)\bigr\}$ and $\Var\bigl\{ F^*(Y_i,T_i,Z_i,X_i)\bigr\}$, respectively.  Further, we show that (any linear combination of) the pathwise derivatives of the numerator and denominator of $\theta_L$ is contained in the appropriate tangent space and that the same is true for $\theta^*$. What this means is that $\Var\bigl\{ F_L(Y_i,Z_i,X_i)\bigr\}$ and $\Var\bigl\{ F^*(Y_i,T_i,Z_i,X_i)\bigr\}$ are in fact the semiparametric efficiency bounds of $\theta_L$ and $\theta^*$, respectively.  We summarize these implications in the following theorem.

\begin{theorem}\label{thm:inference}
    Suppose that the conditions stated in \Cref{thm:efficiency} are satisfied. If $\hat \theta_L$ and $\hat\theta^*$ are $\sqrt{n}$--consistent and asymptotically linear, then their asymptotic variances are given by 
    \[
        V_L = \Exp\bigl\{ F_L^2(Y_i,Z_i,X_i) \bigr\}
        \quad \textrm{and}\quad 
        V^* = \Exp\bigl\{ F^{*2}(Y_i,T_i,Z_i,X_i)\bigr\}.  
    \]    
    Further, $V_L$ and $V^*$ are the semiparametric efficiency bounds for estimating $\theta_L$ and $\theta^*$, respectively. 	
\end{theorem}

By using the formulas of $F_L$ and $F^*$ provided in \Cref{lem:pathwise theta_L,lem:pathwise theta*}, we can consistently estimate $V_L$ and $V^*$.  Alternatively, one can simply rely on some resampling techniques such as the bootstrap.  Once we obtain the estimates $\hat V_L$ and $\hat V^*$ of the asymptotic variances, we can conduct inference on $\theta_L$ and $\theta^*$. In fact, inference on $\theta_L$ can be naturally extended to that of $\theta_\ate$.  Consider the simplest case where there is no uncertainty in the upper bound: i.e.,\ we only have data on $(Y_i, Z_i, X_i^\tr)^\tr$. In this case, the sharp identified bounds of $\theta_\ate$ is simply $[\theta_L,\ 1]$, the asymptotically valid confidence interval for $\theta_\ate$ with the shortest length will be $\bigl[ \hat \theta_L - z_{\alpha} \times \sqrt{\hat V_L/n},\ 1\bigr]$, where $z_{\alpha}$ is the one--sided critical value from the standard normal distribution (e.g.,\ when $\alpha = 0.05$, we have $z_{\alpha} = 1.645$).   When we observe all of $(Y_i,T_i,Z_i,X_i^\tr)^\tr$, $\theta^*$ is point identified and therefore the shortest confidence interval of $\theta^*$ can be obtained by using the usual two--sided critical value as in $\hat \theta^* \pm z_{\alpha/2}\times \sqrt{\hat V^*/n}$.

\subsection{Revisiting the NTV Example}

We revisit the NTV example in \Cref{sec:NTV} to illustrate the results of \Cref{subsec:inference}. For simplicity, we focus on estimating the lower bound $\theta_L$ using a two--step parametric approach. For this exercise, we first create a binary instrument
\begin{equation*}
Z_i := \one\bigl\{ \text{Signal Power}_i > \text{median(Signal Power)} \bigr\}
\end{equation*}
by using the original continuous instrument. The conditional lower bound $\tilde\theta_L(x)$ is estimated using probit models that are linear in covariates used in \Cref{sec:NTV} and is averaged out with respect to covariates by sample survey weight.
The standard error is obtained by replacing population quantities of $F_L(Y_i,Z_i,X_i)$ in \Cref{lem:pathwise theta_L} with parametric estimates.

\begin{table}[htbp]
\centering
\caption{Persuasion Rates: NTV Effects Using a Binary Instrument \label{NTV-results-more} }
\begin{tabular}{lcc}
\hline \hline
                              & (1)                      & (2) \\
                              & Not voting for  & Voting for  \\
                              & Unity & OVR \\
\hline
Point estimate of the lower bound      & 0.203 &    0.056   \\
Standard error of the lower bound & 0.072 &    0.024  \\
One-sided 95\% confidence interval for $\theta_\ate$ & [0.084,1] & [0.017,1]  \\
\hline
\end{tabular}

\end{table}%

The estimation results are summarized in \Cref{NTV-results-more}.
Both lower bounds are significantly different from zero; however, they are far smaller than the estimates of $\theta_\ate$ based on
the original continuous instrument. This again illustrates the limitation of the identifying power of a binary instrument.

\section*{\Large Part IV. Proofs}

\section{Proofs}\label{app:proofs}

\subsection{Fr\'{e}chet--Hoeffding Inequalities}

\begin{lemma}\label{lem:frechet-hoeffding}
	For any events $A,B$ and for any probability measure $\Pr^*$, we have
	\[
	\max\{ 0, \Pr^*(A) - \Pr^*(B^c) \}
	\leq
	\Pr^*( A\cap B)
	\leq
	\min\{\Pr^*(A),\ \Pr^*(B) \},
	\]
	where the bounds are sharp in that $\Pr^*(A\cap B)$ can be anything between the bounds without changing $\Pr^*(A)$ and $\Pr^*(B)$.
	\proof
	This is a version of the Fr\'{e}chet--Hoeffding bounds. 
	We prove it here for self-containment.
	The upper bound is trivially true. For the lower bound, simply note that
	\[
	\Pr^*(A)\leq \Pr^*(A\cap B) + \Pr^*(B^c)
	\quad\text{and}\quad
	\Pr^*(B) \leq \Pr^*(A\cap B) + \Pr^*(A^c),
	\]
	where $\Pr^*(A) - \Pr^*(B^c) = \Pr^*(B) - \Pr^*(A^c)$. For sharpness, note that the upper bound is achieved when $A\subset B$ or $B\subset A$. Also, the lower bound is achieved when $A\cap B=\emptyset$ or $A\cup B=\Omega$, where $\Omega$ is the entire sample space. To show that anything between the bounds can be achieved,  consider the canonical probability space without loss of generality: i.e.,\ $\Omega = [0,1]$ and $\Pr^*$ be the Lebesgue measure on the Borel $\sigma$--algebra on $\Omega$.  Choose $p_A, p_B \in [0,1]$, where we assume that $p_A\geq p_B$ without loss of generality; the other case is symmetric. Let $A = [0,\ p_A]$ and $B=x+[p_A-p_B,\ p_A]$, where $0\leq x\leq 1-p_A$. So, $A,B$ are events in $\Omega$ with $\Pr^*(A)=p_A, \Pr^*(B)=p_B$ for all $x \in [0, 1-p_A]$. Now, note that $\Pr^*(A\cap B) = \max(p_B-x,\ 0)$ is a continuous function in $x\in[0, 1-p_A]$, where  its maximum and the minimum are given by $p_B$ and $\max(p_B-1+p_A,0)$.  \qed
\end{lemma}

\subsection{Proof of \texorpdfstring{\Cref{lem:direction}}{lem:direction}}

First of all, applying \Cref{lem:frechet-hoeffding} with $P^*(A) = \Pr\{ Y(1)=1\}$ and $P^*(B) = \Pr\{Y(0) = 0\}$ yields \Cref{lemma1-result0}. For \Cref{lemma1-result}, we show that $\Pr\{Y_i(1)=1\} - \Pr\{ Y_i(0)=1\} = \Pr\{ Y_i(1)=1,Y_i(0) = 0 \}$ if and only if $Y_i(1)\geq Y_i(0)$ with probability one. First, suppose that $Y_i(1)\geq Y_i(0)$. Then, $Y_i(1) - Y_i(0) = \one\{ Y_i(1) = 1, Y_i(0) = 0 \}$ with probability one, and hence taking the expectation on both sides shows the claim. Conversely, if $\Pr\{Y_i(1)=1\} - \Pr\{ Y_i(0)=1\} = \Pr\{ Y_i(1)=1,Y_i(0) = 0 \}$, then the fact that $\Pr\{Y_i(1) = 1\} = \Pr\{Y_i(1)=1,Y_i(0)=1\} + \Pr\{Y_i(1)=1,Y_i(0)=0\}$ and $\Pr\{Y_i(0) = 1\} = \Pr\{Y_i(1)=1,Y_i(0)=1\} + \Pr\{Y_i(1)=0,Y_i(0)=1\}$ shows that $\Pr\{Y_i(1)=0,Y_i(0)=1\} = 0$. \qed

\subsection{Proof of \texorpdfstring{\Cref{lem:direction:app}}{lem:direction:app}}

If $q_i(0) \leq q_i(1)$, then $Y_i(0) = 1$ and $Y_i(1) = 0$ cannot happen. Now, conversely, suppose that $Y_i(0) \leq Y_i(1)$. If $\Pr\{ q_i(1) < q_i(0) \} >0$, then \Cref{ass:dist} implies that 	$\Pr\{ q_i(1)< 1/(1+U_i) < q_i(0) \} > 0$. This contradicts $\Pr\{ Y_i(0)\leq Y_i(1)\} = 1$. \qed

\subsection{Proof of \texorpdfstring{\Cref{thm:sharp}}{thm:sharp}}
By \Cref{ass:instrument},
\begin{equation*}
\Pr\{ Y_i(z)=1\}
=
\Pr( Y_i(z)=1 \mid Z_i = z)
=
\Pr( Y_i = 1\mid T_i = z)
=
\Pr( Y_i = 1\mid Z_i = z).
\end{equation*}
So, the assertion follows from \Cref{lem:direction} and the definition of $\theta_L$. \qed

\subsection{Using \texorpdfstring{\Cref{ass:binary-monotone,ass:instrument,ass:fuzzy-best}}{ass:fuzzy-best}}

\begin{lemma}\label{lem:no compliers}
	Suppose that \Cref{ass:instrument,ass:fuzzy-best} hold. For $z = 0,1$, $\Pr\{ Y_i(1) = 1, V_i \leq e(z) \}$ is identified by
	\begin{equation*}
		\Pr(Y_i = 1, T_i = 1 \mid Z_i = z).
	\end{equation*}
	Similarly, $\Pr\{ Y_i(0) = 1, V_i > e(z) \}$ is identified by
	\begin{equation*}
		\Pr(Y_i = 1, T_i = 0\mid Z_i = z).
	\end{equation*}
\proof
It follows from \Cref{ass:instrument}. \qed
\end{lemma}

\begin{lemma}\label{lem:compliers}
	Suppose that \Cref{ass:instrument,ass:fuzzy-best} hold. 
	Then, $\Pr\{ Y_i(1) = 1 \mid e(0) < V_i \leq e(1) \}$ is identified by
	\begin{equation*}
		\dfrac{\Pr(Y_i = 1, T_i = 1\mid Z_i = 1) - \Pr(Y_i = 1, T_i = 1\mid Z_i = 0) }{e(1) - e(0)}.
	\end{equation*}
	Similarly, $\Pr\{ Y_i(0) = 1 \mid e(0) < V_i \leq e(1) \}$ is identified by
	\begin{equation*}
		\dfrac{\Pr(Y_i = 1, T_i = 0\mid Z_i = 0) - \Pr(Y_i = 1, T_i = 0\mid Z_i = 1) }{e(1) - e(0)}.
	\end{equation*}
\proof
The first assertion follows from \Cref{lem:no compliers} because
\begin{align*}
\Pr\{ Y_i(1) = 1, e(0) < V_i \leq e(1) \}
= \Pr\{ Y_i(1) = 1, V_i \leq e(1) \} - \Pr\{ Y_i(1) = 1, V_i \leq e(0) \}.
\end{align*}
The second statement is similar. \qed
\end{lemma}

\begin{lemma}\label{lem:general}
 Suppose that \Cref{ass:instrument,ass:fuzzy-best} hold. Then,
	\begin{align*}
		\Pr\{ Y_i(1) = 1\} 
		&=
		\Pr(Y_i = 1, T_i = 1\mid Z_i=1) + \Pr\{ Y_i(1) = 1, V_i > e(1) \},\\
		\Pr\{ Y_i(0) = 1\} 
		&=
		\Pr(Y_i = 1, T_i = 0\mid Z_i=0) + \Pr\{ Y_i(0) = 1, V_i \leq e(0) \}.
		\end{align*}
\proof 
For the first assertion, note that 
\begin{multline}\label{eq:Manski}
	\Pr\{ Y_i(1) = 1\}
	=
	\Pr\{ Y_i(1) = 1 \mid e(0) < V_i \leq e(1) \} \{ e(1) - e(0) \} \\
	+
	\Pr\{ Y_i(1) = 1, V_i \leq e(0) \}
	+
	\Pr\{ Y_i(1) = 1, V_i > e(1) \}.
\end{multline}
By \Cref{lem:compliers,lem:no compliers}, the first two terms on the right--hand side of \Cref{eq:Manski} are identified, and their sum is equal to $\Pr(Y_i = 1,T_i = 1\mid Z_i = 1)$. The second assertion is similar. \qed
\end{lemma}

\begin{lemma}\label{lem:Manski}
	Suppose that \Cref{ass:binary-monotone,ass:instrument,ass:fuzzy-best} hold. Then, the sharp identified interval of $\Pr\{ Y_i(1) = 1\}$ is given by
	\begin{equation*}
	\bigl[\Pr(Y_i = 1\mid Z_i =1),\   \Pr(Y_i = 1, T_i = 1\mid Z_i = 1) + 1-e(1) \bigr].
	\end{equation*}
	Similarly, the sharp identified interval of $\Pr\{ Y_i(0) = 1\}$ is given by
	\begin{equation*}
	\bigl[ \Pr(Y_i = 1, T_i = 0\mid Z_i = 0),\  \Pr(Y_i = 1 \mid Z_i = 0) \bigr].
	\end{equation*}
\proof
For the first assertion, recall from \cref{lem:general} that 
\begin{equation} \label{eq:Manski-edited}
	\Pr\{ Y_i(1) = 1\} 
	=
	\Pr(Y_i = 1, T_i = 1\mid Z_i=1) + \Pr\{ Y_i(1) = 1, V_i > e(1) \},
\end{equation} 
where 
\[
    \Pr(Y_i = 1, T_i = 0\mid Z_i = 1)
    =
    \Pr\{ Y_i(0) = 1, V_i > e(1) \}
    \leq
    \Pr\{ Y_i(1) = 1, V_i > e(1) \}
    \leq
    1-e(1).
\] 
Here, the first inequality is sharp under \Cref{ass:binary-monotone} and so is the second one by \Cref{lem:frechet-hoeffding}. Therefore, the sharp bounds of the second term on the right--hand side of \Cref{eq:Manski-edited} is the interval between $\Pr(Y_i = 1, T_i = 0\mid Z_i = 1)$ and $1-e(1)$.  Combining all these proves the first assertion. The second assertion is similar.  \qed
\end{lemma}

\subsection{Proof of \texorpdfstring{\Cref{thm:fuzzy-best}}{thm:fuzzy-best}}

Let $a = \Pr\{ Y_i(1) = 1\}$ and $b = \Pr\{ Y_i(0) = 1\}$: so, $\theta_\ate = (a-b)/(1-b)$. Let $m_a, M_a$ be the lower and upper bounds of $a$ provided in \Cref{lem:Manski}. Similarly, let $m_b, M_b$ be the bounds of $b$ given in \Cref{lem:Manski}.  By \Cref{lem:Manski} and the fact that the dependence between $Y_i(0)$ and $Y_i(1)$ is unrestricted except that $Y_i(0)\leq Y_i(1)$, the identified upper bound of $\theta_\ate$ can be obtained by
\begin{equation*}
	\max_{a,b}\ \dfrac{a-b}{1-b}
	\quad\text{subject to}\quad
	a \in [m_a,\ M_a],\
	b \in [m_b,\ M_b],\
	a\geq b.
\end{equation*}
We can find the identified lower bound by solving minimization instead of maximization.  Here, note that the restriction $a\geq b$ is redundant, because
\begin{multline*}
	m_a - M_b
	=
	\Pr( Y_i=1\mid Z_i=1) - \Pr(Y_i = 1\mid Z_i=0) \\
	=
	\Pr\{ Y_i(1)=1,\ e(0) < V_i\leq e(1) \} - \Pr\{ Y_i(0)=1,\ e(0) < V_i \leq e(1) \} \geq 0,
\end{multline*}
where the last inequality is from $Y_i(1) \geq Y_i(0)$. So, the minimum is $\theta_L = (m_a - M_b)/(1-M_b)\geq 0$ and the maximum is $\theta_U=(M_a - m_b)/(1-m_b) \leq 1$: both $\theta_L$ and $\theta_U$ are well-defined because $m_b\leq M_b < 1$.  Finally, sharpness follows from the intermediate value theorem because $(a-b)/(1-b)$ varies continuously between $\theta_L$ and $\theta_U$.  \qed

\subsection{Proof of \texorpdfstring{\Cref{thm:fuzzy-best-wo-mono}}{thm:fuzzy-best-wo-mono}}
\Cref{lem:general} shows that
\begin{align}
    \Pr(Y_i = 1, T_i = 1\mid Z_i=1)
    \leq
    \Pr\{ Y_i(1) = 1\} 
    &\leq 
    \Pr(Y_i = 1, T_i = 1\mid Z_i=1) + 1-e(1)
    \leq 
    1,
    \label{eq:general1}
    \\
    \Pr(Y_i = 1, T_i = 0\mid Z_i=0)
    \leq 
    \Pr\{ Y_i(0) = 1\} 
    &\leq
    \Pr(Y_i = 1, T_i = 0\mid Z_i=0) + e(0)
    \leq 
    1 
    \label{eq:general2}
\end{align}
if no restrictions are imposed on $\Pr\{ Y_i(1)= 1, V_i > e(1)\}$ and $\Pr\{ Y_i(0) = 1, V_i \leq e(0)\}$: these bounds are sharp. Therefore, combining \eqref{eq:general1}, \eqref{eq:general2} with \eqref{lemma1-result0} in \Cref{lem:direction} and following the same reasoning as the proof of \Cref{thm:fuzzy-best} shows the first assertion. We remark that the denominator in the upper bound is not equal to zero because $\Pr(Y_i = 1, T_i=0\mid Z_i = 0) \leq \Pr(Y_i = 1\mid Z_i = 0) < 1$. Also, if $\Pr(Y_i = 1, T_i=0\mid Z_i = 0)+ e(0)\uparrow 1$, then the lower bound goes to $\max(0, -\infty) = 0$ because $\Pr(Y_i = 1\mid Z_1 = 1)<1$. Therefore, the lower bound is well-defined as well.  For the second assertion, suppose that the stated stochastic dominance condition holds in addition. Then, it follows from \Cref{lem:general} that 
\begin{align}
    \Pr(Y_i = 1\mid Z_i=1)
    \leq
    \Pr\{ Y_i(1) = 1\} 
    &\leq 
    \Pr(Y_i = 1, T_i = 1\mid Z_i=1) + 1-e(1)
    \leq 
    1,
    \label{eq:general5}
    \\
    \Pr(Y_i = 1, T_i = 0\mid Z_i=0)
    \leq 
    \Pr\{ Y_i(0) = 1\} 
    &\leq
    \Pr(Y_i = 1, \mid Z_i=0)
    \label{eq:general6}
\end{align}
where the bounds are sharp. Using \eqref{eq:general5}, \eqref{eq:general6}, and \eqref{lemma1-result0} in \Cref{lem:direction}, we follow the same reasoning as the proof of \Cref{thm:fuzzy-best} to obtain the second result of the theorem. Again, the bounds are all well-defined because $\Pr(Y_i = 1\mid Z_i = 0) < 1$. \qed

\subsection{Using \texorpdfstring{\Cref{ass:binary-monotone,ass:instrument,ass:fuzzy-e}}{ass:fuzzy-e}}

\begin{lemma}\label{lem:Manski2}
	Suppose that \Cref{ass:binary-monotone,ass:instrument,ass:fuzzy-e} hold. The sharp identified interval of $\Pr\{ Y_i(1)=1\}$ is given by
	\begin{equation*}
		\bigl[ \Pr(Y_i = 1\mid Z_i = 1),\ \min\{ 1, \Pr(Y_i = 1\mid Z_i = 1)+1-e(1) \} \bigr]
	\end{equation*}
	Similarly, the sharp identified interval of $\Pr\{ Y_i(0) = 1\}$ is given by \begin{equation*}
		\bigl[ \max\{0,\ \Pr(Y_i = 1\mid Z_i = 0)-e(0) \},\ \Pr(Y_i = 1\mid Z_i = 0) \bigr].
	\end{equation*}
\proof
We start from \Cref{lem:Manski}, and we apply \Cref{lem:frechet-hoeffding} to address the difference between \Cref{ass:fuzzy-best,ass:fuzzy-e}. \qed
\end{lemma}

\subsection{Proof of \texorpdfstring{\Cref{thm:fuzzy-e}}{thm:fuzzy-e}}

Similarly to the proof of \Cref{thm:fuzzy-best}, we need to consider
\begin{equation*}
	\max_{a,b} \text{ and } \min_{a,b}\ \  \dfrac{a-b}{1-b}
	\quad\text{subject to}\quad
	a \in [m_a, \tilde M_a],\
	b \in [ \tilde m_b, M_b],\
	a\geq b,
\end{equation*}
where $m_a, \tilde M_a, \tilde m_b, M_b$ are given in \Cref{lem:Manski2}. Follow the same reasoning as \Cref{thm:fuzzy-best}. \qed

\subsection{Proof of \texorpdfstring{\Cref{thm:fuzzy-worst}}{thm:fuzzy-worst}}

Since \Cref{thm:fuzzy-e} uses more information but its lower bound only depends on the distribution of $(Y_i, Z_i)$, it suffices to focus on the upper bound. From \Cref{thm:fuzzy-e}, we can find the sharp upper bound in this case by 
\begin{equation*}
	\max_{0 \leq e(0) \leq e(1) \leq 1 }\ 
    \theta_{U_e}
    =
    \dfrac{ \min\{ 1, \Pr(Y_i = 1\mid Z_i = 1)+1-e(1) \} - \max\{0,\ \Pr(Y_i = 1\mid Z_i = 0)-e(0) \}}{1 - \max\{0,\ \Pr(Y_i = 1\mid Z_i = 0)-e(0) \}}.
\end{equation*}
Note that setting $e(0) = \Pr(Y_i = 1\mid Z_i = 0)\leq \Pr(Y_i = 1\mid Z_i = 1) = e(1)$ yields the maximum value $1$. Sharpness follows from the fact that $\theta_{U_e}$ is continuous in $\bigl(e(0), e(1) \bigr)$.
\qed

\subsection{For the Compliers}\

\noindent
\textbf{Proof of \Cref{thm:local}: }
For part \textit{\ref{part1-local}}, note that
\[
	\Pr(Y_i = 1 \mid Z_i = z)
	=
	\Pr\{ Y_i(1) = 1,  V_i \leq e(z) \}
	+
	\Pr\{ Y_i(0) = 1, V_i > e(z) \},
\]
from which it follows that
\begin{equation}\label{eq:lateproof}
\theta_\local
=
\dfrac{ \Pr(Y_i = 1 \mid Z_i = 1) - \Pr(Y_i = 1\mid Z_i = 0)}{\{ e(1) - e(0) \} \Pr\{ Y_i(0) = 0 \mid e(0) < V_i \leq e(1) \}}.
\end{equation}
Finally, note that the denominator on the right--hand side of \Cref{eq:lateproof} is equal to
\[
\Pr\{ Y_i(0) = 0, e(0) < V_i \leq e(1) \}
=
\Pr\{ Y_i(0) = 0, e(0) < V_i  \}
-
\Pr\{ Y_i(0) = 0, e(1) < V_i  \},
\]
where $\Pr\{ Y_i(0) =0, e(z) < V_i \} = \Pr\{ Y_i =0, T_i = 0  \mid Z_i = z\}$.

For part \textit{\ref{part2-local}}, we look for sharp bounds for $\Pr\{ Y_i(0) = 0, e(0) < V_i \leq e(1) \}$ under \Cref{ass:fuzzy-e}. Using the fact that the sharp bounds of $\Pr(A\cap B\cap C)$ when $\Pr(A\cap B), \Pr(B\cap C)$, and $\Pr(C\cap A)$ are given are equal to the interval between $0$ and $\min\{ \Pr(A\cap B), \Pr(B\cap C), \Pr(C\cap A) \}$, we know that
\begin{multline}\label{eq:late-e1}
0
\leq
\Pr\{ Y_i(0) = 0, e(0) < V_i \leq e(1) \} \\
\leq
\min\bigl[ \Pr\{ Y_i(0) = 0, V_i> e(0) \},\ e(1) - e(0),\ \Pr\{ Y_i(0) = 0, V_i\leq e(1) \} \bigr],
\end{multline}
where it suffices to look for the sharp upper bound of the expression on the utmost right--hand side. First,
\[
\Pr\{ Y_i(0) = 0, V_i>e(0)\}
=
\Pr( Y_i = 0, T_i = 0 \mid Z_i = 0 )
\leq
\Pr(Y_i = 0\mid Z_i = 0),
\]
where the inequality holds with equality when $\Pr(Y_i = 0, T_i = 1\mid Z_i = 0) = 0$. Second, note that
\[
\Pr\{ Y_i(0) = 0, V_i\leq e(1) \}
=
\Pr\{ Y_i(0) = 0, T_i = 1 \mid Z_i = 1 \}
\]
is totally unidentified. So, we conclude that the sharp upper bound of the term on the right--hand side of \Cref{eq:late-e1} is
\begin{equation*}
\min\bigl\{ \Pr(Y_i = 0\mid Z_i = 0), e(1) - e(0) \bigr\}.
\end{equation*}
The bound in part \textit{\ref{part3-local}} corresponds to the case where $e(1) - e(0) = 1$.   \qed

\noindent
\textbf{Proof of \Cref{thm:mte}: }
By the same reasoning as \Cref{lem:direction}, we have
\begin{equation*}
	\theta_\mte (v)
	=
	\frac{\Exp\{ Y_i(1) - Y_i(0) \mid V_i = v \}}{\Pr( Y_i(0) = 0 \mid  V_i = v \} }.
\end{equation*}
Then as shown in \citet{Heckman/Vytlacil:05},
\[
\Exp\{ Y_i(1) - Y_i(0)\mid V_i = v \} = \dfrac{ \partial \Pr \{ Y_i = 1 \mid e(Z_i) = e \}}{\partial e} \bigg|_{e=v}.
\]
Also, by the same argument as in \citet{Heckman/Vytlacil:05},
\[
\Exp\{  Y_i(0) \mid V_i = v \} = - \dfrac{ \partial \Pr\{ Y_i = 1, T_i = 0 \mid e(Z_i) = e \}}{\partial e} \bigg|_{e=v}.
\]
The desired result follows immediately.  \qed

\subsection{Proofs with Nonbinary Outcomes}\

\noindent
\textbf{Proof of \Cref{thm:mult}: } Fixing $Y_{i0}(t) = 0$ for $t=0,1$, there are only four possibilities, where one of them can be ruled out by $Y_{i1}(1) \geq Y_{i1}(0)$: this is illustrated in the following table, where the outcomes in the third row (those with $^\ast$)  have probability zero.
\begin{table}[htbp]
	\begin{tabular}{l|l}
		$Y(1)$					& 		$Y(0)$\\ \hline
		$(0,1,0)$ 	 			&		$(0,1,0)$ \\
		$(0,1,0)$ 	 			&		$(0,0,1)$ \\
		$(0,0,1)^\ast$ 	 		&		$(0,1,0)^\ast$  \\
		$(0,0,1)$ 	 			&		$(0,0,1)$ \\ \hline	
	\end{tabular}
\end{table}

Therefore,
\begin{equation*}
\Pr\{ Y_i(1) = (0,1,0), Y_i(0) = (0,0,1) \} = \Pr\{ Y_i(1) = (0,1,0) \} - \Pr\{ Y_i(0) = (0,1,0) \},
\end{equation*}
which takes care of the numerator of the conditional probability of $\theta_\mathrm{mult}$.  For the denominator, note that
\begin{equation*}
\Pr\{ Y_i(0) = (0,0,1) \}
=
\Pr\{ Y_{i,-1}(0) = 1 \}
=
1 - \Pr\{ Y_{i0}(0) = 1\} - \Pr\{ Y_{i1}(0) = 1\}. \qedhere
\end{equation*}

\begin{lemma}\label{lem:compliers mult}
	For $j\in \{0,1,-1\}$, we have
	\[
	\Pr\{ Y_{ij}(1) = 1 \mid e(0) < V_i \leq e(1) \}
	=
	\frac{p_j(1,1\mid 1) - p_j(1,1\mid 0)}{e(1) - e(0)}.
	\]
	Similarly,
	\[
	\Pr\{ Y_{ij}(0) = 1 \mid e(0) < V_i \leq e(1) \}
	=
	\frac{ p_j(1,0\mid 0) - p_j(1,0\mid 1)}{e(1) - e(0)}.
	\]
	\proof
	It follows from the same proof as \Cref{lem:compliers}.   \qed
\end{lemma}

\begin{lemma}\label{lem:no compliers mult}
	For $z = 0,1$ and $j\in \{0,1,-1\}$, we have
	\[
	\Pr\{ Y_{ij}(1) = 1, V_i \leq e(z) \}
	=
	p_j(1,1\mid z).
	\]
	Similarly,
	\[
	\Pr\{ Y_{ij}(0) = 1, V_i > e(z) \}
	=
	p_j(1,0\mid z).
	\]
	\proof
	It follows from the same proof as \Cref{lem:no compliers}. \qed
\end{lemma}

\begin{lemma}\label{lem:sharp region}
	The sharp identified region of $\bigl(\Pr\{ Y_{i1}(1) = 1\}, \Pr\{ Y_{i0}(0) = 1\}, \Pr\{ Y_{i1}(0) = 1\}\bigr)$ is given by $\mathcal{S}_1(1) \times \mathcal{S}_{01}(0)$,
	where $\mathcal{S}_1(1) = \bigl[ p_1(1\mid 1),\  p_1(1,1\mid 1) + 1-e(1) \bigr]$ is the sharp identified interval of $\Pr\{ Y_{i1}(1) = 1\}$ and
	\begin{multline*}
	\mathcal{S}_{01}(0)
	=
	\Big\{
	(x,y)\in [0,1]^2:\
	p_0(1,0\mid 0) \leq x \leq p_0(1,0\mid 0) + e(0), \\
	p_1(1,0\mid 0) \leq y \leq p_1(1\mid 0),\quad
	 x+y \leq \sum_{j=0}^1 p_j(1,0\mid 0) + e(0)
	\Bigr\} \neq \emptyset
	\end{multline*}
	is the sharp identified region of $\bigl(\Pr\{ Y_{i0}(0) = 1\}, \Pr\{ Y_{i1}(0) = 1\}\bigr)$.
	\proof
	Note that
	\begin{multline*}
	\Pr\{ Y_{i1}(1) = 1\}
	=
	\Pr\{ Y_{i1}(1) = 1, V_i\leq e(0)\} 
	+
	\Pr\{ Y_{i1}(1) = 1, e(0) < V_i\leq e(1)\} \\
	+
	\Pr\{ Y_{i1}(1) = 1, V_i > e(1)\},
	\end{multline*}
	where the first two terms on the right-hand side are identified by \Cref{lem:compliers mult,lem:no compliers mult}. Specifically,
	\begin{equation}\label{eq1:mult lemma}
	\Pr\{ Y_{i1}(1) = 1\}
	=
	p_1(1,1\mid 1)
	+
	\Pr\{ Y_{i1}(1) = 1, V_i > e(1)\}.
	\end{equation}
	Similarly, for $j\in \{0,1,-1\}$, we obtain
	\begin{equation}\label{eq2:mult lemma}
	\Pr\{ Y_{ij}(0) = 1\}
	=
	p_j(1,0\mid 0)
	+		
	\Pr\{ Y_{ij}(0) = 1, V_i \leq e(0)  \}.
	\end{equation}
	Here, note that there is no restriction on the relationship between $\tilde a:=\Pr\{ Y_{i1}(1) = 1, V_i > e(1)\}$ and $(\tilde b,\tilde c):=\bigl(\Pr\{ Y_{i0}(0) = 1, V_i \leq e(0)  \},\ \Pr\{ Y_{i1}(0) = 1, V_i \leq e(0)  \} \bigr)$.
	Therefore, it suffices to consider $\tilde a$ and $(\tilde b,\tilde c)$ separately, after which we take the Cartesian product of the two sharp identified sets. For the sharp interval of $\tilde a$, we simply combine \Cref{eq1:mult lemma} with
	\begin{equation} \label{eq:mult atilde}
    \Pr( Y_{i1} = 1, T_i = 0 \mid Z_i = 1 )
    =
    \Pr\{ Y_{i1}(0) = 1, V_i > e(1) \}
    \leq
    \Pr\{ Y_{i1}(1) = 1, V_i > e(1) \}
    \leq
    1-e(1),
    \end{equation}
	which yields $	\Pr\{ Y_{i1}(1) = 1\} \in \mathcal{S}_1(1)$.  For the sharp region of $(\tilde b,\tilde c)$, we use \Cref{eq2:mult lemma} with the fact that
	\begin{equation*}
	0
	\leq
	\Pr\{ Y_{i1}(0) = 1, V_i \leq e(0)\}
	\leq
	\Pr\{ Y_{i1}(1) = 1, V_i \leq e(0)\}
	=
	\Pr( Y_{i1} = 1, T_i = 1 \mid Z_i = 0 ).
	\end{equation*}
	Therefore, it follows that
	\begin{align}
	\Pr\{ Y_{i0}(0) = 1\} &\in \bigl[ p_0(1,0\mid 0),\  p_0(1,0\mid 0) + e(0) \bigr],  \label{multbound1} \\	
	\Pr\{ Y_{i1}(0) = 1\} &\in \bigl[ p_1(1,0\mid 0),\ p_1(1\mid 0) \bigr],   \label{multbound2}\\
	\Pr\{ Y_{i,-1}(0) = 1\} &\in \bigl[ p_{-1}(1,0\mid 0),\  p_{-1}(1,0\mid 0) + e(0) \bigr],  \label{multbound3}
	\end{align}
	where we must have
	\begin{equation}\label{eq:all choices}
	\sum_{j\in \{0,1,-1\}} \Pr\{ Y_{ij}(0) = 1 \} = 1
	\quad
	\text{and}
	\quad
	\sum_{j\in\{0,1,-1\}} p_j(1,0\mid 0) = 1-e(0).
	\end{equation}
	by \Cref{ass:mult1}.\footnote{The upper end point of the bounds in \Cref{multbound2} is no larger than the lower end point of the interval $\mathcal{S}_1(1)$ because 		
		\[
		p_1(1\mid 1) - p_1(1\mid 0)
		=
		\Pr\{ Y_{i1}(1) = 1, e(0) < V_i\leq e(1)\}
		-
		\Pr\{ Y_{i1}(0) = 1, e(0) < V_i\leq e(1)\}
		\geq
		0,
		\]
		where the inequality is by $Y_{i1}(1)\geq Y_{i1}(0)$.  \label{fn:redundancy} }  Therefore, we can rewrite \Cref{multbound3} by using \Cref{eq:all choices}.   Specifically, \Cref{multbound3} can be written as
\begin{multline*}
	\Pr\{ Y_{i,-1}(0) = 1\}
	=
	1-	\Pr\{ Y_{i0}(0) = 1\} - \Pr\{ Y_{i1}(0) = 1\} \\
\in
\bigl[
1-e(0) - p_0(1,0\mid 0) - p_1(1,0\mid 0),\ 1-e(0)- p_0(1,0\mid 0) - p_1(1,0\mid 0) + e(0)
\bigr],
\end{multline*}
which implies
\begin{equation}\label{eq:the sum}
	\Pr\{ Y_{i0}(0) = 1\} + \Pr\{ Y_{i1}(0) = 1\}
	\in
	\Bigl[
	\sum_{j=0}^1 p_j(1,0\mid 0),\ \sum_{j=0}^1 p_j(1,0\mid 0)+ e(0)
	\Bigr].
\end{equation}
But the lower bound in \Cref{eq:the sum} is implied by \Cref{multbound1,multbound2}, whereas the upper bound is not redundant. Note that $\mathcal{S}_{01}(0)$ is not empty, unless
\[
-p_0(1,0\mid 0) + \sum_{j=0}^1 p_j(1,0\mid 0) + e(0) < p_1(1,0\mid 0),
\]
which is not possible.   \qed
\end{lemma}

\begin{figure}[htbp]
	\centering
	\caption{The set of the feasible values of $(b,c)$  \label{fig:mult}}
	\begin{threeparttable}
	\begin{tabularx}{17cm}{XcX}\hspace*{3.3cm}
			\begin{tabular}{c}
				\includegraphics[scale=.55]{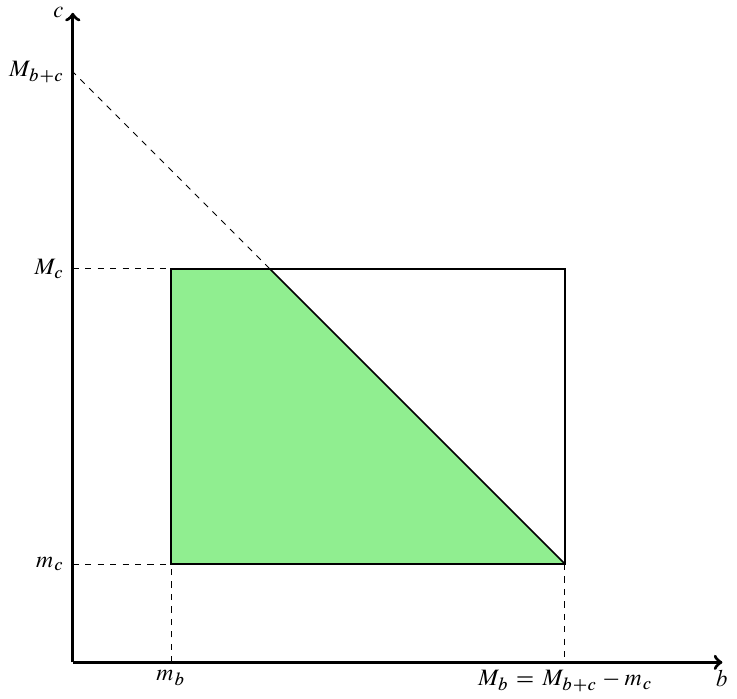}
			\end{tabular}
	\end{tabularx}
	\begin{tablenotes}[online]
			\item[Note: ] $M_{b+c} - m_b = p_0(1,0\mid 0)+p_1(1,0\mid 0)+e(0) - p_0(1,0\mid 0) \geq p_1(1,0\mid 0) + p_1(1,1\mid 0) = M_c$.
			\item[Note: ] $M_{b+c} - m_c = p_0(1,0\mid 0)+e(0) = M_b$.
			\item[Note: ] $M_{b+c}-M_b = p_0(1,0\mid 0)+p_1(1,0\mid 0) +e(0)-p_0(1,0\mid 0)-e(0) \leq  p_1(1,0\mid 0) + p_1(1,1\mid 0) = M_c$.
	\end{tablenotes}
	\end{threeparttable}
\end{figure}

\noindent
\textbf{Proof of \Cref{thm:mult id}: } Let $a:=\Pr\{ Y_{i1}(1) = 1 \}, b := \Pr\{ Y_{i0}(0) = 1 \}$, and $c := \Pr\{ Y_{i1}(0) = 1 \}$, and we have
\[
\theta_\mathrm{mult}=
\frac{a-c}{1-b-c},
\]
where the sharp identified region of $(a,b,c)$ is $\mathcal{S}_1(1)\times\mathcal{S}_{01}(0)$ by \Cref{lem:sharp region}.  Therefore, we can obtain the sharp identified bounds of $\theta_\mathrm{mult}$ by solving constrained maximization/minimization problems:
\begin{equation}\label{eq:max and min}
\max/\min_{a,b,c}\quad \frac{a-c}{1-b-c}\qquad\textrm{s.t.}\qquad
\left\{
\begin{aligned}
m_a \leq a &\leq M_a,\\
m_b\leq b&\leq M_b,\\
m_c\leq c&\leq M_c,\\
b+ c&\leq M_{b+c},
\end{aligned}
\right.
\end{equation}
where $m_a, M_a,\cdots, M_{b+c}$ are given in the definitions of $\mathcal{S}_1(1)$ and $\mathcal{S}_{01}(0)$. Here, the restriction of $a\geq c$ is automatic, because $m_a = p_1(1\mid 1) \geq p_1(1\mid 0) = M_c$ as we explained in \Cref{fn:redundancy}.  Further, note that $m_b + m_c \leq M_{b+c}$ because $e(0) \geq 0$; therefore, the set of $(b,c)$ that satisfies the constraints is not empty. In the following arguments, the set of the feasible values of $(b,c)$ is important, which is illustrated in \Cref{fig:mult}.

Consider the minimization first, for which it suffices to solve
\begin{equation*}
\min_{b,c}\quad \frac{m_a-c}{1-b-c}\qquad\textrm{s.t.}\qquad
\left\{
\begin{aligned}
m_b\leq b&\leq M_b,\\
m_c\leq c&\leq M_c,\\
b+ c&\leq M_{b+c}.
\end{aligned}
\right.
\end{equation*}
Here, the objective function is monotonically increasing in $b$ for every $c$. Therefore,
\begin{equation}\label{eq:min remove b}
\min\Bigl(\frac{m_a-c}{1-m_b-c},\ 1\Bigr) \leq \min\Bigl(\frac{m_a-c}{1-b-c},\ 1 \Bigr) \leq \theta_\textrm{mult} \leq 1
\end{equation}
for all $m_c\leq c\leq \min(M_{b+c} - m_b,\ M_c) = M_c$, where the utmost left--hand side inequality in \Cref{eq:min remove b} holds with equality when $b= m_b$.

Now, we consider two possibilities: i.e.,\ $m_a+m_b \geq 1$ and $m_a+m_b <1$.  If $m_a+m_b \geq 1$, then the utmost left--hand side of \Cref{eq:min remove b} becomes $1$ for any $c \leq 1-m_b$. But, taking $c=m_c$ leads to $m_c \leq M_{b+c} - m_b \leq 1-m_b$ by
\begin{equation*}
M_{b+c}
=
p_0(1,0\mid 0) + p_1(1,0\mid 0) + e(0)
=
1-e(0) - p_{-1}(1,0\mid 0) + e(0)
\leq 1.
\end{equation*}
Therefore, we conclude $\theta_\textrm{mult} = 1$ in this case, which is achieved when $a = m_a, b = m_b$, and $c = m_c$.

  If $m_a + m_b < 1$, then $(m_a - c)/(1-m_b-c)$ is decreasing in $c$, and therefore, we simply take the largest value of $c$ to achieve the minimum: i.e.,
\begin{equation*}
\frac{m_a-M_c}{1-m_b-M_c}
\leq
\theta_\mathrm{mult}.
\end{equation*}
This is indeed the sharp lower bound, because $(a-c)/(1-b-c)$ is a continuous function in $(a,b,c)$ and the lower bound is achieved when $a = m_a, b=m_b$, and $c = M_c$. In particular, \Cref{ass:mult1} ensures that $m_a- M_c \geq 0$, and since we assume that $p_{-1}(1\mid 0) >0$ together with \Cref{ass:mult1}, we know that $0\leq m_b + M_c = p_0(1,0\mid 0) + p_1(1\mid 0)\leq p_0(1\mid 0) + p_1(1\mid 0) = 1-p_{-1}(1\mid 0) < 1$. 

The sharp upper bound can be similarly obtained by solving the maximization problem in \Cref{eq:max and min}, for which it suffices to consider
\begin{equation*}
\max_{b,c}\quad \frac{M_a-c}{1-b-c}\qquad\textrm{s.t.}\qquad
\left\{
\begin{aligned}
m_b\leq b&\leq M_b,\\
m_c\leq c&\leq M_c,\\
b+ c&\leq M_{b+c}.
\end{aligned}
\right.
\end{equation*}
Since the objective function is increasing in $b$, it suffices to have $b$ take its largest value within the feasible set, i.e.,\ $b = \min( M_{b+c}-m_c,\ M_b ) = M_b$.

Then, we have
\begin{equation}\label{eq:max remove b}
\theta_\textrm{mult}\leq \frac{M_a -c}{1-b-c}\leq \frac{M_a-c}{1- M_b - c},
\end{equation}
for all $m_c \leq c \leq \min(M_{b+c}-m_b,\ M_c) = M_c$, where the utmost right--hand side inequality in \Cref{eq:max remove b} holds with equality when $b=M_b$.

Now, we consider two possibilities: i.e.,\ $M_a+M_b \geq 1$ and $M_a+M_b < 1$. First, suppose that $M_a + M_b\geq1$. Then, the utmost right--hand side of \Cref{eq:max remove b} is no smaller than $1$ for any $c\leq 1- M_b$. Therefore, the sharp upper bound of $\theta_\textrm{mult}$ is trivial and equal to $1$ in this case.  Suppose that $M_a+M_b < 1$. Then, the utmost right--hand side expression in \Cref{eq:max remove b} is decreasing in $c$, and therefore
\begin{equation*}
\theta_\textrm{mult}\leq \frac{M_a -c}{1-b-c}\leq \frac{M_a-m_c}{1-M_b-m_c}.
\end{equation*}
Sharpness follows from the fact that $(a-c)/(1-b-c)$ is continuous in $(a,b,c)$ and the upper bound is achieved when $a=M_a, b = M_b$, and $c = m_c$. We note that this bound is smaller than $1$, because we have $M_a - m_c < 1-M_b-m_c$ in this case, where $M_a - m_c = p_1(1,1\mid 1) + 1-e(1) - p_1(1,0\mid 0) \geq 1-e(1) + \Pr\{ Y_{i1}(0) = 1, e(0)< V_i \leq e(1) \} > 0$. 

Finally, since $m_a + m_b \leq M_a+M_b$, what we have shown so far yields the sharp upper and lower bounds in all of the three possible cases, i.e., (i) $1\leq m_a + m_b$, (ii) $m_a+m_b < 1 \leq M_a+M_b$, and (iii) $m_a+m_b\leq M_a+M_b < 1$.    \qed

\noindent
\textbf{Proof of \Cref{thm:mult id2}: } We use \Cref{lem:sharp region,lem:frechet-hoeffding} to obtain the sharp identified region of $\bigl( \Pr\{ Y_{i1}(1)=1 \}, \Pr\{ Y_{i0}(0)=1  \}, \Pr\{ Y_{i1}(0)=1  \}  \bigr)$, after which we follow the same reasoning as in the proof of \Cref{thm:mult id}.  \qed

\noindent
\textbf{Proof of \Cref{thm:mult id3}: }  The case of $p_1(1\mid 1) = 1$ is trivial, because we can simply take $e(0) = p_0(1\mid 0)$; this leads us to case (i) in \Cref{thm:mult id2}, but the claimed bounds yield the same conclusion.  Suppose that $p_1(1\mid 1)<1$.  Then, we can always choose $e(1) = \max\{ p_0(1\mid 0),\ p_1(1\mid 0) \}$ and take $e(0)$ smaller than but arbitrarily close to $e(1)$, which leads us to case (ii) in \Cref{thm:mult id2} with the claimed bounds.  \qed

\subsection{Proofs for the Results for Efficient Estimation of \texorpdfstring{{$\theta_L$}}{thetaL}}

\noindent
In this part of the appendix we consider the efficiency issues for the $\theta_L$ parameter defined in \Cref{eq:theta_L theta^*}.  So, we assume that the data available to us are i.i.d. observations of $(Y_i, Z_i, X_i^\tr)^\tr$. Below we write $f$ for $F'$, i.e.,\ the density of $X_i$. Further, we use the following notation:
\begin{align*}
\sP_{y_1\mid z_1}(x) := \Pr(Y_i = 1\mid Z_i = 1, X_i = x)
&\quad\text{and}\quad
\sP_{y_0\mid z_1}(x) := 1- \sP_{y_1\mid z_1}(x),\\
\sP_{y_1\mid z_0}(x) := \Pr(Y_i = 1\mid Z_i = 0, X_i = x)
&\quad\text{and}\quad
\sP_{y_0\mid z_0}(x) := 1- \sP_{y_1\mid z_0}(x),\\
\sP_{z_1}(x) := \Pr( Z_i = 1 \mid X_i = x)
&\quad\text{and}\quad
\sP_{z_0}(x) := 1 - \sP_{z_1}(x).
\end{align*}
Then, we can write the likelihood function as follows.
\begin{equation*}
q_L(y,z,x)
=
f(x) \prod_{j\in\{1,0\}} \Bigl[ \sP_{z_j}(x)\bigl\{ \sP_{y_1\mid z_j}(x)^y\sP_{y_0\mid z_j}(x)^{1-y} \bigr\} \Bigr]^{\tilde z_j},
\end{equation*}
where $\tilde z_1 = z$ and $\tilde z_0 = 1-z$. Therefore, the loglikelihood function is given by
\begin{equation}\label{eq:loglikelihood theta_L}
\log q_L(y,z,x)
=
\log f(x)
+
\sum_{j\in \{1,0\}}\bigl\{
\tilde z_j \log \sP_{z_j}(x) + \tilde z_j y\log \sP_{y_1\mid z_j}(x) + \tilde z_j (1-y) \sP_{y_0\mid z_j}(x)
\bigr\}.
\end{equation}

\begin{lemma}\label{lem:tangent:theta_L}
	The tangent space for $\theta_L$ is given by
	\begin{equation*}
	\mathcal{T}_L
	=\Bigl\{
	\alpha_f(x)
	+ \{z-\sP_{z_1}(x)\}\alpha_z(x)
	+ \sum_{j\in\{0,1\}} \tilde z_j \{y - \sP_{y_1\mid z_j}(x)\} \alpha_{y\mid z_j}(x)
	\Bigr\},
	\end{equation*}
	where $\alpha_f$ is any square--integrable function with $\Exp\{\alpha_f(X_i)\} = 0$ and $\alpha_z, \alpha_{y\mid z_j}$ are any square--integrable functions of $x$.
	\proof
	Let $\sP_{z_j}(x\mid \gamma), \sP_{y_k\mid z_j}(x\mid \gamma)$ denote regular parametric submodels indexed by $\gamma$:\footnote{There is no loss of generality in assuming that $\gamma$ is a scalar.} we will denote the true value by $\gamma_0$. Then, it follows from \Cref{eq:loglikelihood theta_L} that the score is given by
	\begin{equation}\label{eq:score-lower}
	s(y,z,x\mid \gamma_0)
	=
	s_X(x\mid \gamma_0)
	+
	s_{Z\mid X}(z,x\mid \gamma_0)
	+
	s_{Y\mid Z,X}(y,z,x\mid \gamma_0),
	\end{equation}
	where
	\begin{align*}
	s_X(x\mid \gamma_0)
	&=
	\frac{1}{f(x)}\frac{\partial f(x\mid \gamma)}{\partial \gamma}, \\
	s_{Z\mid X}(z,x\mid \gamma_0)
	&=
	\Bigl\{ \frac{z}{\sP_{z_1}(x)} - \frac{1-z}{\sP_{z_0}(x)} \Bigr\} \frac{\partial \sP_{z_1}(x\mid \gamma)}{\partial \gamma}, \\
	s_{Y\mid Z,X}(y,z,x\mid \gamma_0)
	&=
	\sum_{j\in\{1,0\}} \tilde z_j\Bigl\{
	\frac{y}{\sP_{y_1\mid z_j}(x)} - \frac{1-y}{\sP_{y_0\mid z_j}(x)} \Bigr\}
	\frac{\partial \sP_{y_1\mid z_j}(x\mid \gamma)}{\partial \gamma},
	\end{align*}
	where all the derivatives are evaluated at $\gamma_0$. The conclusion follows from the fact that all the derivatives are unrestricted here. \qed
\end{lemma}

For $j=0,1$, let $\beta_j(\gamma) :=\int \sP_{y_1\mid z_j}(x\mid \gamma) f(x\mid \gamma) dx$, and let $\beta_{j0} = \beta_j(\gamma_0)$.   
\begin{lemma}\label{lem:F1F0 eff}
	For any $(c_0,c_1)\neq (0,0)$, $c_0 \beta_0(\gamma)+c_1\beta_1(\gamma)$ is pathwise differentiable, and its pathwise derivative is given by $c_0 F_0(Y,Z,X) + c_1F_1(Y,Z,X)$, where for $j=0,1$,
\[
F_j(Y,Z,X)	
:=
\left( \frac{Z \bigl\{Y - \sP_{y_1\mid z_1}(X)  \bigr\}}{\sP_{z_1}(X)} \right)^j
\left( \frac{(1-Z)\bigl\{Y - \sP_{y_1\mid z_0}(X)  \bigr\}}{\sP_{z_0}(X)} \right)^{1-j}
+
\sP_{y_1\mid z_j}(X) - \beta_{j0}
\]
\proof 
First, 
\begin{equation*}
\frac{\partial \beta_j(\gamma_0)}{\partial \gamma}
=
\int \frac{\partial \sP_{y_1\mid z_j}(x\mid \gamma_0)}{\partial \gamma} f(x) dx 
+
\int \sP_{y_1\mid z_j}(x) \frac{\partial f(x\mid \gamma_0)}{\partial \gamma} dx. 
\end{equation*}
Now, note that $s(Y,Z,X\mid \gamma_0)$ given in \Cref{eq:score-lower} is the sum of Bernoulli scores, and it follows that 
\[
	\frac{\partial \beta_j(\gamma_0)}{\partial \gamma}
	=
	\Exp\bigl\{ F_j(Y,Z,X) s(Y,Z,X\mid \gamma_0)  \bigr\}.  \qedhere
\] 
\end{lemma}

\begin{lemma}\label{lem:pathwise theta_L}
The efficient influence function of $\theta_L$ is given by 
\[
F_L(Y,Z,X)	
:=
\frac{1}{1-\beta_{00}}
\left\{  
F_1(Y,Z,X)
-
\frac{ 1 - \beta_{10}}{1-\beta_{00}} F_0(Y,Z,X)
\right\}.
\]
\proof 
For any $(c_1,c_0) \neq (0,0)$, $c_1 F_1 + c_0 F_0$ is in the tangent space $\mathcal{T}_L$: i.e., its projections onto $\mathcal{T}_L$ is just itself. Also, the scores given in \Cref{eq:score-lower} can approximate any mean zero random variable with an arbitrarily small mean squared error. Therefore, Theorem 2.1 in \citet{newey1994asymptotic} and \cref{lem:F1F0 eff} ensure that the efficient influence function of $\sum_{j=0}^1 c_j \beta_{j0}$ is given by $\sum_{j=0}^1 c_j F_j(Y,Z,X)$. So, the efficient influence function of $[\beta_{10}\ \beta_{00}]^\tr$ is given by $\bigl[ F_1(Y,Z,X)\ F_0(Y,Z,X) \bigr]^\tr$. We now note that $\theta_L = (\beta_{10} - \beta_{00})/\{1-\beta_{00} \}$, and we apply the delta method.    \qed
\end{lemma}

\subsection{Proofs for the Results for Efficient Estimation of 
\texorpdfstring{{$\theta^*$}}{thetastar}}

\noindent
We now derive the efficient influence function of the local persuasion parameter defined in \Cref{eq:theta_L theta^*} when an i.i.d. sample of $(Y_i, T_i, Z_i,X_i)$ is available. Similarly to the previous subsection we use the following notation:
\begin{equation*}
\sP_{z_1}(x) := \Pr(Z_i=1\mid X_i=x)
\quad \text{and}\quad
\sP_{z_0} := 1-\sP_{z_1}(x),
\end{equation*}
\begin{align*}
&\sP_{t_1\mid z_1}(x) := \Pr(T_i=1\mid Z_i = 1,X_i=x)
\quad \text{and}\quad
\sP_{t_0\mid z_1}(x) := 1- \sP_{t_1\mid z_1}(x), \\
&\sP_{t_1\mid z_0}(x) := \Pr(T_i=1\mid Z_i = 0,X_i=x)
\quad \text{and}\quad
\sP_{t_0\mid z_0}(x) := 1-\sP_{t_1\mid z_0}(x),
\end{align*}
\begin{align*}
&\sP_{y_1\mid t_1,z_1}(x) := \Pr(Y_i=1\mid T_i=1,Z_i = 1,X_i=x)
\quad \text{and}\quad
\sP_{y_0\mid t_1,z_1}(x) := 1-\sP_{y_1\mid t_1,z_1}(x),\\
&\sP_{y_1\mid t_0,z_1}(x) := \Pr(Y_i=1\mid T_i=0,Z_i = 1,X_i=x)
\quad \text{and}\quad
\sP_{y_0\mid t_0,z_1}(x) := 1-\sP_{y_1\mid t_0,z_1}(x),\\
&\sP_{y_1\mid t_1,z_0}(x) := \Pr(Y_i=1\mid T_i=1,Z_i = 0,X_i=x)
\quad \text{and}\quad
\sP_{y_0\mid t_1,z_0}(x) := 1-\sP_{y_1\mid t_1,z_0}(x),\\
&\sP_{y_1\mid t_0,z_0}(x) := \Pr(Y_i=1\mid T_i=0,Z_i = 0,X_i=x)
\quad \text{and}\quad
\sP_{y_0\mid t_0,z_0}(x) := 1-\sP_{y_1\mid t_0,z_0}(x).
\end{align*}
Using this notation, the likelihood function can be written as
\begin{multline*}
q^*(y,t,z,x)
=
f(x)
\prod_{j=\{1,0\}}
\Bigl[
\sP_{z_j}(x)
\prod_{k=\{1,0\}}
\bigl\{
\sP_{t_k\mid z_j}(x) \prod_{\ell = \{1,0\}} \sP_{y_\ell \mid t_k,z_j}(x)^{\tilde y_\ell}
\bigr\}^{\tilde t_k}
\Bigr]^{\tilde z_j} \\
=
f(x)
\prod_{j=\{1,0\}}
\Bigl[
\sP_{z_j}(x)^{\tilde z_j}
\prod_{k=\{1,0\}}
\bigl\{
\sP_{t_k\mid z_j}(x)^{\tilde t_k \tilde z_j} \prod_{\ell = \{1,0\}} \sP_{y_\ell\mid t_k,z_j}(x)^{\tilde y_\ell \tilde t_k \tilde z_j }
\bigr\}
\Bigr],
\end{multline*}
where $\tilde z_1 = z, \tilde z_0 = 1-z$, $\tilde t_1 = t, \tilde t_0 = 1-t$, and $\tilde y_1 = y, \tilde y_0 = 1-y$. Therefore, the loglikelihood function is given by
\begin{multline}\label{eq:loglikelihood theta*}
\log q^*(y,t,z,x)
=
\log f(x)
+
\sum_{j\in\{1,0\}} \tilde z_j \log \sP_{z_j}(x) \\
+
\sum_{j\in\{1,0\}}\sum_{k\in\{0,1\}} \tilde t_k \tilde z_j \log \sP_{t_k\mid z_j}(x)
+
\sum_{\ell\in\{1,0\}} \sum_{j\in\{1,0\}}\sum_{k\in\{0,1\}} \tilde y_\ell \tilde t_k \tilde z_j \log \sP_{y_\ell\mid t_k,z_j}(x).
\end{multline}

\begin{lemma}
	The tangent space for $\theta^*$ is given by
	\begin{multline*}
	\mathcal{T}^* 
    =
	\Bigl\{
		\alpha_f(x)
		+
		\{ z-\sP_{z_1}(x) \}\alpha_z(x)
		+
		\sum_{j\in\{0,1\}} \tilde z_j\{t-\sP_{t_1\mid z_j}(x)\}\alpha_{t\mid z_j}(x) \\
		+
		\sum_{j\in\{0,1\}}\sum_{k\in\{0,1\}}\tilde t_k\tilde z_j\{ y-\sP_{y_1\mid t_k,z_j}(x)\}\alpha_{y\mid t_kz_j}(x)
	\Bigr\},
	\end{multline*}
	where $\alpha_f$ is any square--integrable function with $\Exp\{ \alpha_f(X_1)\} = 0$, and $\alpha_z, \alpha_{t\mid z_j}, \alpha_{y\mid t_k,z_j}$ are any square--integrable functions of $x$.
	\proof
	Let $\sP_{z_j}(x\mid \gamma), \sP_{t_k\mid z_j}(x\mid \gamma), \sP_{y_\ell\mid t_k,z_j}(x\mid \gamma)$ denote regular parametric submodels indexed by $\gamma$: as in the proof of \Cref{lem:tangent:theta_L}, $\gamma$ is a scalar--valued parameter and its true value is denoted by $\gamma_0$. From the loglikelihood function described in \Cref{eq:loglikelihood theta*}, we know that the score of the regular parametric submodel can be written as follows:
	\begin{multline}\label{eq:score-local}
	s^*(y,t,z,x\mid \gamma_0)
	=
	\frac{1}{f(x)}\frac{\partial f(x\mid \gamma)}{\partial\gamma}
	+
	\Bigl\{ \frac{\tilde z_1}{\sP_{z_1}(x)} - \frac{\tilde z_0}{\sP_{z_0}(x)}   \Bigr\}\frac{\partial \sP_{z_1}(x\mid \gamma)}{\partial \gamma}  \\
	+
	\sum_{j\in\{1,0\}}
	\tilde z_j \Bigl\{ \frac{\tilde t_1}{\sP_{t_1\mid z_j}(x)} - \frac{\tilde t_0}{\sP_{t_0\mid z_j}(x)}   \Bigr\}\frac{\partial \sP_{t_1\mid z_j}(x\mid \gamma)}{\partial \gamma} \\
	+\sum_{j\in\{1,0\}}\sum_{k\in\{1,0\}}
	\tilde t_k \tilde z_j \Bigl\{ \frac{\tilde y_1 }{\sP_{y_1\mid t_k,z_j}(x)} - \frac{\tilde y_0}{\sP_{y_0\mid t_k,z_j}(x)}   \Bigr\}\frac{\partial \sP_{y_1\mid t_k,z_j}(x\mid \gamma)}{\partial \gamma},
	\end{multline}
	where all the derivatives are evaluated at $\gamma_0$.  Here, we can do further algebra by using
	\begin{align*}
	\frac{\partial \sP_{z_1}(x\mid \gamma)}{\partial\gamma}
	&=
	- \frac{\partial \sP_{z_0}(x\mid \gamma)}{\partial\gamma},\\
	\frac{\partial \sP_{t_1\mid z_j}(x\mid \gamma)}{\partial\gamma}
	&=
	- \frac{\partial \sP_{t_0\mid z_j}(x\mid \gamma)}{\partial\gamma} \quad \text{for $j=0,1$}\\
	\frac{\partial \sP_{y_1\mid t_k,z_j}(x\mid \gamma)}{\partial\gamma}
	&=
	- \frac{\partial \sP_{y_0\mid t_k,z_j}(x\mid \gamma)}{\partial\gamma}\quad \text{for $j,k=0,1$}.
	\end{align*}
	Therefore, the score in \cref{eq:score-local} can be rewritten as follows:
	\begin{equation}\label{eq:score}
	s^*(y,t,z,x\mid \gamma_0)
	=
	s_X(x\mid \gamma_0)
	+
	s_{Z\mid X}(z,x\mid \gamma_0)
	+
	s_{T\mid Z,X}(t,z,x\mid \gamma_0)
	+
	s_{Y\mid T,Z,X}(y,t,z,x\mid \gamma_0),
	\end{equation}
	where
	\begin{align*}
	s_X(x\mid \gamma_0)
	&:=
	\frac{1}{f(x)}\frac{\partial f(x\mid \gamma)}{\partial\gamma}, \\
	s_{Z\mid X}(z,x\mid \gamma_0)
	&:=
	\frac{z - \sP_{z_1}(x)}{ \sP_{z_1}(x)\bigl\{ 1-\sP_{z_1}(x)\bigr\}}
	\frac{\partial \sP_{z_1}(x\mid \gamma)}{\partial \gamma}, \\
	s_{T\mid Z,X}(t,z,x\mid \gamma_0)
	&:=
	\sum_{j\in\{1,0\}}
	\frac{\tilde z_j \bigl\{ t-\sP_{t_1\mid z_j}(x) \bigr\}}{\sP_{t_1\mid z_j}(x)\bigl\{1-\sP_{t_1\mid z_j}(x)\bigr\}}
	\frac{\partial \sP_{t_1\mid z_j}(x\mid \gamma)}{\partial \gamma},\\
	s_{Y\mid T,Z,X}(y,t,z,x\mid \gamma_0)
	&:=
	\sum_{j\in\{1,0\}}\sum_{k\in\{1,0\}}
	\frac{\tilde t_k \tilde z_j \bigl\{ y -\sP_{y_1\mid t_k,z_j}(x) \bigr\}}{\sP_{y_1\mid t_k,z_j}(x) \bigl\{1-\sP_{y_1\mid t_k,z_j}(x) \bigr\}}
	\frac{\partial \sP_{y_1\mid t_k,z_j}(x\mid \gamma)}{\partial \gamma}.
	\end{align*}
	Interpretation of each term should be straightforward. For example, $s_{T\mid Z,X}(t,z,x\mid \gamma_0)$ is the score of $T$ at $t$ conditional on $Z=z,X=x$. Finally, the conclusion follows from \Cref{eq:score} and the fact that all the derivatives here are unrestricted. \qed
	
\end{lemma}
Now, define 
\begin{align*} 
\beta_N^* 
&:=
\int Q_N(x) f(x) dx, \\
\beta_D^*
&:=
\int Q_D(x) f(x) dx,
\end{align*}
where
\begin{align*}
	Q_N(x)
	&:=
	\sum_{k\in\{1,0\}}\sP_{y_1\mid t_k,z_1}(x)\sP_{t_k\mid z_1}(x)
	-
	\sum_{k\in\{1,0\}}\sP_{y_1\mid t_k,z_0}(x)\sP_{t_k\mid z_0}(x), \\
	Q_D(x)
	&:=
	\sP_{y_0\mid t_0,z_0}(x)\sP_{t_0\mid z_0}(x)
	-
	\sP_{y_0\mid t_0,z_1}(x)\sP_{t_0\mid z_1}(x). 
\end{align*}
Then, $\theta^*$ given in \Cref{eq:theta_L theta^*} is equal to $\beta_N^*/\beta_D^*$. Below we derive the efficient influence functions of $\beta_N^*$ and $\beta_D^*$ jointly, after which we will use the delta method to obtain that of $\theta^*$. Let $\beta_N^*(\gamma), \beta_D^*(\gamma), Q_N(x\mid \gamma)$, and $Q_D(x\mid \gamma)$ be the perturbed versions of $\beta_N^*, \beta_D^*, Q_N(x)$, and $Q_D(x)$ along regular parametric submodels, respectively. 

Define the following functions: 
\begin{align*} 
F_{1N}^*(Y,T,Z,X)
&=
\frac{TZ}{\sP_{t_1\mid z_1}(X) \sP_{z_1}(X)}\bigl\{ Y - \sP_{y_1\mid t_1,z_1}(X)   \bigr\} \sP_{t_1\mid z_1}(X)
\\
F_{2N}^*(Y,T,Z,X)
&=
\frac{(1-T)Z}{\sP_{t_0\mid z_1}(X) \sP_{z_1}(X)}\bigl\{ Y - \sP_{y_1\mid t_0,z_1}(X)   \bigr\} \sP_{t_0\mid z_1}(X)
\\
F_{3N}^*(Y,T,Z,X)
&=
-
\frac{T(1-Z)}{\sP_{t_1\mid z_0}(X) \sP_{z_0}(X)}\bigl\{ Y - \sP_{y_1\mid t_1,z_0}(X)   \bigr\} \sP_{t_1\mid z_0}(X) 
\\
F_{4N}^*(Y,T,Z,X)
&=
-
\frac{(1-T)(1-Z)}{\sP_{t_0\mid z_0}(X) \sP_{z_0}(X)}\bigl\{ Y - \sP_{y_1\mid t_0,z_0}(X)   \bigr\} \sP_{t_0\mid z_0}(X)
\\
F_{5N}^*(Y,T,Z,X)
&=
\frac{Z}{\sP_{z_1}(X)} \bigl\{ T - \sP_{t_1\mid z_1}(X) \bigr\}
		\bigl\{ \sP_{y_1\mid t_1,z_1}(X) - \sP_{y_1\mid t_0,z_1}(X)  \bigr\} 
\\
F_{6N}^*(Y,T,Z,X)
&=
-
\frac{1-Z}{\sP_{z_0}(X)}\bigl\{ T-\sP_{t_1\mid z_0}(X)  \bigr\}
		\bigl\{ \sP_{y_1\mid t_1,z_0}(X) - \sP_{y_1\mid t_0,z_0}(X)	\bigr\}
\\
F_{7N}^*(Y,T,Z,X) 
&= 
Q_N(X) - \beta_N^*
\end{align*}
Also, define
\begin{align*} 
	F_{1D}^*(Y,T,Z,X)
	&=
	\frac{(1-T)Z}{\sP_{t_0\mid z_1}(X) \sP_{z_1}(X)}\bigl\{ Y - \sP_{y_1\mid t_0,z_1}(X) \bigr\} \sP_{t_0\mid z_1}(X)
	\\
	F_{2D}^*(Y,T,Z,X)
	&=
	-
	\frac{(1-T)(1-Z)}{\sP_{t_0\mid z_0}(X) \sP_{z_0}(X)}\bigl\{ Y - \sP_{y_1\mid t_0,z_0}(X) \bigr\} \sP_{t_0\mid z_0}(X)
	\\
	F_{3D}^*(Y,T,Z,X)
	&=
	\frac{Z}{\sP_{z_1}(X)}\bigl\{ T - \sP_{t_1\mid z_1}(X) \bigr\}\sP_{y_0\mid t_0,z_1}(X)
	\\
	F_{4D}^*(Y,T,Z,X)
	&= 
	-
	\frac{1-Z}{\sP_{z_0}(X)}\bigl\{ T - \sP_{t_1\mid z_0}(X) \bigr\} \sP_{y_0\mid t_0,z_0}(X)
	\\
	F_{5D}^*(Y,T,Z,X)
	&=
	Q_D(X) - \beta_D^*.
\end{align*}

\begin{lemma}\label{lem:FnFd eff}
For any $(c_N, c_D) \neq (0,0)$, $c_N\beta_N^*(\gamma) + c_D \beta_D^*(\gamma)$ is pathwise differentiable, and its pathwise derivative is given by $c_N F_N^*(Y,T,Z,X) + c_D F_D^*(Y,T,Z,X)$, where $F_N^*(Y,T,Z,X) := \sum_{j=1}^7 F_{jN}^*(Y,T,Z,X)$ and $F_D^*(Y,T,Z,X):= \sum_{j=1}^5 F_{jD}^*(Y,T,Z,X)$. 
\proof 
We first calculate the derivatives of the $Q$ functions with respect to $\gamma$ (evaluated at $\gamma_0$). For this calculation, there are only seven relevant derivatives, i.e.,
\begin{align*}
	&\frac{\partial\sP_{y_1\mid t_1,z_1}(x\mid \gamma)}{\partial\gamma},
	\frac{\partial\sP_{y_1\mid t_0,z_1}(x\mid \gamma)}{\partial\gamma},
	\frac{\partial\sP_{y_1\mid t_1,z_0}(x\mid \gamma)}{\partial\gamma},
	\frac{\partial\sP_{y_1\mid t_0,z_0}(x\mid \gamma)}{\partial\gamma}, \\
	&\frac{\partial\sP_{t_1\mid z_1}(x\mid \gamma)}{\partial\gamma},
	\frac{\partial\sP_{t_1\mid z_0}(x\mid \gamma)}{\partial\gamma}, \\
	&\frac{\partial f(x\mid \gamma)}{\partial\gamma}.
\end{align*}
Indeed, 
\begin{align*}
	\frac{\partial Q_N(x\mid \gamma)}{\partial \gamma}
	&=
	\sum_{k\in \{1,0\}}
	\frac{\partial \sP_{y_1\mid t_k,z_1}(x\mid \gamma)}{\partial \gamma}\sP_{t_k\mid z_1}(x)
	+
	\Bigl\{ \sP_{y_1\mid t_1,z_1}(x) - \sP_{y_1\mid t_0,z_1}(x)  \Bigr\} \frac{\partial \sP_{t_1\mid z_1}(x\mid \gamma)}{\partial \gamma} \\
	&\hspace{1.5cm}
	-\sum_{k\in \{1,0\}}
	\frac{\partial \sP_{y_1\mid t_k,z_0}(x\mid \gamma)}{\partial \gamma}\sP_{t_k\mid z_0}(x)
	-
	\Bigl\{ \sP_{y_1\mid t_1,z_0}(x) - \sP_{y_1\mid t_0,z_0}(x)  \Bigr\} \frac{\partial \sP_{t_1\mid z_0}(x\mid \gamma)}{\partial \gamma},
\end{align*} 
and 
\begin{align*}
	\frac{\partial Q_D(x\mid \gamma)}{\partial \gamma}
	&=
	\frac{\partial \sP_{y_1\mid t_0,z_1}(x\mid \gamma)}{\partial \gamma}\sP_{t_0\mid z_1}(x)
	-\frac{\partial \sP_{y_1\mid t_0,z_0}(x\mid \gamma)}{\partial \gamma}\sP_{t_0\mid z_0}(x) \\
	&\hspace{5cm}
	+ \sP_{y_0\mid t_0,z_1} \frac{\partial \sP_{t_1\mid z_1}(x\mid \gamma)}{\partial \gamma}
	- \sP_{y_0\mid t_0,z_0} \frac{\partial \sP_{t_1\mid z_0}(x\mid \gamma)}{\partial \gamma}.
\end{align*}

We now consider
\begin{equation*}
	\partial \beta_N^*(\gamma_0)/\partial\gamma
    =
    \int \frac{\partial Q_N(x\mid \gamma)}{\partial \gamma} f(x) dx, 
	+
    \int Q_N(x) \frac{\partial f(x)}{\partial \gamma} dx,
\end{equation*}
which is the sum of the following terms: 
\begin{align*} 
	& \int \sP_{t_1\mid z_1}(x) f(x) \frac{\partial \sP_{y_1\mid t_1,z_1}(x\mid \gamma)}{\partial \gamma} dx, \\
	& \int \sP_{t_0\mid z_1}(x) f(x) \frac{\partial \sP_{y_1\mid t_0,z_1}(x\mid \gamma)}{\partial \gamma} dx, \\
  - & \int \sP_{t_1\mid z_0}(x) f(x) \frac{\partial \sP_{y_1\mid t_1,z_0}(x\mid \gamma)}{\partial \gamma} dx, \\
  - & \int \sP_{t_0\mid z_0}(x) f(x) \frac{\partial \sP_{y_1\mid t_0,z_0}(x\mid \gamma)}{\partial \gamma} dx, \\
	&
	\int
		\bigl\{ \sP_{y_1\mid t_1,z_1}(x) - \sP_{y_1\mid t_0,z_1}(x)  \bigr\} 
	 f(x) \frac{\partial \sP_{t_1\mid z_1}(x\mid \gamma)}{\partial \gamma} dx,\\
  -&
  	\int
		\bigl\{ \sP_{y_1\mid t_1,z_0}(x) - \sP_{y_1\mid t_0,z_0}(x)	\bigr\}
	  f(x) \frac{\partial \sP_{t_1\mid z_0}(x\mid \gamma)}{\partial \gamma} dx, \\
	& \int Q_N(x) \frac{\partial f(x\mid \gamma_0)}{\partial \gamma} dx.
\end{align*}
Therefore, $F_N^*(Y,T,Z,X) := \sum_{j=1}^N F_{jN}^*(Y,T,Z,X)$ satisfies 
\[
	\partial \beta_N^*(\gamma_0)/\partial\gamma
	=
	\Exp\{ 	F_N^*(Y,T,Z,X)s^*(Y,T,Z,X\mid \gamma_0) \},
\]
where $s^*$ is defined in \Cref{eq:score}: in fact, this equation can be seen immediately from the fact that $s^*$ is the sum of Bernoulli scores.\footnote{To see this point, it is helpful to consider a simple binary example. For instance, consider a generic binary variable $B$ such that $\Pr(B=1\mid X,\gamma) = p_B(X\mid \gamma)$. Suppose that the parameter $\beta$ satisfies $\partial\beta(\gamma)/\partial \gamma = \int A(x) \partial p_B(X\mid \gamma)/\partial\gamma f(x) dx = \Exp\{ A(X) \partial p_B(X\mid \gamma)/\partial\gamma \}$. Here, the score of $B$ given $X$ is
\[
s_B(B\mid X) = \frac{B - p_B(X)}{ p_B(X)\{ 1-p_B(X)\} } \frac{\partial p_B(X\mid \gamma)}{\partial \gamma}.
\]
Now, we are looking for the function $G^*(B,X)$ such that
\[
\Exp\{ G^*(B,X)s_B(B\mid X) \} = \Exp\Bigl\{ A(X) \frac{\partial p_B(X\mid \gamma)}{\partial\gamma} \Bigr\},
\]
where
\[
\Exp\{ G^*(B,X) s_B(B\mid X) \}
=
\Exp\Bigl[ \bigl\{ G^*(1,X)-G^*(0,X)\bigr\} \frac{\partial p_B(X\mid \gamma)}{\partial\gamma} \Bigr].
\]
Therefore, we immediately see that $G^*(B,X) = \{B - p_B(X)\} A(X)$ does the job. }

Now, we turn to
\[
\partial \beta_D^*(\gamma_0)/\partial \gamma
=
\int \frac{\partial Q_D(x\mid \gamma)}{\partial \gamma} f(x) dx
+
\int Q_D(x) \frac{\partial f(x\mid \gamma_0)}{\partial \gamma} dx,
\] 
which is the sum of the following terms: 
\begin{align*} 
	& \int \sP_{t_0\mid z_1}(x) f(x) 
		\frac{\partial \sP_{y_1\mid t_0,z_1}(x\mid \gamma_0)}{\partial \gamma} dx \\
  - & \int \sP_{t_0\mid z_0}(x) f(x) 
  		\frac{\partial \sP_{y_1\mid t_0,z_0}(x\mid \gamma_0)}{\partial \gamma} dx \\
	& \int \sP_{y_0\mid t_0,z_1}(x) f(x) 
		  \frac{\partial \sP_{t_1\mid z_1}(x\mid \gamma_0)}{\partial \gamma} dx  \\ 
  -	& \int \sP_{y_0\mid t_0,z_0}(x) f(x) 
  		\frac{\partial \sP_{t_1\mid z_0}(x\mid \gamma_0)}{\partial \gamma} dx \\
	& \int Q_D(x) \frac{\partial f(x\mid \gamma_0)}{\partial \gamma} dx.
\end{align*} 
Therefore, $F_D(Y,T,Z,X) := \sum_{j=1}^5 F_{jD}^*(Y,T,Z,X)$ satisfies 
\[
	\frac{\partial \beta_D^*(\gamma_0)}{\partial \gamma}
	=
	\Exp\bigl\{ F_D^*(Y,T,Z,X) s^*(Y,T,Z,X\mid \gamma_0) \bigr\},
\]
where $s^*$ is defined in \Cref{eq:score}. \qed 
\end{lemma}

\begin{lemma}\label{lem:pathwise theta*}
  The efficient influence function of $\theta^*$ is given by 
  \[
	F^*(Y,T,Z,X)
	:=
	\frac{1}{\beta_D^*}
	\bigl\{ F_N^*(Y,T,Z,X) - \theta^* F_D^*(Y,T,Z,X) \bigr\}
  \]
\proof 
For any $(c_N,c_D) \neq 0$, $c_N F_N^* + c_D F_D^*$ is in the tangent space $\mathcal{T}^*$: i.e., its projection onto $\mathcal{T}^*$ is just itself. Also, the scores given in \Cref{eq:score} can approximate any mean zero random variable with an arbitrarily small mean squared error. Therefore, Theorem 2.1 in \citet{newey1994asymptotic} and \cref{lem:FnFd eff} ensure that the efficient influence function of $c_N \beta_N^* + c_D \beta_D^*$ is given by $c_N F_N^*(Y,T,Z,X) + c_D F_D^*(Y,T,Z,X)$. So, the efficient influence function of $[\beta_N\ \beta_D]^\tr$ is given by $\bigl[ F_N(Y,T,Z,X)\ F_D(Y,T,Z,X) \bigr]^\tr$. We now note that $\theta^* = \beta_N^*/\beta_D^*$, and we apply the delta method.     \qed
\end{lemma}

\noindent
\textbf{Proof of \Cref{thm:efficiency}} It follows from \Cref{lem:F1F0 eff,lem:FnFd eff}. \qed

\noindent
\textbf{Proof of \Cref{thm:inference}} It follows from \Cref{lem:pathwise theta_L,lem:pathwise theta*}.  \qed

%

\bibliographystyle{econometrica}
\bibliography{persuasion}

\end{document}